\documentclass[final,onefignum,onetabnum]{siamonline220329}

\pdfoutput=1

\usepackage{lipsum}
\usepackage{seqsplit, xstring}

\usepackage{mauraisMacros}
\usepackage{enumitem}
\usepackage{booktabs} 
\usepackage{amsfonts, amsmath, amssymb, mathrsfs, mathtools}
\usepackage{graphicx}
\usepackage{epstopdf}
\usepackage{bold-extra}
\usepackage{float} 
\usepackage{algorithmic}
\usepackage{arydshln}
\ifpdf
  \DeclareGraphicsExtensions{.eps,.pdf,.png,.jpg}
\else
  \DeclareGraphicsExtensions{.eps}
\fi

\newcommand{\rmT}{\mathrm{T}}

\newcommand{\Shi}{{S_{\rm hi}}}
\newcommand{\Slo}{{S_{\rm lo}}}

\renewcommand{\chi}{{c_{\rm hi}}}

\newcommand{\Xhi}{{X_{\rm hi}}}
\newcommand{\Xlo}{{X_{\rm lo}}}

\newcommand{\Sigmahi}{{\Sigma_{\rm hi}}}
\newcommand{\Sigmalo}{{\Sigma_{\rm lo}}}

\newcommand{\calShi}{{\calS_{\rm hi}}}
\newcommand{\calSlo}{{\calS_{\rm lo}}}
\newcommand{\calShihat}{{\hat{\calS}_{\rm hi}}}

\newcommand{\Sigmahihat}{{\hat{\Sigma}_{\text{hi}}}}
\newcommand{\Sigmalohat}{{\hat{\Sigma}_{\text{lo}}}}

\newcommand{\SloOne}{{S^1_{\text{lo}}}}
\newcommand{\SloTwo}{{S^2_{\text{lo}}}}
\newcommand{\Slobar}{{\bar{S}_{\text{lo}}}}

\newcommand{\bfSreal}{\text{\large\textbf{\texttt{S}}}}
\newcommand{\Sreal}{\text{\large\texttt{S}}}

\newcommand{\SrealloOne}{{\Sreal^1_{\text{lo}}}}
\newcommand{\SrealloTwo}{{\Sreal^2_{\text{lo}}}}

\newcommand{\sfI}{\mathsf{I}}
\newcommand{\bfsfI}{\bm{\sfI}}

\newcommand{\frakD}{\mathfrak{D}}

\DeclareMathSymbol{\Rho}{\mathalpha}{operators}{"50}
\DeclareMathAlphabet{\mathpzc}{OT1}{pzc}{m}{it}
\DeclareMathSymbol{\shortminus}{\mathbin}{AMSa}{"39}

\newcommand{\indep}{\perp \!\!\! \perp}

\newsiamremark{remark}{Remark}
\newsiamremark{hypothesis}{Hypothesis}
\newsiamremark{example}{Example}
\crefname{hypothesis}{Hypothesis}{Hypotheses}
\newsiamthm{claim}{Claim}

\headers{Multifidelity Covariance Estimation via Regression}{} %

\title{Multifidelity Covariance Estimation via Regression on the Manifold of Symmetric Positive Definite Matrices\thanks{Submitted to the editors 8 August 2023.
\funding{AM and YM were supported by the Office of Naval Research, SIMDA (Sea Ice Modeling and Data Assimilation) MURI, award number N00014-20-1-2595 (Dr.~Reza Malek-Madani and Dr.~Scott Harper). AM was additionally supported by the NSF Graduate Research Fellowship under Grant No.\ 1745302. TA and BP were supported by AFOSR under Award Number FA9550-21-1-0222 (Dr.~Fariba Fahroo)}}}

\author{Aimee Maurais\thanks{Center for Computational Science and Engineering, MIT, Cambridge, MA 
  (\email{maurais@mit.edu}, \email{ymarz@mit.edu}).}
\and Terrence Alsup\thanks{Courant Institute of Mathematical Sciences, NYU, New York, NY 
  (\email{alsup.terrence@gmail.com}, \email{pehersto@cims.nyu.edu}).}
\and Benjamin Peherstofer\footnotemark[3] 
\and Youssef Marzouk\footnotemark[2]}

\usepackage{amsopn}

\usepackage[textsize=tiny]{todonotes}
\definecolor{darkgreen}{rgb}{.15,.55,0}

\newcommand{\rev}[1]{\textcolor{black}{#1}}

\makeatletter
\newcommand*{\addFileDependency}[1]{%
  \typeout{(#1)}%
  \@addtofilelist{#1}%
  \IfFileExists{#1}{}{\typeout{No file #1.}}%
}
\makeatother

\graphicspath{{figures/}}

\ifpdf
\hypersetup{
  pdftitle={Multifidelity Covariance Estimation via Regression on the Manifold of Symmetric Positive Definite Matrices},
  pdfauthor={A.\@ Maurais, T.\@ Alsup, B.\@ Peherstorfer, Y.\@ Marzouk}
}
\fi

\makeatletter 
\@mparswitchfalse%
\makeatother
\normalmarginpar

\begin{document}

\maketitle
\begin{abstract} 
We introduce a multifidelity estimator of covariance matrices formulated as the solution to a regression problem on the manifold of symmetric positive definite matrices. The estimator is positive definite by construction, and the Mahalanobis distance minimized to obtain it possesses properties enabling practical computation. We show that our manifold regression multifidelity (MRMF) covariance estimator is a maximum likelihood estimator under a certain error model on manifold tangent space. More broadly, we show that our Riemannian regression framework encompasses existing multifidelity covariance estimators constructed from control variates.  We demonstrate via numerical examples that the MRMF estimator can provide significant decreases, up to one order of magnitude, in squared estimation error relative to both single-fidelity and other multifidelity covariance estimators. Furthermore, preservation of positive definiteness ensures that our estimator is compatible with downstream tasks, such as data assimilation and metric learning, in which this property is essential.
\end{abstract}



\section{Introduction}
In the words of \cite{ledoit2022power}, the covariance matrix is ``arguably the second most important object in all of statistics.'' Covariance matrices are key objects in portfolio theory \cite{markowitz1952portfolio} and spatial statistics \cite{cressie2015statistics}. In Bayesian inference, covariance matrices are the essential elements of prior distributions for inverse problems \cite{kaipio2006statistical}, and they dictate the extent to which predictive models are corrected by observations in Kalman-type data assimilation \cite{evensenekf, kalmanfilter} and inversion \cite{iglesias2013ensemble} schemes. In data science and machine learning, covariance matrices underlie principal component analysis \cite{ringner2008principal}, perhaps the most canonical method for dimension reduction, and arise in downstream tasks such as metric learning \cite{kulis2013metric}.
 
Covariance matrices are usually estimated from data, and often the biggest hurdle to doing so effectively is insufficient sample information: 
in many applications, data are expensive and the number of samples we can practically obtain may be on the order of the parameter dimension. For this reason there has been extensive development of regularization methods for covariance estimation in the small-sample regime, including 
shrinkage \cite{ledoit2022power,ledoit2012nonlinear,donohogavishjohnstone}, enforcement of sparsity in the precision matrix \cite{bickellevina,graphicallasso}, and localization/tapering \cite{gasparicohn,localizationEnKF}. 

Small-sample covariance estimators generally assume access to a low number of identically distributed samples, which in the context of computational and data science we might associate with the output of a single computational model. We refer to this model as the \textit{high fidelity} model and assume that, in addition to being costly to evaluate, it retains a high level of veracity to the underlying process it seeks to capture. For example, a high-fidelity model may correspond to computationally intensive dynamic simulations as encountered in numerical weather prediction and aerodynamic modeling. 
In such physically-driven applications, however, we seldom have only \textit{one} model at our disposal. Rather, we may additionally have any number of \emph{lower-fidelity models} obtained, e.g., via coarser discretizations, machine learning surrogates, or reduced physics approximations of the high-fidelity model. Lower-fidelity models are generally cheaper to sample but less accurate than the high-fidelity model. 

Scarcity of high-fidelity samples is an issue for many tasks in computational and data science, and for this reason a wide range of \textit{multifidelity} and \textit{multilevel} methods \cite{giles2015multilevel,cliffe_multilevel_2011,NME:NME4761,PWK16MFMCAsymptotics,GORODETSKY2020109257} have been developed to exploit the model hierarchies that exist in applications. The idea is simple: rather than devoting all computational resources to high-fidelity model evaluations, we judiciously allocate our budget among evaluations of high \textit{and} lower-fidelity models, and in doing so achieve better performance for the same cost. The scope and range of applications of multifidelity methods are vast, and we do not attempt to summarize them here; see \cite{multifidelityReview} for a comprehensive review. Among current multifidelity approaches, the best linear unbiased estimator (BLUE) framework of \cite{schaden2020multilevel,schaden2021asymptotic} is an inspiration for our effort; it uses generalized linear regression to obtain multilevel estimates of scalar quantities of interest.

To our knowledge, multifidelity methods have only recently been brought to bear on covariance estimation. Multifidelity covariance estimation is particularly challenging because covariance matrices have geometric properties that an estimator must respect: namely, symmetry and positive semi-definiteness. 
Straightforward application of techniques designed for Euclidean data, including multilevel Monte Carlo \cite{giles2015multilevel,cliffe_multilevel_2011}, multifidelity Monte Carlo \cite{NME:NME4761,bpkwmg} and multilevel BLUEs \cite{schaden2020multilevel}, to covariance matrices may yield results which are not positive semidefinite, and therefore not covariance matrices. 

\subsection{Multifidelity covariance estimation: literature review} 
\label{sec:relatedWork}
Multifidelity and multilevel covariance estimators in the current literature are most often specialized to the one-dimensional case; i.e., they are multifidelity and multilevel estimators of \textit{scalar} variances and covariances. 
Convergence of multilevel Monte Carlo variance estimators is discussed in \cite{bierig2015convergence} and similar analysis of multilevel Monte Carlo estimation of scalar covariances 
can be found in \cite{mycek2019multilevel};  both employ control variates and typical multilevel assumptions on the rates of error decay and cost increase with increasing model ``level.'' Multifidelity estimators of variance and Sobol sensitivity indices are developed in \cite{qian2018multifidelity}; the framework employed therein is rate-free
and formulated in terms of correlations between model fidelities.

The earliest approaches to multilevel and multifidelity covariance \textit{matrix} estimation are largely embedded in works on multifidelity and multilevel data assimilation, 
in which low-fidelity samples are used to improve an estimate of a quantity-of-interest depending on a high-fidelity covariance matrix, such as the Kalman gain operator. For instance, at each step of the multilevel EnKF (MLEnKF) \cite{mlenkf} a multilevel covariance estimate is constructed using the standard  
``telescoping sum'' of multilevel Monte Carlo, which, due to the presence of subtraction, can induce loss of positive definiteness.  
Loss of definiteness in the MLEnKF is corrected in a post-hoc manner by rounding negative eigenvalues up to zero, but 
the authors note that ``it would be of independent interest to devise multilevel [covariance] estimators which preserve positivity without such an imposition.'' 

There has been some development to this end, namely the positive-definite multifidelity covariance estimators of \cite{maurais2023logEuclidean}, constructed using control variates in the log-Euclidean geometry \cite{arsigny2006log} for symmetric positive definite (SPD) matrices; we will compare to these estimators in the present work. 
Other recent approaches to multifidelity/multilevel covariance estimation, such as the hierarchical ($\mathcal{H}^2$)-formatted multilevel covariance estimator of \cite{dolzDataSparseMultilevel2023}  
and the multivariate generalization of scalar multilevel BLUEs in \cite{destouches2023multivariate}, rely on the Euclidean geometry for symmetric matrices and hence do not ensure positive definite results.

\subsection{Contributions}
In this paper, we formulate multifidelity covariance estimation as a regression problem on the manifold of SPD matrices equipped with the affine-invariant geometry \cite{bhatiaposdef}. We take our inspiration from the regression framework of \cite{schaden2020multilevel} but operate within a {Riemannian}, rather than Euclidean, geometry for SPD matrices and thus obtain guaranteeably positive-definite results. Our manifold regression multifidelity (MRMF) estimator can further be seen as a generalization of control-variate type multifidelity estimators, including those in \cite{maurais2023logEuclidean}; we show that such estimators can be obtained as simplifications of the regression framework we present here. 
We discuss the numerical implementation of our estimator, introducing regularization schemes and a parameterization enabling the use of unconstrained optimization methods. We show via numerical examples that our estimator can yield significant reductions in covariance estimation error and improved performance in downstream tasks, such as metric learning, relative to single-fidelity {and} existing multifidelity estimators.
\medskip

The rest of the paper is organized as follows. \Cref{sec:bg} reviews some necessary background. In \Cref{sec:formulation} we introduce our estimator. In \Cref{sec:analysis} we discuss its properties, connections to existing multifidelity estimators, and generalizations. We discuss computational considerations in \Cref{sec:computation} and demonstrate the estimator's performance in two numerical examples in \Cref{sec:numerics}; then we close and provide some outlook in  \Cref{sec:conclusion}. 

\section{Background}
\label{sec:bg}

\subsection{The manifold of SPD matrices}
\label{sec:spdManifold}
The set of $d \times d$ symmetric positive definite matrices, which we denote by $\bbP_d$, forms a \textit{Riemannian manifold} embedded in the vector space of $d \times d$ symmetric matrices $\bbH_d$. The manifold $\bbP_d$ is locally similar to $\bbH_d$ at each point $A \in \bbP_d$, and at each $A \in \bbP_d$ we define the \textit{tangent space} $\rmT_A\bbP_d  \subseteq \bbH_d$ with a unique inner product. In the development of our estimator (\Cref{sec:formulation}) we make use of this inner product along with its corresponding outer product, geodesics, and geodesic distance. We introduce these concepts here briefly and direct the reader interested in a more rigorous treatment to \cite{bhatiaposdef}. 

Let $A \in \bbP_d$, and $U, V \in \mathrm{T}_A\bbP_d \subseteq \bbH_d$. The \textbf{inner product} on $\rmT_A\bbP_d$, $g_A(\cdot, \cdot): \mathrm{T}_A\bbP_d\times \mathrm{T}_A\bbP_d \to \R$, is defined as a weighted Frobenius inner product $\langle \cdot, \cdot \rangle$ for symmetric matrices, 
\begin{equation}
g_A(U, V) = \langle U, \; V \rangle_A = \langle U, \; A\inv VA\inv \rangle = \trace{U A\inv V A\inv}.
\label{eq:spd_ip}
\end{equation}
This inner product gives rise to a corresponding \textbf{outer product} on $\rmT_A \bbP_d$, 
\begin{equation*}
	U \otimes_A V = U \otimes (A\inv V A\inv),
\end{equation*}
where $\otimes$ is the Euclidean outer product for symmetric matrices, and \rev{$U \otimes_A V$ may be interpreted as a (rank-one) linear operator on $\rmT_A \bbP_d$. From a geometric perspective, $U\otimes_A V$ is a 1-covariant, 1-contravariant tensor, and application of the transformation $V \mapsto A\inv V A\inv$ to the second argument of $\otimes_A$ ensures that $\otimes_A$ is consistent with $\langle \cdot, \cdot \rangle$, i.e., that $\trace{U \otimes_A V} = \langle U, \; V \rangle_A$.} 

For $A \in \bbP_d$ there exist diffeomorphic \textbf{logarithmic} and \textbf{exponential} mappings between $\bbP_d$ and $\rmT_A \bbP_d$. Let $A, B \in \bbP_d$. The logarithmic mapping $\log_A: \bbP_d \to \mathrm{T}_A\bbP_d \subseteq \bbH_d$ is given by
\begin{equation}
\log_A(B) = A^{\frac{1}{2}}\log(A^{-\frac{1}{2}}BA^{-\frac{1}{2}})A^{\frac{1}{2}} = A\log(A\inv B),
\label{eq:logA}
\end{equation}
Now let $X \in \mathrm{T}_A\bbP_d$. The exponential mapping $\exp_A: \mathrm{T}_A\bbP_d \to \bbP_d$ is given by
\begin{equation}
\exp_A(X) = A^{\frac{1}{2}}\exp(A^{-\frac{1}{2}}XA^{-\frac{1}{2}})A^{\frac{1}{2}} = A\exp(A\inv X).
\label{eq:expA}
\end{equation}
\rev{Here $\exp(\cdot)$ is the matrix exponential and $\log(\cdot)$ is the principal matrix logarithm on $\bbP_d$ \cite{highamCh11}.}
The first forms of \cref{eq:logA} and \cref{eq:expA} make explicit the fact that $\log_A$ and $\exp_A$ produce symmetric outputs, while the second can be advantageous in analysis and computation.

The inner-product \eqref{eq:spd_ip} defines a {natural metric} on $\bbP_d$, giving rise to notions of geodesics and distance. For $A, B \in \bbP_d$, the \textbf{geodesic}, or shortest path, on $\bbP_d$ between $A$ and $B$ is
\begin{equation}
\gamma(t) = A^{1/2}(A^{-1/2}BA^{-1/2})^tA^{1/2}, \quad t\in [0,1].
\label{eq:geodesic}
\end{equation}
The \textbf{intrinsic distance} between $A$ and $B$ is equal to the length of this geodesic and is %
\begin{equation}
	d(A,B) =  \sqrt{\langle \log_A B,\; \log_A B\rangle_A }
	= ||\log(A^{-1/2}BA^{-1/2})||_{\rm F} 
	= \left(\sum_{i=1}^d\log^2\lambda_i(A\inv B) \right)^{\frac{1}{2}}. 
	\label{eq:intrinsicdist}
\end{equation} 

In defining our multifidelity covariance estimator we primarily work with \textit{product manifolds} of SPD matrices, i.e., $\bbP_d^K = \bbP_d \times \cdots \times \bbP_d$ ($K$ times) where $K \in \Z^+$. $\bbP_d^K$ is itself a Riemannian manifold with geometry obtained by extension of the geometry of $\bbP_d$; see \cref{app:prodMan_geom} for details. 

\subsection{Statistics on the manifold}  
\label{sec:intrinsicStatistics}
Utilizing definitions in \cite{pennec2006intrinsic} with the geometry above, we obtain notions of mean, variance, and covariance for $\bbP_d$-valued random matrices. As with geometry, the extension of these statistics to product-manifolds is straightforward (\cref{app:prodMan_stats}).

Let $S \in \bbP_d$ be random. We define the \textbf{expectation} of $S$ to be the \textit{Fr\'{e}chet mean} of $S$, 
\begin{equation}
\begin{aligned}
	\bfE[S] &= \argmin_{Y \in \bbP_d} \E\left[d^2(Y, S)\right]
	= \argmin_{Y \in \bbP_d} \E\left[||\log(Y^{-1/2}SY^{-1/2})||_{\rm F}^2\right] \equiv \Sigma.
\end{aligned}
\label{eq:frechet_mean}
\end{equation}
Because $\bbP_d$ is a complete, non-positively curved manifold \cite{bhatiaposdef}, this mean is unique \cite{pennec2006intrinsic}. \rev{Note that we use distinct notation $\bfE[\cdot]$ to indicate Frechet means, while conventional expectations are denoted by $\bbE[\cdot]$.} The \textbf{variance} of $S$ is the expected squared distance between $S$ and its mean $\Sigma = \bfE[S]$\footnote{In other words, the variance $\sigma^2_S$ is the {minimum} over $Y \in \bbP_d$ of $\E\left[||\log(Y^{-1/2}SY^{-1/2})||_{\rm F}^2\right]$, while $\Sigma = \bfE[S]$ is the corresponding minimizer.},
\begin{equation*}
	\sigma^2_{S} = \E[d^2(\Sigma, S)]  = \E\left[||\log(\Sigma^{-1/2}S\Sigma^{-1/2})||_{\rm F}^2\right]. 
\end{equation*}

Next we introduce a notion of covariance for $S \in \bbP_d$. Recall that for random $x \in \R^n$ with mean $\mu$, the covariance of $x$ is the expected outer product of $x - \mu$ with itself, 
\begin{equation*}
\Cov[x] = \E[(x - \mu)(x - \mu)\t] = \E[(x - \mu) \otimes (x - \mu)]. 
\end{equation*}
Because $\R^n$ is a vector space, the vector difference $x - \mu$ is an element of $\R^n$ and $\Cov[x] \in \R^{n \times n}$ defines a symmetric positive definite linear operator on $\R^n$. 
The SPD manifold is \textit{not} a vector space, so the {covariance} of $S \in \bbP_d$ cannot be defined directly on $\bbP_d$. Thus we define the \textbf{covariance} of $S$ on the {tangent space}  to $\bbP_d$ at $\Sigma$, setting 
\begin{equation}
	\Cov[S] = \E[\log_\Sigma S \otimes_\Sigma \log_\Sigma S] \equiv \Gamma_S.
	\label{eq:gammaS_def}
\end{equation}
$\Cov[S]$ \cref{eq:gammaS_def} possesses the structure of a traditional vector covariance in that it is an expected outer product of a function of $S$ and its mean $\Sigma$. This function, $\log_\Sigma S$, is a mapping of $S \in \bbP_d$ onto $\rmT_\Sigma \bbP_d$, and we interpret it as the ``vector difference'' between $S$ and $\Sigma$ \cite{pennec2006intrinsic, pennec2006riemannian}: if $S = \Sigma$ then $\log_\Sigma S = 0^{d\times d}$, and if $S \neq \Sigma$ then $S$ has a nonzero image under $\log_\Sigma(\cdot)$. 

$\Gamma_\bfS$ is a symmetric  positive semidefinite linear operator on $\rmT_\Sigma \bbP_d \subseteq \bbH_d$.
Note that the trace of $\Gamma_S$ is indeed the variance $\sigma^2_S$,
\begin{equation*}
\begin{aligned}
\trace{\Gamma_S} &= \trace{\E[\log_\Sigma S \otimes_\Sigma \log_\Sigma S ]} = \E[\trace{\log_\Sigma S \otimes_\Sigma \log_\Sigma S }] \\
 &= \E[\langle \log_\Sigma S,\; \log_\Sigma S \rangle_\Sigma] = \E[d^2(\Sigma, S)] \equiv \sigma^2_S.
\end{aligned}
\end{equation*}

\rev{Under the assumption that the covariance of $S$ is invertible}, we define a notion of (squared) \textbf{Mahalanobis distance} between the random variable $S$ and a deterministic point $Y \in \bbP_d$,
\begin{equation}
	d^2_S(Y;\, \Sigma) = \langle \log_\Sigma Y,\; \Gamma_S\inv \log_\Sigma Y \rangle_\Sigma.
	\label{eq:mdist_def}
\end{equation}
The Mahalanobis distance is a $\Gamma_S\inv$-weighted version of the intrinsic distance \eqref{eq:intrinsicdist} between $\Sigma$ and $Y$ and is analogous to the Mahalanobis distance for vector-valued random variables. In writing \cref{eq:mdist_def} we have chosen to explicitly highlight the dependence on $\bfE[S] = \Sigma$, because for the remainder of our development $\Sigma$ will generally be unknown.

\section{Estimator formulation}
\label{sec:formulation}
In this section we introduce the basic formulation of our estimator, encompassing assumptions on how data are sampled, a model for the data on SPD product manifolds, and the optimization we solve to obtain multifidelity covariance estimates. 

\subsection{Problem setup}
Let $[S_0]$ denote an equivalence class of random matrices $S \in \bbP_d$ such that $\bfE[S] = \Sigma_0$, i.e., $[S_0] = \{S \in \bbP_d: \bfE[S] = \Sigma_0 \}$.\footnote{We introduce these classes because in practice it is common to employ covariance estimators which possess the same mean but have different distributions, e.g., due to different sample sizes. The key structure of our formulation involves grouping such estimators by their means, corresponding to different levels of fidelity.
} 
Suppose that we are able to sample elements of $[S_0]$ at high computational cost and would like to estimate the unknown mean matrix $\Sigma_0 \in \bbP_d$. At the same time we are able to sample from a number of related \textit{low-fidelity} equivalence classes, $[S_i] = \{S \in \bbP_d: \bfE[S] = \Sigma_i\}$, $i \in \{1, \dots, L\}$, at comparatively lower computational costs. The low-fidelity mean-matrices $\Sigma_1, \dots, \Sigma_L$ are also unknown and may be of some interest to estimate, but our primary objective is to estimate $\Sigma_0$. 

We assume that we can obtain \textit{statistically coupled} samples from any combination of the equivalence classes $[S_0], \dots, [S_L]$. Specifically, letting $F = (F^k)_{k=1}^K \subseteq 2^{\{0, \ldots, L\}}$ represent $K$ subsets of the indices $\{0, \dots, L\}$, our data consist of $K$ collections of samples from $[S_0], \dots, [S_L]$,
\begin{equation}
\left(S_i^{(k)}: i \in F^k \right) , \quad k = 1, \dots, K, 
\label{eq:Sdata}
\end{equation} 
which accordingly have expectations 
\[
\bfE \left[(S_i^{(k)}: i \in F^k)\right] = (\Sigma_i: i \in F^k), \quad k = 1, \dots, K.
\] 
The collections $(S_i^{(k)}: i \in F^k)$ are generated such that for each $k \in \{1, \dots, K\}$ the random matrices $S_i^{(k)}$, $i \in F^k$ are {correlated} with each other, but $S_i^{(k)}$ is independent of $S^{(\ell)}_j$ (written as $S_i^{(k)} \indep S^{(\ell)}_j$) for any $\ell \neq k$, $i, j \in \{1, \dots, L\}$. Note that for $i \in \{0, \dots, L\}$ and $j \neq k$ we do \textit{not} assume that $S_i^{(j)} \overset{d}{=} S_i^{(k)}$; rather we only assume equivalence of means $\bfE[S_i^{(j)}] = \bfE[S_i^{(k)}] = \Sigma_i$. A convenient way to visualize this equivalence-class/coupling structure is via a table, which we illustrate in \cref{tab:dataStructure} for an example with $L = 3$.

\begin{table}[h]
    \centering
    \begin{tabular}{c|c|c|c|c}
         & $[S_0]$ & $[S_1]$ & $[S_2]$ & $[S_3]$  \\
         \hline
      $k = 1$   &  $S_0^{(1)}$ & $S_1^{(1)}$ & & \\ 
      $k = 2$   &  & $S_1^{(2)}$ &  $S_2^{(2)}$ &  \\
      $k = 3$   &  & $S_1^{(3)}$ & $S_2^{(3)}$ & $S_3^{(3)}$ \\
      $k = 4$   & & & & $S_3^{(4)}$ 
    \end{tabular}
    \caption{Example data \cref{eq:Sdata} corresponding to $L = 3$ with $F^1 = \{0, 1\}$, $F^2 = \{1, 2\}$, $F^3 = \{1, 2, 3\}$, and $F^4 = \{3\}$. Matrices within the same \textit{column} of the table have the same mean, $\bfE[S_i^{(k)}] = \bfE[S_i^{(j)}] = \Sigma_i$, while matrices within the same \textit{row} are statistically coupled with each other, $S_i^{(k)} \not\!\perp\!\!\!\perp S_\ell^{(k)}$.}
    \label{tab:dataStructure}
\end{table}

\subsection{Manifold regression estimator} \label{sec:sRV} 
In a similar vein to \cite{schaden2020multilevel}, we define our manifold regression multifidelity covariance estimator by interpreting the data \cref{eq:Sdata} as \rev{realizations of} a random variable. For $k \in \{1, \dots, K\}$ denote $\bfS^{(k)} = (S_i^{(k)}: i \in F^k)$ and $\bfSigma^{(k)} = (\Sigma_i: i \in F_k)$; it follows  
that $\bfE[\bfS^{(k)}] = \bfSigma^{(k)}$. We model the data \cref{eq:Sdata} by ``stacking'' $\bfS^{(1)}, \dots, \bfS^{(K)}$ into an $N$-vector of matrices, where $N = \sum_{k=1}^K \left| F^k \right| = \sum_{k=1}^K N_k$. We denote the mean \rev{$\bfE[\bfS]$} of the $\bbP_d^{N}$-valued random variable $\bfS$ by $\mu_\bfS(\Sigma_0, \dots, \Sigma_L)$ and the Riemannian covariance of $\bfS$ by $\Gamma_\bfS$ and write 
\begin{equation}
\bfS = \begin{bmatrix} 
\bfS^{(1)} \\ \vdots \\ \bfS^{(K)}
\end{bmatrix} \sim \left(\mu_\bfS(\Sigma_0, \dots, \Sigma_L) =  \begin{bmatrix}
\bfSigma^{(1)} \\ \vdots \\ \bfSigma^{(K)}
\end{bmatrix} \equiv \bfSigma,\quad \Gamma_\bfS = \E[\log_\bfSigma \bfS \otimes_\bfSigma \log_\bfSigma \bfS ] \right).
\label{eq:S_rv}
\end{equation}

\subsubsection{Covariance of $\bfS$}
\label{sec:covS}
The covariance of $\bfS$ is 
\begin{equation}
	\Gamma_\bfS = \E\left[ \begin{bmatrix}
		\log_{\bfSigma^{(1)}} \bfS^{(1)} \\ \vdots \\ \log_{\bfSigma^{(K)}} \bfS^{(K)}
	\end{bmatrix} \!\otimes_\bfSigma \!\begin{bmatrix}
		\log_{\bfSigma^{(1)}} \bfS^{(1)} \\ \vdots \\ \log_{\bfSigma^{(K)}} \bfS^{(K)}
	\end{bmatrix} \right] = \E\left[ \begin{bmatrix}
		\log_{\bfSigma^{(1)}} \bfS^{(1)} \\ \vdots \\ \log_{\bfSigma^{(K)}} \bfS^{(K)}
	\end{bmatrix} \!\otimes\! \begin{bmatrix}
		G_{\bfSigma^{(1)}}\log_{\bfSigma^{(1)}} \bfS^{(1)} \\ \vdots \\ G_{\bfSigma^{(K)}}\log_{\bfSigma^{(K)}} \bfS^{(K)}
	\end{bmatrix} \right]\!,
\label{eq:gammaS}
\end{equation}
where $G_{\bfSigma^{(k)}}$ is the linear transformation mapping on $\bbP_d^{N_k}$ mapping
\[
\bfA = \left(A_1, \dots, A_K \right) \mapsto \left(\Sigma_{k_1}\inv A_1\Sigma_{k_1}\inv, \dots, \Sigma_{k_{N_k}}\inv A_{N_k} \Sigma_{k_{N_k}}\inv \right) = G_{\bfSigma^{(k)}}\bfA.
\]
The transformations $G_{\bfSigma^{(1)}}, \dots, G_{\bfSigma^{(K)}}$ arise from the affine-invariant metric on $\bbP_d$.

$\Gamma_\bfS$ is a symmetric positive semidefinite linear operator on $\rmT_\bfSigma \bbP_d^{N} = \bbH_d^{N}$. Due to the coupling and independence structure in our data $\bfS$ \cref{eq:Sdata}, $\Gamma_\bfS$ has ``block diagonal'' structure,
\begin{equation} 
\Gamma_\bfS = 
\begin{bmatrix}
\Gamma_\bfS^{(1)} \\
& \ddots \\
& & \Gamma_\bfS^{(K)}
\end{bmatrix},
\label{eq:gamma_blockDiag}
\end{equation} 
where $\Gamma_{\bfS^{(k)}} = \E[\log_{\bfSigma^{(k)}} \bfS^{(k)} \otimes_{\bfSigma^{(k)}} \log_{\bfSigma^{(k)}} \bfS^{(k)}]$ is the Riemannian covariance of $\bfS^{(k)}$ and a symmetric positive semidefinite linear operator from $\bbH_d^{(N_k)}$ to $\bbH_d^{(N_k)}$, $k \in \{1, \dots, K\}$. 

Within this random variable model for $\bfS \sim (\bfSigma, \Gamma_{\bf S})$ \rev{and under the assumption that $\Gamma_\bfS$ is invertible}, we estimate the mean of $\bfS$ by \textit{minimizing Mahalanobis distance}.

\begin{definition}[Manifold Regression Multifidelity (MRMF) Covariance Estimator]
Given a realization of the random variable $\bfS \sim (\bfSigma, \Gamma_\bfS)$ \cref{eq:S_rv} we estimate the true covariance matrices $\Sigma_0, \dots, \Sigma_L$ which parameterize $\mu_\bfS(\Sigma_0, \dots, \Sigma_L) = \bfSigma$ by minimizing squared \textit{Mahalanobis distance} with respect to $\bfSigma$,  
\begin{equation}
	(\hat\Sigma_0,\dots, \hat\Sigma_L) = \argmin_{\Sigma_0, \dots,\Sigma_L \in \bbP_d} \langle \log_{\bfSigma} \bfS,\; \Gamma_\bfS\inv \log_{\bfSigma} \bfS \rangle_{\bfSigma} \quad \text{s.t. } \bfSigma = \mu_\bfS(\Sigma_0, \dots, \Sigma_L).
	\label{eq:mdist_min}
\end{equation}
\label{sec:mdistMin}
\end{definition}
\subsection{Running example} 
\label{sec:example_s3}
As a concrete illustration of the ideas in this section, we consider an example with three data matrices.
Take the model of \cref{eq:S_rv} with $L = 1$ and $F = \{\{0, 1\}, \{1\}\} = \{F^1, F^2\}$  such that our data are 
\[
\bfS = (S_0^{(1)}, S_1^{(1)}, S_1^{(2)}) \equiv (\Shi, \SloOne, \SloTwo),
\]
where $\Shi$ and $\SloOne$ are correlated with each other but $\SloTwo \indep (\Shi, \SloOne)$. This structure may arise, for example, if $\Shi$ and $\SloOne$ are sample covariance matrices (SCMs) computed from statistically coupled realizations of random \textit{vectors} $\Xhi, \Xlo \in \R^d$ and $\SloTwo$ is a sample covariance matrix computed from independent realizations of $\Xlo$. Specifically, suppose that we have %
\begin{equation} 
\begin{aligned}
\left\{(X_{\text{hi}}^i, X_{\text{lo}}^i)\right\}_{i=1}^{M_1}, & \quad \text{\textbf{statistically coupled} sample pairs of $X_{\rm hi}$, $X_{\rm lo} \in \R^d$} \\
\left\{X_{\text{lo}}^i \right\}_{i=M_1+1}^{M_1 + M_2}, & \quad \text{\textbf{independent} samples of $X_{\rm lo} \in \R^d$}.
\end{aligned}
\label{eq:exampleData}
\end{equation} 
The \textit{samples} $X_{\text{hi}}^i$ and $X_{\text{lo}}^i$ are {correlated} for the same $i$, but the \textit{pairs} $\{(X_{\text{hi}}^i, X_{\text{lo}}^i)\}_{i=1}^{M_1}$ are independent and identically distributed (i.i.d.). %
Likewise, the additional low-fidelity samples $\{X_{\text{lo}}^i \}_{i=M_1+1}^{M_1 + M_2}$ are i.i.d.\ and independent of the pairs $\left\{(X_{\text{hi}}^i, X_{\text{lo}}^i)\right\}_{i=1}^{M_1}$. In this setting we take %
\begin{equation*}
		\Shi \equiv \widehat\Cov[\{X_{\rm hi}^i\}_{i=1}^{M_1} ], \quad \SloOne \equiv \widehat\Cov[\{X_{\rm lo}^i\}_{i=1}^{M_1} ], \quad  \SloTwo \equiv \widehat\Cov[\{X_{\rm lo}^i\}_{i=M_1 + 1}^{M} ],
\end{equation*}
where we have defined $M = M_1 + M_2$. Due to the coupling and independence structure in the data \cref{eq:exampleData}, $\Shi$ and $\SloOne$ are correlated with each other while $\SloTwo$ is independent of $(\Shi, \SloOne)$. Furthermore, $\bfE[\SloOne] = \bfE[\SloTwo] = \Sigmalo$ but $\SloOne \overset{d}{\neq} \SloTwo$ because $\SloOne$ and $\SloTwo$ are constructed from different numbers of samples of $\Xlo$.\footnote{As noted in \cite{smith2005covariance}, sample covariance matrices are only \textit{asymptotically} unbiased in the intrinsic metric, i.e., even though $\E[\SloOne] = \E[\SloTwo]$ in the Euclidean sense, in general $\bfE[\SloOne] \neq \bfE[\SloTwo]$. However, in the absence of intrinsically unbiased sample covariance estimators we make the modeling assumption that $\bfE[\SloOne] = \bfE[\SloTwo] = \Sigmalo$ and have obtained good results in practice from doing so; see \Cref{sec:numerics}.}
$\bfS$ takes values in $\bbP_d^3$ with mean $\bfSigma$ and covariance $\Gamma_\bfS$, 
\begin{equation}
\bfS = \begin{bmatrix}
	\Shi \\ \SloOne \\ \SloTwo \end{bmatrix} \sim \left(\bfSigma= \begin{bmatrix}
	\Sigmahi \\ \Sigmalo \\ \Sigmalo
\end{bmatrix}, \Gamma_\bfS  = \E[\log_\bfSigma \bfS \otimes_\bfSigma \log_\bfSigma \bfS] \right),
\label{eq:threemat_model}
\end{equation}
where
\begin{equation*}
	\Gamma_\bfS = \E\left[ \begin{bmatrix}
		\log_\Sigmahi(\Shi) \\ \log_\Sigmalo(\SloOne) \\ \log_\Sigmalo(\SloTwo) \end{bmatrix} \otimes_\bfSigma \begin{bmatrix} \log_\Sigmahi(\Shi) \\ \log_\Sigmalo(\SloOne) \\ \log_\Sigmalo(\SloTwo) \end{bmatrix} \right] 
\end{equation*}
is the Riemannian covariance of $\bfS$, a symmetric positive semidefinite linear operator on $\rmT_\bfSigma \bbP_d^3 = \bbH_d^3$. Because $\SloTwo$ is independent of $\Shi$ and $\SloOne$, $\Gamma_\bfS$ has block structure 
\begin{equation} 
\Gamma_{\bfS} = 
\begin{bmatrix}
\Gamma_{\rm hi} & \Gamma_{\rm lo, hi} & \mathbf{0} \\
\Gamma_{\rm hi, lo} & \Gamma_{\rm lo, 1} & \mathbf{0} \\ 
\mathbf{0} & \mathbf{0} & \Gamma_{\rm lo, 2}
\end{bmatrix}.
\label{eq:gamma_blockdiag_ex}
\end{equation}
The nonzero blocks of $\Gamma_\bfS$ are the auto-covariances of $\Shi$, $\SloOne$, and $\SloTwo$,
\begin{equation*}
\begin{gathered}
 \Gamma_{\rm hi} = \E[\log_\Sigmahi(\Shi) \otimes_\Sigmahi \log_\Sigmahi (\Shi)], \quad \Gamma_{\rm lo, 1} = \E[\log_\Sigmalo(\SloOne) \otimes_\Sigmalo \log_\Sigmalo(\SloOne)], \\ \Gamma_{\rm lo, 2} = \E[\log_\Sigmalo(\SloTwo) \otimes_\Sigmalo \log_\Sigmalo(\SloTwo)]
 \end{gathered}
 \end{equation*}
 and the cross-covariances between $\Shi$ and $\SloOne$,
\begin{equation*}
\Gamma_{\rm lo, hi} = \E[\log_\Sigmahi \Shi \otimes_\Sigmalo \log_\Sigmalo \SloOne] \quad \text{and} \quad \Gamma_{\rm hi, lo} = \E[\log_\Sigmalo \SloOne \otimes_\Sigmahi \log_\Sigmahi \Shi]. 
\end{equation*}
The squared Mahalanobis distance minimization we solve to estimate $\Sigmahi$ and $\Sigmalo$ is  
 \begin{equation} 
	\begin{aligned}
		\left(\Sigmahihat, \Sigmalohat \right) 		& = \argmin_{\Sigmahi, \Sigmalo \in \bbP_d} \langle\log_\bfSigma \bfS,\; \Gamma_\bfS\inv \log_\bfSigma \bfS \rangle_\bfSigma &\text{s.t. } \bfSigma = (\Sigmahi, \Sigmalo, \Sigmalo)&  \\
        &= \argmin_{\Sigmahi, \Sigmalo \in \bbP_d} \left\langle \begin{bmatrix} \log_\Sigmahi(\Shi) \\ \log_\Sigmalo(\SloOne) \\ \log_\Sigmalo(\SloTwo) \end{bmatrix}, \; \Gamma_\bfS\inv \begin{bmatrix} \log_\Sigmahi(\Shi) \\ \log_\Sigmalo(\SloOne) \\ \log_\Sigmalo(\SloTwo) \end{bmatrix}  \right\rangle_{\bfSigma} &\text{s.t. } \bfSigma = (\Sigmahi, \Sigmalo, \Sigmalo)&  \\
	\end{aligned}
 \label{eq:mdist_min_ex}
 \end{equation} 

\section{Analysis, simplification, and interpretation of the manifold regression estimator}
\label{sec:analysis}
In this section we analyze the regression estimator \cref{eq:mdist_min}, demonstrating useful properties, simplifications, and interpretations. In \Cref{ss:mdistProperties} we show that the Mahalanobis distance we minimize is affine-invariant and \rev{independent of the basis used to represent the tangent space in which} it is defined; moreover, it can be endowed with a {maximum likelihood} interpretation. In \Cref{sec:example_s4} we demonstrate how these properties are useful in practice. In \Cref{sec:fixSigmalo} we present a simplification of our estimator wherein $\Sigmalo$ is fixed and find that, in addition to being computationally advantageous, this choice leads to greater analytical tractability. In particular, we show in \Cref{sec:mfGeneralGeom} that the fixed-$\Sigmalo$ simplification yields a surprising link to control variates, uniting our work here with many existing multifidelity estimators. Proofs of results in this section can be found in \cref{app:proofs}.
\subsection{Properties of Mahalanobis distance} 
\label{ss:mdistProperties}
In what follows here we demonstrate three properties of the Mahalanobis distance. The first property, \rev{basis independence} (\cref{thm:ts_agnostic}), simplifies computation of the estimator by eliminating dependence on the $\bfSigma$-specific weightings defining $\langle \cdot, \cdot \rangle_\bfSigma$ and $\otimes_\bfSigma$. The second, affine-invariance (\cref{thm:affineinvariance}), enables use of stabilizing preconditioners. The third provides a maximum likelihood interpretation under a Gaussian model for the error $\log_\bfSigma \bfS = \bm{\calE} \in \bbH_d^N$. 

\subsubsection{Basis independence}
\label{sec:tsAgnostic}
As formulated in \cref{eq:mdist_min}, the squared Mahalanobis distance between $\bfSigma$ and $\bfS$ depends {highly} non-trivially on $\bfSigma$: not only do we have to contend with the ``vector difference'' $\log_\bfSigma \bfS$, but the very {operators} $\otimes_\bfSigma$ and $\langle \cdot, \cdot \rangle_\bfSigma$ of $\rmT_\bfSigma \bbP_d^{N}$ defining the covariance and Mahalanobis distance depend on $\bfSigma$.  \cref{ex:threemat_mdist}. 

While we could perhaps compute with \cref{eq:mdist_min} directly and find a way to estimate $\Gamma_\bfS$ as defined with $\otimes_\bfSigma$, it would be convenient to remove the dependence of $\Gamma_\bfS$ and the Mahalanobis distance on the $\bfSigma$-dependent weightings of \rev{$\langle \cdot, \cdot \rangle_{\bfSigma}$ and $\otimes_\bfSigma$}.  
Equivalently, we would like our estimator to behave ``the same'' \rev{regardless of the particular basis used to represent the tangent space $\bbH_d^N$.}
In this instance we indeed get our wish: Mahalanobis distance is \rev{invariant to the choice of $\bfSigma$ used to define $\langle \cdot, \cdot \rangle_{\bfSigma}$ and $\otimes_\bfSigma$, hence we can take $\bfSigma = \mathbf{I}$ and compute with the \textit{unweighted} operators $\langle \cdot, \cdot \rangle$ and $\otimes$.} 
\begin{proposition}
Let $\bfS$ and $\bfSigma = \mu_\bfS(\Sigma_0,\dots, \Sigma_L)$ be as in \cref{eq:S_rv}. The squared Mahalanobis distance objective of \cref{eq:mdist_min} is independent of the \rev{basis used to represent the tangent space $\bbH_d^N$, that is, independent of the $\bfSigma$-specific weightings in $\langle \cdot, \cdot \rangle_\bfSigma$ and $\otimes_\bfSigma$},
\begin{equation}
\mathfrak{D}^2_\bfS(\bfSigma) 	
  \coloneqq d^2_\bfS(\bfS;\, \bfSigma) = \langle \log_\bfSigma \bfS, \; \Gamma_\bfS\inv \log_\bfSigma \bfS \rangle_\bfSigma = \langle \log_\bfSigma \bfS, \; \Gamma_{\bfS, \bfI} \inv \log_\bfSigma \bfS \rangle,
\end{equation}
where $\Gamma_{\bfS, \bfI} = \E[\log_\bfSigma \bfS \otimes \log_\bfSigma \bfS]$ is the covariance of $\bfS$ computed using the standard (unweighted) outer product, and $\langle \cdot, \cdot \rangle$ denotes the unweighted Frobenius inner-product on $\bbH_d^{N}$. 
	\label{thm:ts_agnostic}
\end{proposition}
This result follows from the fact that the linear transformation used to define $\langle \cdot, \cdot \rangle_\bfSigma$ in the Mahalanobis distance is canceled by its own inverse when $\Gamma_\bfS\inv$ is applied as a weighting.

\subsubsection{Affine invariance} 
\label{ss:affine_invariance}
A salient property of the intrinsic metric on $\bbP_d$ is that it is \textit{affine-invariant} 
in the sense that if for $A, B \in \bbP_d$ we define $\tilde A = Y\inv AY\inv$ and $\tilde B = Y\inv B Y\inv$ with $Y \in \bbP_d$, then it holds that $d(\tilde A, \tilde B) = d(A, B)$ \cite{bhatiaposdef}.
Affine-invariance of the intrinsic metric on $\bbP_d$ immediately gives affine-invariance on $\bbP_d^N$: if $\bfA = (A_1, \dots, A_N)$, $\bfB = (B_0, \dots, B_N) \in \bbP_d^N$, and we define $\tilde\bfA = (Y_1\inv A_1 Y_1\inv, \dots, Y_N\inv A_N Y_N\inv)$ and $\tilde\bfB = (Y_1\inv B_0 Y_1\inv, \dots, Y_N\inv B_N Y_N\inv)$ for some $\bfY \in \bbP_d^N$, then one can easily show that 
\begin{equation*}
d^2(\tilde \bfA, \tilde \bfB) = d^2(\bfA, \bfB).
\end{equation*}

In this section we show that the affine-invariance property of the intrinsic metric on $\bbP_d^N$ extends to the Mahalanobis distance \cref{eq:mdist_min} defining our multifidelity covariance estimator. 

\begin{proposition}
	Consider the random variable $\bfS$ in \cref{eq:S_rv} and the Mahalanobis distance 
	\begin{equation}
	\frakD^2_\bfS(\bfSigma) = \langle \log_\bfSigma \bfS, \; \Gamma_\bfS\inv \log_\bfSigma \bfS \rangle_\bfSigma. 
	\label{eq:mdist_thm}
	\end{equation}
	Let $\tilde \bfS = G_\bfY \bfS$, where $\bfY \in \bbP_d^N$ and $G_\bfY: \bbH_d^N \to \bbH_d^N$ is the linear operator mapping 
 \[
\bfC = (C_1, \dots, C_N) \mapsto  (Y_1 \inv C_1 Y_1 \inv, \dots, Y_N \inv C_{N} Y_{N} \inv) = G_\bfY\bfC.
 \]
 $\tilde\bfS$ is a linear transformation of $\bfS$ with corresponding mean $\tilde \bfSigma = G_\bfY \bfSigma$ and covariance $\Gamma_{\tilde \bfS} = \E[\log_{\tilde \bfSigma} \tilde \bfS \otimes_{\tilde \bfSigma} \log_{\tilde \bfSigma} \tilde \bfS] $. It holds that 
	\begin{equation*}
	\frakD^2_\bfS(\bfSigma) = \langle \log_\bfSigma \bfS, \; \Gamma_\bfS\inv \log_\bfSigma \bfS \rangle_\bfSigma = \langle \log_{\tilde \bfSigma} \tilde \bfS, \; \Gamma_{\tilde \bfS}\inv \log_{\tilde \bfSigma} \tilde \bfS \rangle_{\tilde \bfSigma} = \frakD^2_{\tilde \bfS}(\tilde \bfSigma).
	\end{equation*}
	Furthermore, $\Gamma_{\tilde \bfS, \bfI} = \rev{G_\bfY \circ \Gamma_{\bfS, \bfI} \circ G_\bfY}$. 
\label{thm:affineinvariance}
\end{proposition}

\cref{thm:affineinvariance} is useful for computation, as it enables us to apply stabilizing affine preconditioners. In particular, we can transform the data $\bfS$ to a vector of identity matrices, significantly simplifying the form of $\log_\bfSigma \bfS$. We demonstrate this technique in \Cref{sec:example_s4}. 

\subsubsection{Maximum likelihood interpretation}
\label{sec:modelOnTS}
The Mahalanobis distance minimization in \cref{eq:mdist_min} can be viewed as nonlinear regression for $\Sigma_0, \dots, \Sigma_L$, corresponding to an additive noise model for $\log_\bfSigma \bfS$ on tangent space $\rmT_\bfSigma \bbP_d^{N}$. %
We have defined the random variable $\bfS$ via 
\begin{equation*}
\bfS \sim \left( \bfE[\bfS] = \bfSigma,\; \Cov[\bfS] = \Gamma_\bfS  \right),
\end{equation*}
with $\bfE[\bfS] = \bfSigma$ taking values on the manifold $\bbP_d^{N}$ and the covariance $\Gamma_\bfS$ defined on the \textit{tangent space} to the manifold $\rmT_\bfSigma \bbP_d^{N} \subseteq \bbH_d^{N}$. $\bbH_d^{N}$ is a Euclidean vector space, so we can interpret $\log_\bfSigma \bfS$ as a new random variable on that space and view
\begin{equation*}
\Gamma_{\bfS, \bfI} = \E[\log_\bfSigma \bfS \otimes \log_\bfSigma \bfS] = \E[(\log_\bfSigma \bfS  - \mathbf{0}) \otimes (\log_\bfSigma \bfS - \mathbf{0})]
 \end{equation*} 
as the Euclidean covariance of $\log_\bfSigma \bfS$. An additive noise model for the variation of $\log_\bfSigma \bfS$ on $\bbH_d^{N}$ corresponding to $\Gamma_\bfS$ would then be 
\begin{equation}
	\log_\bfSigma\bfS = \log_\bfSigma\bfSigma + \bm{\calE} = \bm{\calE}, 
	\label{eq:tangentspacemodel}
\end{equation}
where $\bm{\calE}\equiv \log_\bfSigma \bfS$ is a $\bbH_d^{N}$-valued, mean-zero random variable with covariance $\Gamma_{\bm{\calE}} = \Gamma_{\bfS, \bfI}$. The additive noise model \cref{eq:tangentspacemodel} on tangent space suggests an exponential model on the manifold, 
\begin{equation}
\begin{gathered}
	\log_\bfSigma \bfS = \bm{\calE} \\
	\Updownarrow\\
	\bfS = \exp_\bfSigma \bm{\calE},
\end{gathered}
\label{eq:dualmodels}
\end{equation}
wherein we see that the mean-zero, symmetric-matrix-valued perturbations in $\bm{\calE}$ are transformed by $\exp_\bfSigma(\cdot)$ to define an inherently positive definite $\bbP_d^{N}$-valued random variable. 

The relationship \cref{eq:dualmodels} is an example of an ``exponential-wrapped distribution'' \cite{chevallier2022exponential} for symmetric positive definite matrices. In the particular case where the elements of $\bm{\calE}$ are symmetric-matrix-Gaussian, one obtains ``canonical log-normal'' distributions for each element of $\bfS$ \cite{schwartzmanLognormalDistributionsGeometric2016}.  
In fact, \rev{as shown similarly in \cite{fletcher2013geodesic}}, solving \cref{eq:mdist_min} is equivalent to performing maximum likelihood estimation in the case that the elements of $\bm{\calE}$ have a centered Gaussian distribution on $\bbH_d^N$. 
\begin{proposition}
    Suppose that $\log_\bfSigma \bfS = \bm{\calE} \in \bbH_d^N$ has a Gaussian distribution on $\bbH_d^N$, 
    \begin{equation}
    \bm{\calE} \sim \calN_{\bbH_d^N}(\mathbf{0}, \Gamma_{\bm{\calE}}).
    \label{eq:gaussianE}
    \end{equation}
   Then the solution to \cref{eq:mdist_min} is a maximum likelihood estimate.
\label{prop:mle}
\end{proposition}
While the Gaussian model \cref{eq:gaussianE} for $\log_\bfSigma \bfS$ does lead to a satisfying statistical interpretation, this distributional assumption is \textit{not} a requirement. In the same way that ordinary least squares estimation (based only on first and second moments) is justified 
even when scalar data do not satisfy a Gaussian noise model, our Mahalanobis distance minimization estimator \cref{eq:mdist_min} 
is applicable
to data $\bfS$ possessing a variety of error distributions, as we demonstrate in our numerical examples (\Cref{sec:numerics}). 

\subsection{Running example} 
\label{sec:example_s4}
Continuing with the setup of \Cref{sec:example_s3} with $\bfS \!=\! (\Shi, \SloOne, \SloTwo)$ and $\bfE[\bfS] = \bfSigma = (\Sigmahi, \Sigmalo, \Sigmalo)$ we demonstrate here how the properties discussed so far 
apply to that particular model. For clarity we use \rev{$\bfSreal$} to denote the specific realization of the random variable $\bfS$ which appears in our estimator.
Owing to the \rev{basis independence} described in \cref{thm:ts_agnostic}, we can minimize the Mahalanobis distance \cref{eq:mdist_min_ex} by equivalently solving 
\begin{equation}
(\Sigmahihat, \Sigmalohat) = \argmin_{\Sigmahi, \Sigmalo \in \bbP_d} \left\langle\log_\bfSigma \rev{\bfSreal},\, \Gamma_{\bfS, \bfI}\inv \log_\bfSigma \rev{\bfSreal} \right\rangle \quad \text{s.t. } \bfSigma = (\Sigmahi, \Sigmalo, \Sigmalo),
\label{eq:mdist_min_ex_tsa}
\end{equation}
which is formulated with the standard Euclidean inner- and outer-products, where $\Gamma_{\bfS, \bfI} = \E[\log_\bfSigma \bfS \otimes \log_\bfSigma \bfS]$.
Thanks to \cref{thm:affineinvariance} we can further simplify numerics by applying an affine preconditioner: let $\bfY = (\rev{\Sreal}_{\rm hi}^{\frac{1}{2}},\, (\rev{\SrealloOne})^{\frac{1}{2}},\, (\rev{\SrealloTwo})^{\frac{1}{2}})$ and $G_\bfY: \bbH_d^3 \to \bbH_d^3$ be the mapping
\[
\bfA = (A_1, A_2, A_3) \mapsto \left(\rev{\Sreal}_{\rm hi}^{-\frac{1}{2}}A_1 \rev{\Sreal}_{\rm hi}^{-\frac{1}{2}},\; (\rev{\SrealloOne})^{-\frac{1}{2}}A_2(\rev{\SrealloOne})^{-\frac{1}{2}},\; (\rev{\SrealloTwo})^{-\frac{1}{2}}A_3(\rev{\SrealloTwo})^{-\frac{1}{2}} \right) = G_\bfY \bfA,
\]
with which we transform \rev{$\bfSreal$}, $\bfSigma$ and $\Gamma_{\bfS, \bfI}$, obtaining
\begin{equation*}
\begin{aligned}
	\rev{\bfSreal} &\mapsto \rev{\tilde\bfSreal} = (\mathrm{I},\,\mathrm{I},\, \mathrm{I})  = \bfsfI \\
	\bfSigma &\mapsto \tilde\bfSigma = \left(\rev{\Sreal}_{\rm hi}^{-\frac{1}{2}}\Sigmahi \rev{\Sreal}_{\rm hi}^{-\frac{1}{2}},\; (\rev{\SrealloOne})^{-\frac{1}{2}}\Sigmalo (\rev{\SrealloOne})^{-\frac{1}{2}},\; (\rev{\SrealloTwo})^{-\frac{1}{2}}\Sigmalo (\rev{\SrealloTwo})^{-\frac{1}{2}} \right) \\
	\Gamma_{\bfS, \bfI} &\mapsto \Gamma_{\tilde\bfS, \bfI} = \rev{G_\bfY \circ \Gamma_{\bfS, \bfI} \circ G_\bfY}.
\end{aligned}
\end{equation*}
Defining $B = (\rev{\SrealloTwo})^{-\frac{1}{2}}(\rev{\SrealloOne})^{\frac{1}{2}}$, instead of \cref{eq:mdist_min_ex_tsa} we can alternately solve 
\begin{equation}
	\left(\widehat{ \tilde{\Sigma}}_{\rm hi}, \widehat{ \tilde{\Sigma}}_{\rm lo} \right) = \argmin_{\tilde \Sigma_{\rm hi}, \tilde \Sigma_{\rm lo} \in \bbP_d} \langle\log_{\tilde\Sigma}\bfsfI,\; \rev{(G_\bfY \inv \circ \Gamma_{\bfS, \bfI}\inv \circ G_\bfY \inv)} \log_{\tilde \Sigma } \bfsfI \rangle \quad \text{s.t. } \tilde\Sigma = (\tilde\Sigma_{\rm hi},\, \tilde\Sigma_{\rm lo},\, B\tilde\Sigma_{\rm lo} B\t )
	\label{eq:mdist_iddata}
\end{equation}
and transform the resulting minimizers to obtain 
\[
\Sigmahihat = \rev{\Sreal}_{\rm hi}^{\frac{1}{2}}\widehat{ \tilde{\Sigma}}_{\rm hi}\rev{\Sreal}_{\rm hi}^{\frac{1}{2}} \quad \text{and} \quad\Sigmalohat = (\rev{\SrealloOne})^{\frac{1}{2}}\widehat{\tilde\Sigma}_{\rm lo}(\rev{\SrealloOne})^{\frac{1}{2}}.
\]

The fact that $\rev{\tilde\bfSreal} = \bfsfI$ simplifies the form of $\log_{\tilde\bfSigma} \bfsfI$ relative to that of $\log_\bfSigma \rev{\bfSreal}$. Consider, for example, the first components of $\log_{\tilde\bfSigma} \bfsfI$ and $\rev{\log_\bfSigma\bfSreal}$, involving $\Sigmahi$ and $\tilde\Sigma_{\rm hi}$. We have  
\begin{equation*}
\log_\Sigmahi \rev{\Sreal_{\text{hi}}}  
= \Sigmahi \log\left(\Sigma_{\rm hi}\inv \rev{\Sreal_{\text{hi}}} \right), \quad \text{while} \quad \log_{\tilde{\Sigma}_{\rm hi}} \mathrm{I} = -\tilde{\Sigma}_{\rm hi} \log \left(\tilde{\Sigma}_{\rm hi} \right).
\end{equation*}

Within the framework of \Cref{sec:modelOnTS} the linear model \cref{eq:tangentspacemodel} for $\log_\bfSigma \bfS$ on $\rmT_\bfSigma \bbP_d^3$ suggests the following exponential model for the random variable $\bfS$ on $\bbP_d^3$, 
\begin{equation}
    \bfS = \exp_\bfSigma(\bm{\calE}) = \begin{bmatrix}
        \Sigma_{\rm hi}^{\frac{1}{2}}\exp(\Sigma_{\rm hi}^{-\frac{1}{2}}\calE_{\rm hi}^1 \Sigma_{\rm hi}^{-\frac{1}{2}})\Sigma_{\rm hi}^{\frac{1}{2}} \\[0.2cm]
\Sigma_{\rm lo}^{\frac{1}{2}}\exp(\Sigma_{\rm lo}^{-\frac{1}{2}}\calE_{\rm lo}^1 \Sigma_{\rm lo}^{-\frac{1}{2}})\Sigma_{\rm lo}^{\frac{1}{2}} \\[0.2cm] 
\Sigma_{\rm lo}^{\frac{1}{2}}\exp(\Sigma_{\rm lo}^{-\frac{1}{2}}\calE_{\rm lo}^2 \Sigma_{\rm lo}^{-\frac{1}{2}})\Sigma_{\rm lo}^{\frac{1}{2}}
    \end{bmatrix},
    \label{eq:manifoldmodel}
\end{equation}
where $\calE_{\rm hi}^1$, $\calE_{\rm lo}^1$, $\calE_{\rm lo}^2$ are mean-zero, symmetric-matrix valued perturbations. We see that when $\bm{\calE} = \begin{bmatrix}\mathbf{0}^{d\times d} & \mathbf{0}^{d\times d} & \mathbf{0}^{d\times d}\end{bmatrix}\t$ we indeed have $\bfS = \bfSigma$.

\subsection{Fixed-$\Sigmalo$ simplification} 
\label{sec:fixSigmalo}
In this section we consider the regression problem specifically with the setup of \Cref{sec:example_s3,sec:example_s4},
\begin{equation*}
	\bfS = \begin{bmatrix}
		\Shi \\ \SloOne \\ \SloTwo \end{bmatrix} \sim \left(\bfSigma= \begin{bmatrix}
		\Sigmahi \\ \Sigmalo \\ \Sigmalo
	\end{bmatrix}, \Gamma_\bfS  = \E[\log_\bfSigma \bfS \otimes_\bfSigma \log_\bfSigma \bfS] \right),
\end{equation*}
with $\Shi$ and $\SloOne$ correlated and $\SloTwo \indep (\Shi, \SloOne)$. We motivate our development by the setting in which $\Shi$ and $\SloOne$ are sample covariance matrices constructed from $M_1$ coupled pairs of $(\Xhi, \Xlo)$ and $\SloTwo$ a sample covariance matrix constructed from an additional $M_2$ i.i.d.\ samples of $\Xlo$, as discussed in \Cref{sec:example_s3}. In instances when the total number of low-fidelity samples $M = M_1 + M_2$ is high relative to $d$, which may occur if sampling $\Xlo$ is cheap, the sample covariance matrix $\Slobar = \widehat{\Cov}[\{X_{\rm lo}^{(i)}\}_{i=1}^{M}]$ may be a good estimate of $\Sigmalo$ {on its own}, absent any multifidelity correction. Indeed, we have seen in practice that the estimate $\Sigmalohat$ resulting from solving \cref{eq:mdist_min} often does not differ greatly from $\Slobar$ when $M \gg d$. 

A reasonable and cost-effective approach to solving \cref{eq:mdist_min} in this setting is to fix $\Sigmalo = \Slobar$ in the squared Mahalanobis distance and obtain a simplified multifidelity estimator for $\Sigmahi$,
\begin{equation} 
	\Sigmahihat = \argmin_{\Sigmahi \in \bbP_d} \langle \log_{\bfSigma} \bfS,\; \Gamma_{\bfS,
 \bfI}\inv \log_{\bfSigma} \bfS \rangle  \quad \text{s.t. } \bfSigma = (\Sigmahi, \Slobar, \Slobar). 
 \label{eq:fixlo_three}
\end{equation} 

When $\Sigmalo$ is fixed in this manner, the objective function simplifies and we effectively solve 
\begin{equation} 
\begin{aligned}
\Sigmahihat = \argmin_{\Sigmahi \in \bbP_d} \left\langle \log_{\bfSigma_{1:2}}\bfS_{1:2}, \; \Gamma_{\bfS_{1:2}, \bfI}\inv \log_{\bfSigma_{1:2}}\bfS_{1:2} \right\rangle \quad \text{s.t. } \bfSigma_{1:2} = (\Sigmahi, \bar S_{\rm lo}), 
\end{aligned}
\label{eq:twomat_mdist}
\end{equation} 
where $\bfS_{1:2} = (\Shi, \SloOne)$ and $\Gamma_{\bfS_{1:2}, \bfI} = \E[\log_{\bfSigma_{1:2}}(\bfS_{1:2}) \otimes \log_{\bfSigma_{1:2}}(\bfS_{1:2})]$ is the upper ``block'' of $\Gamma_{\bfS, \bfI}$ corresponding to the variables $\Shi$ and $\SloOne$. Thus we do not need to include $\SloTwo$ in our optimization for $\Sigmahi$ when $\Sigmalo$ is fixed \textit{a priori}; rather, in the case that $\SloOne$ and $\SloTwo$ are SCMs, we only combine the samples of $\Xlo$ that would correspond to $\SloTwo$ with those involved in $\SloOne$ to construct $\bar S_{\rm lo} \approx \Sigmalo$. In the remainder of this section we thus use $\bfS$ to refer to the $\bbP_d^2$-valued random variable $\bfS = (\Shi, \SloOne) \equiv (\Shi, \Slo)$ with mean and covariance 
\begin{equation}
\bfS \sim \left( \bfE[\bfS] = (\Sigmahi, \Sigmalo) = \bfSigma, \; \Gamma_\bfS = \E\left[\log_\bfSigma \bfS \otimes \log_\bfSigma \bfS\right] \right)
\label{eq:twoMatrixRV}
\end{equation}  
and understand $\bar S_{\rm lo}$ to refer to a very good \textit{a priori} estimate of $\Sigmalo$. In writing \cref{eq:twoMatrixRV} we have taken $\Gamma_{\bfS} \equiv \Gamma_{\bfS, \bfI}$ and in the following will drop the dependence of Mahalanobis distance on $\langle \cdot, \cdot \rangle_\bfSigma$, as allowed by \cref{thm:ts_agnostic}. 

Beyond being computationally convenient, the simplification \cref{eq:twomat_mdist} is more analytically tractable than the full regression problem \cref{eq:mdist_min_ex}. For instance,  \cref{eq:twomat_mdist} has a closed-form expected minimum Mahalanobis distance in the case that $\Sigmalo$ is known exactly: 
\begin{proposition}
Suppose that $\Sigmalo$ is fixed at its \textit{true value} in \cref{eq:twomat_mdist}. 
The expected value of the corresponding minimum Mahalanobis distance is 
\begin{equation} 
\E_{(\Shi, \Slo)} \left[\argmin_{\Sigmahi \in \bbP_d} \left\langle \begin{bmatrix} \log_\Sigmahi(\Shi) \\ \log_\Sigmalo(\Slo) \end{bmatrix}, \; \Gamma_{S} \inv \begin{bmatrix} \log_\Sigmahi(\Shi) \\ \log_\Sigmalo(\Slo) \end{bmatrix}  \right\rangle \right] = \sdfrac{d(d+1)}{2}.
 \label{eq:fixlo_mdist}
\end{equation} 
\label{prop:expectedMdist}
\end{proposition}

Knowing the expected minimum of \cref{eq:twomat_mdist} can be useful when implementing regularization schemes, as we discuss in \Cref{sec:regularization}. 
Furthermore, the \textit{minimizer} of \cref{eq:twomat_mdist} satisfies a nonlinear equation interpretable as a control-variate estimator in the affine-invariant geometry for $\bbP_d$. We make this result explicit and discuss consequent connections in \Cref{sec:mfGeneralGeom}.

\subsection{Multifidelity estimation in general geometries} 
\label{sec:mfGeneralGeom}
In this section we unify the multifidelity covariance estimators of \cite{maurais2023logEuclidean}, which employ the Euclidean and log-Euclidean geometries for $\bbP_d$, with our regression estimator \cref{eq:mdist_min}, formulated using the affine-invariant geometry, and discuss broader implications for multifidelity estimation of covariance matrices. Our discussion centers on a striking result arising from the fixed-$\Sigmalo$ simplification \cref{eq:twomat_mdist} of the regression estimator: 
the solution to the Mahalanobis distance minimization problem in this setting satisfies a \textit{nonlinear control variate equation}.

\begin{proposition}
Consider the regression problem 
\begin{equation} 
\begin{aligned}
	\hat{\Sigma}_{\rm hi} &= \argmin_{\Sigmahi \in \bbP_d} \langle \log_{\bfSigma} S,\; \Gamma_{\bfS }\inv \log_{\bfSigma} S \rangle  \quad \text{s.t. } \Sigma = (\Sigmahi, \bar{S}_{\rm lo}) \\
	&= \argmin_{\Sigmahi \in \bbP_d} \left\langle \begin{bmatrix} \log_\Sigmahi(\Shi) \\ \log_{\bar{S}_{\rm lo}}(\Slo) \end{bmatrix}, \; \Gamma_{\bfS} \inv \begin{bmatrix} \log_\Sigmahi(\Shi) \\ \log_{\bar{S}_{\rm lo}}(\Slo) \end{bmatrix}  \right\rangle.  
\end{aligned}
\label{eq:twomat_reg}
\end{equation}     
$\hat{\Sigma}_{\rm hi}$ satisfies 
\begin{equation}
\log_{\hat\Sigma_{\rm hi}}(\Shi) = \rev{(\Gamma_{\rm lo, hi} \circ \Gamma_{\rm lo}\inv)} \log_{\bar{S}_{\rm lo}}(\Slo),
\label{eq:regest_nleq}
\end{equation} 
where $\Gamma_{\rm lo} = \E[\log_\Sigmalo (\Slo) \otimes \log_\Sigmalo (\Slo)]$ and $\Gamma_{\rm lo, hi} = \E[\log_\Sigmahi (\Shi) \otimes \log_\Sigmalo(\Slo)]$.
\label{prop:nleq}
\end{proposition}

The linear operator $\rev{\Gamma_{\rm lo, hi} \circ \Gamma_{\rm lo}\inv}$ is identifiable as the {optimal gain} between $\log_\Sigmahi(\Shi)$ and $\log_\Sigmalo(\Slo)$ and appears, e.g., in the context of vector-valued control variates \cite{rubinstein1985efficiency} and Kalman-type filtering schemes \cite{kalmanfilter, evensenbook}. 
\Cref{prop:nleq} reveals a satisfying connection: control-variate type multifidelity estimators can be viewed as a {special case} of the Riemannian multifidelity regression framework we develop here.

\subsubsection{Interpretation of fixed-$\Sigmalo$ estimator as control variates}
If we fix $\Sigmalo$ at $\Slobar$ and minimize Mahalanobis distance over $\Sigmahi$ alone \cref{eq:fixlo_three}, we obtain $\Sigmahihat \equiv \Sigmahihat^{\rm MRMF}$ satisfying 
\begin{equation}
\log_{\Sigmahihat^{\rm MRMF}}(\Shi) = \rev{(\Gamma_{\rm lo, hi} \circ \Gamma_{\rm lo}\inv)} \log_{\Slobar}(\Slo), 
\label{eq:reg_as_cv}
\end{equation}
where $\Gamma_{\rm lo, hi}  = \E[\log_\Sigmahi(\Shi) \otimes \log_\Sigmalo(\Slo)]$ is the Riemannian cross-covariance between $\Shi$ and $\Slo$ and $\Gamma_{\rm lo} = \E[\log_\Sigmalo(\Slo) \otimes \log_\Sigmalo(\Slo)]$ is the Riemannian auto-covariance of $\Slo$. 
As discussed in \Cref{sec:intrinsicStatistics},  the Riemannian logarithm $\log_A B = A^{\frac{1}{2}}\log(A^{-\frac{1}{2}} B A^{\frac{1}{2}}) A^{\frac{1}{2}}$ can be interpreted as a ``difference'' between $A, B \in \bbP_d$.  
In \eqref{eq:reg_as_cv} we see that the ``difference'' between our Mahalanobis distance-minimizing estimate of $\bfE[\Shi] = \Sigmahi$ and our sample of $\Shi$ is equal to the ``difference'' between $\Slobar \approx \Sigmalo = \bfE[\Slo]$ and our sample of $\Slo$, multiplied by the {optimal gain} \rev{$\Gamma_{\rm lo, hi}\circ \Gamma_{\rm lo}\inv$}. This relationship has the form of an optimal control variate equation analogous to those employed in \cite{maurais2023logEuclidean}:

\paragraph{Euclidean control variate estimator}
The Euclidean multifidelity (EMF) covariance estimator of \cite{maurais2023logEuclidean} is given in the form
\begin{equation}
\Sigmahihat^{\rm EMF} = \Shi + \alpha(\Slobar - \Slo), \quad \alpha \in \R 
\label{eq:sigmaLCV}
\end{equation} 
where we have specialized to the bifidelity case and $\Shi$, $\Slo$, and $\Slobar$ are as in \Cref{sec:fixSigmalo}. The optimal scalar value for $\alpha$ is $\frac{\trace{\Psi_{\rm lo, hi}}}{\trace{\Psi_{\rm lo}}}$ \cite{maurais2023logEuclidean},
where $\Psi_{\rm lo, hi} = \E[(\Shi - \Sigmahi) \otimes (\Slo - \Sigmalo)]$ and $\Psi_{\rm lo} = \E[(\Slo - \Sigmalo) \otimes (\Slo - \Sigmalo)]$ are Euclidean covariances. More generally, if we allow $\alpha$ to be {linear operator}-valued, the optimal LCV estimator satisfies 
\begin{equation}
\Sigmahihat^{\rm EMF} - \Shi = \rev{(\Psi_{\rm lo, hi} \circ \Psi_{\rm lo}\inv)} (\Slobar - \Slo).
\label{eq:LCV_rearr}
\end{equation}
\Cref{eq:LCV_rearr} has the same form as \cref{eq:reg_as_cv}: the difference (computed via subtraction) between $\Sigmahihat^{\rm EMF}$ and $\Shi$ is equal to the difference between $\Slobar$ and $\Slo$ scaled by \rev{$\Psi_{\rm lo, hi}\circ \Psi_{\rm lo}\inv$}, 
the Euclidean analogue of \rev{$\Gamma_{\rm lo, hi} \circ \Gamma_{\rm lo}\inv$}. 

\paragraph{Log-linear control variate estimator}
As a positive-definiteness-preserving alternative to \cref{eq:sigmaLCV}, the authors \cite{maurais2023logEuclidean} propose the log-Euclidean multifidelity (LEMF) estimator,
\begin{equation}
\log \Sigmahihat^{\rm LEMF} = \log \Shi + \alpha(\log\Slobar - \log\Slo), 
\label{eq:LEMF}
\end{equation}
which is a linear control variate estimator in the log-Euclidean geometry for $\bbP_d$ \cite{arsigny2006log}. If we seek to minimize the \textit{log-Euclidean} MSE,  $\E[||\log\Sigmahihat^{\rm LEMF} - \rev{\log \Sigmahi} ||_{\rm F}^2]$, and once again allow $\alpha$ to be a linear operator, then the resulting optimal log-Euclidean control variate estimator satisfies  
\begin{equation}
	\log \Sigmahihat^{\rm LEMF} - \log \Shi = \rev{(\Phi_{\rm lo, hi} \circ \Phi_{\rm lo}\inv)} (\log \Slobar - \log \Slo),
	\label{eq:le_est}
\end{equation}
where $\Phi_{\rm lo, hi}$ and $\Phi_{\rm lo}$ are {log-Euclidean} covariances
\[ 
\Phi_{\rm lo, hi} = \E[(\log \Shi - \log \Sigmahi) \otimes (\log \Slo - \log \Sigmalo)], \quad  \Phi_{\rm lo} = \E[(\log \Slo - \log \Sigmalo) \otimes (\log \Slo - \log \Sigmalo) ].  
\]
The LEMF estimator \eqref{eq:le_est} has the same form as the fixed-$\Sigmalo$ regression estimator \eqref{eq:reg_as_cv} and the LCV estimator \eqref{eq:LCV_rearr}. 
Indeed, each of the three estimators takes the form of a control variate equation in a different geometry for $\bbP_d$:
\begin{align}
	\Sigmahihat^{\rm EMF} - \Shi &= \rev{(\Psi_{\rm lo, hi} \circ \Psi_{\rm lo}\inv)} (\Slobar - \Slo) & \text{Euclidean geometry} \label{eq:cv_euclidean}\\
	\log \Sigmahihat^{\rm LEMF} - \log \Shi &= 
 \rev{(\Phi_{\rm lo, hi} \circ \Phi_{\rm lo}\inv)} (\log \Slobar - \log \Slo) & \text{Log-Euclidean geometry} \label{eq:cv_logEuclidean} \\ 
 	\log_{\Sigmahihat^{\rm MRMF}}(\Shi) &= \rev{(\Gamma_{\rm lo, hi} \circ \Gamma_{\rm lo}\inv)} \log_{\Slobar}(\Slo) & \text{Affine-invariant geometry} \label{eq:cv_affinvar}
\end{align}
Furthermore, the control variate estimators \cref{eq:cv_affinvar,eq:cv_euclidean,eq:cv_logEuclidean} have the form of best linear unbiased estimators (BLUEs) on the tangent spaces corresponding to their respective geometries, as we state in the following theorem. 
\begin{theorem}
    Suppose that $\Sigmalo$ is known and we are given a random variable pair $(\Shi, \Slo)$
    such that $(\Shi, \Slo)$ is an unbiased estimator of $(\Sigmahi, \Sigmalo)$ in the sense of \cite{smith2005covariance} under the (a) Euclidean, (b) log-Euclidean, or (c) affine-invariant geometry for $\bbP_d$. That is, (a) $\E[\Shi - \Sigmahi] = \bf 0$ and $\E[\Slo - \Sigmalo] = \bf 0$, (b) $\E[\log \Shi - \log\Sigmahi] = \bf 0$ and $\E[\log \Slo - \log\Sigmalo] = \bf 0$, or (c) $\E[\log_\Sigmahi \Shi] = \bf 0$ and $\E[\log_\Sigmalo \Slo] = \bf 0$. 
    The following implications hold:
    \begin{enumerate}[label=\Roman*., left=0.05cm]
       \item If (a), then the best unbiased linear estimator of $\Sigmahi$ on $\bbH_d$ equipped with the Frobenius metric satisfies 
        \[
        \hat\Sigma_{\rm hi}^{\rm EMF} - \Shi = \rev{(\Psi_{\rm lo, hi} \circ \Psi_{\rm lo}\inv)} (\Sigmalo - \Slo).
        \]
        \item If (b), then the best unbiased linear estimator of $\Sigmahi$ on $\bbP_d$ equipped with the log-Euclidean geometry and metric satisfies
        \[
        \log \hat\Sigma_{\rm hi}^{\rm LEMF} - \log \Shi = \rev{(\Phi_{\rm lo, hi} \circ \Phi_{\rm lo}\inv)} (\log \Sigmalo - \log \Slo).
        \]
        \item If (c), then the best unbiased linear-on-tangent-space estimator of $\Sigmahi$ on $\bbP_d$ equipped with the affine-invariant geometry satisfies
        \[
        \log_{\hat\Sigma_{\rm hi}^{\rm MRMF}}(\Shi) = \rev{(\Gamma_{\rm lo, hi} \circ \Gamma_{\rm lo}\inv)} \log_{\Sigmalo}(\Slo).
        \]
    \end{enumerate}
\label{prop:BLUEs}
\end{theorem}

\subsubsection{Generality of the regression framework}
As we obtained \cref{eq:cv_affinvar} by applying the simplifying assumption that $\Sigmalo$ is (approximately) known in \cref{eq:mdist_min}, we could have also obtained \cref{eq:cv_euclidean,eq:cv_logEuclidean} by simplifying analogous regression estimators formulated following the structure of \Cref{sec:formulation} but with the Euclidean or log-Euclidean instead of the affine-invariant geometry for $\bbP_d$. Indeed, because the Euclidean and log-Euclidean geometries both feature vector-space structure, the Mahalanobis distance minimization \cref{eq:mdist_min}, which under the affine-invariant geometry defines a nonlinear least-squares problem, would become a {linear} least squares problem, either for $\Sigma_0, \dots, \Sigma_L$ or $\log\Sigma_0, \dots, \log\Sigma_L$, possessing a closed-form solution analogous to that of multilevel scalar BLUEs in \cite{schaden2020multilevel, schaden2021asymptotic}. One could consider a number of other geometries for $\bbP_d$ as well, including Bures-Wasserstein \cite{MalagoEtAl2018, bhatia2019bures} and log-Cholesky \cite{Lin2019}.  
Choice of geometry within the context of multifidelity covariance estimation should depend on a number of factors, including computational complexity, availability of Riemannian logarithmic and exponential maps, preservation of positive-definiteness, and desired interpretation. We discuss implications of this generality for multifidelity estimation more broadly in \Cref{sec:conclusion}.
 
\section{Computational approaches} 
\label{sec:computation}
We now discuss the practicalities of solving 
\begin{equation*}
	(\hat\Sigma_0, \dots, \hat\Sigma_L) = \argmin_{\Sigma_0,\dots, \Sigma_L \in \bbP_d} \langle \log_{\bfSigma} \bfS,\; \Gamma_\bfS\inv \log_{\bfSigma} \bfS \rangle_{\bfSigma} \quad \text{s.t. } \bfSigma = \mu_\bfS(\Sigma_0, \dots, \Sigma_L)
\end{equation*}
in order to estimate the high-fidelity covariance matrix $\Sigma_0$ and (as an added bonus) low-fidelity covariance matrices $\Sigma_1, \dots, \Sigma_L$, where $\bfS$ and $\mu_\bfS(\Sigma_0, \dots, \Sigma_L)$ are as in \cref{eq:S_rv}. As made possible by \cref{thm:ts_agnostic}, in this section we work exclusively with the equivalent problem formulated in terms of the inner- and outer-products of $\rmT_\bfI \bbP_d^{N}$, 
\begin{equation}
	(\hat\Sigma_0, \dots, \hat\Sigma_L) = \argmin_{\Sigma_0,\dots, \Sigma_L \in \bbP_d} \langle \log_{\bfSigma} \bfS,\; \Gamma_{\bfS, \bfI}\inv \log_{\bfSigma} \bfS \rangle \quad \text{s.t. } \bfSigma = \mu_\bfS(\Sigma_0, \dots, \Sigma_L)
	\label{eq:mdist_min_I}.
\end{equation}

\subsection{Regularization in the intrinsic metric}
\label{sec:regularization}
We have noticed empirically that computing the Mahalanobis distance objective can be numerically unstable even when preconditioning is employed. 
A helpful tool for addressing this issue is regularization in the intrinsic metric; instead of solving \cref{eq:mdist_min_I} as written, we solve a \textit{penalized version}
\begin{equation}
\begin{multlined}
(\hat\Sigma_0, \dots, \hat\Sigma_L) = \argmin_{\Sigma_0,\dots, \Sigma_L \in \bbP_d} \langle \log_{\bfSigma} \bfS,\; \Gamma_{\bfS, \bfI}\inv \log_{\bfSigma} \bfS \rangle  + \sum_{\ell = 1}^L \lambda_\ell\, \rev{d^2(\Sigma_\ell, I)} \\ \text{s.t. } \bfSigma = \mu_\bfS(\Sigma_0, \dots, \Sigma_L) 
\label{eq:mdist_min_reg}
\end{multlined}
\end{equation}
where $\lambda_1, \dots, \lambda_L > 0$ are positive regularization parameters. The \rev{intrinsic distance regularization terms are computable as} $d^2(\Sigma_\ell, I) = ||\log(\Sigma_\ell) ||_{\rm F}^2$ and help control the conditioning of $\hat\Sigma_0, \dots, \hat\Sigma_L$, as the intrinsic distance \rev{\cref{eq:intrinsicdist} between the identity and any other matrix will be large if the eigenvalues of that matrix are close to zero. We employ identity matrices as regularization targets for ease of computation, but, more generally, one could penalize the optimization \eqref{eq:mdist_min_reg} with $\sum_{\ell = 1}^L d^2(\Sigma_\ell, \Sigma_\ell^{\rm ref})$, where $\Sigma^{\rm ref}_\ell$, $\ell = 1,\dots,L$ are suitably-chosen, problem-dependent reference covariance matrices.} 
\subsubsection{Regularization parameter selection}
\label{sec:regparselect}
\begin{figure}[H]
	\begin{subfigure}{0.32\linewidth}
		\subcaption*{$d = 3$, $\frac{d(d+1)}{2} = 6$}
		\includegraphics[width=\linewidth]{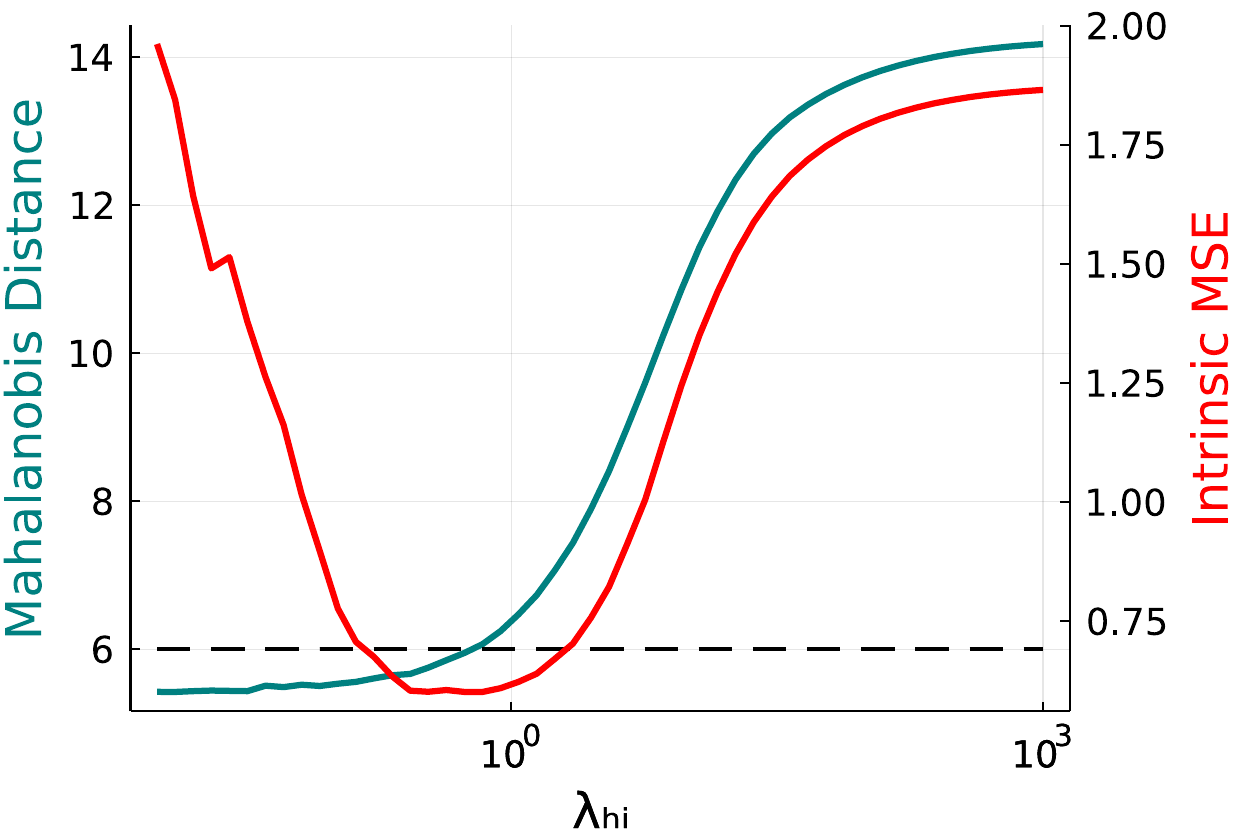}
	\end{subfigure}
	\begin{subfigure}{0.32\linewidth}
		\subcaption*{$d = 4$, $\frac{d(d+1)}{2} = 10$}
		\includegraphics[width=\linewidth]{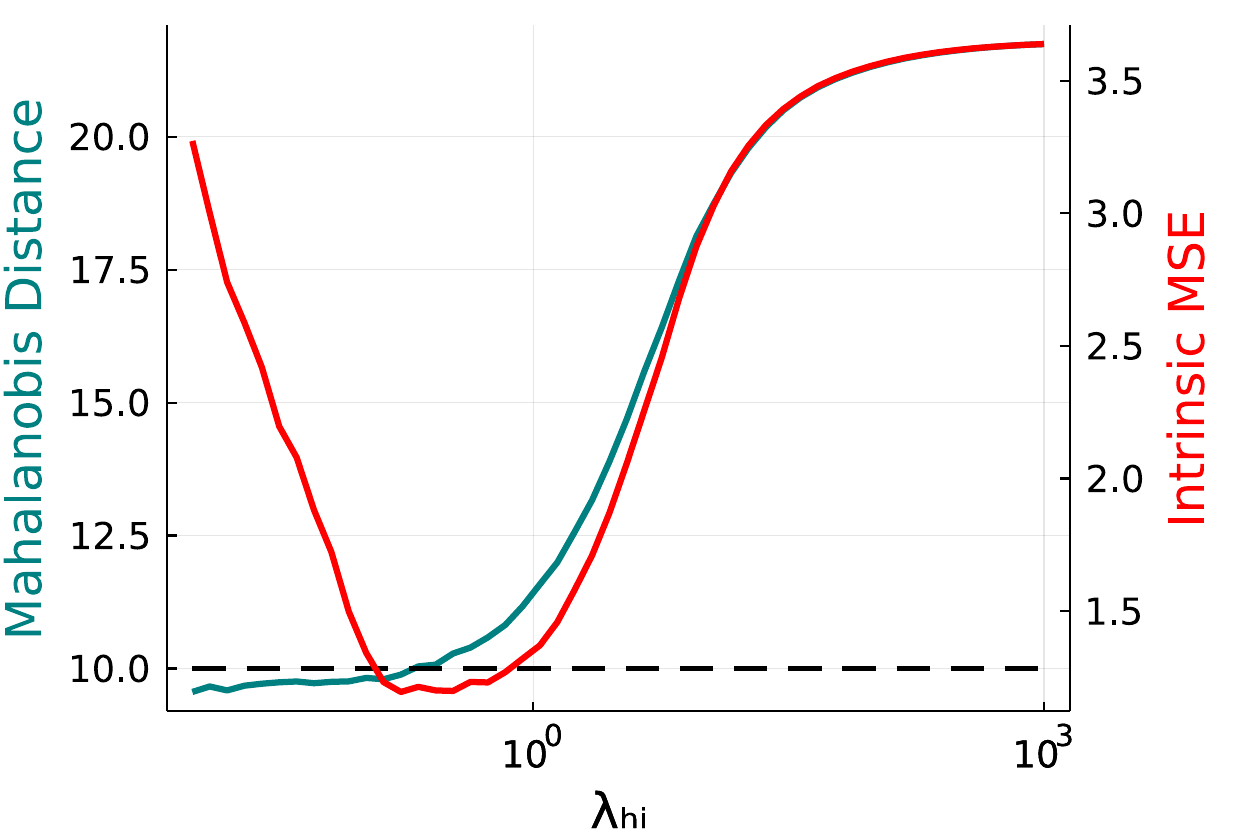}
	\end{subfigure}
	\begin{subfigure}{0.32\linewidth}
		\subcaption*{$d = 5$, $\frac{d(d+1)}{2} = 15$}
		\includegraphics[width=\linewidth]{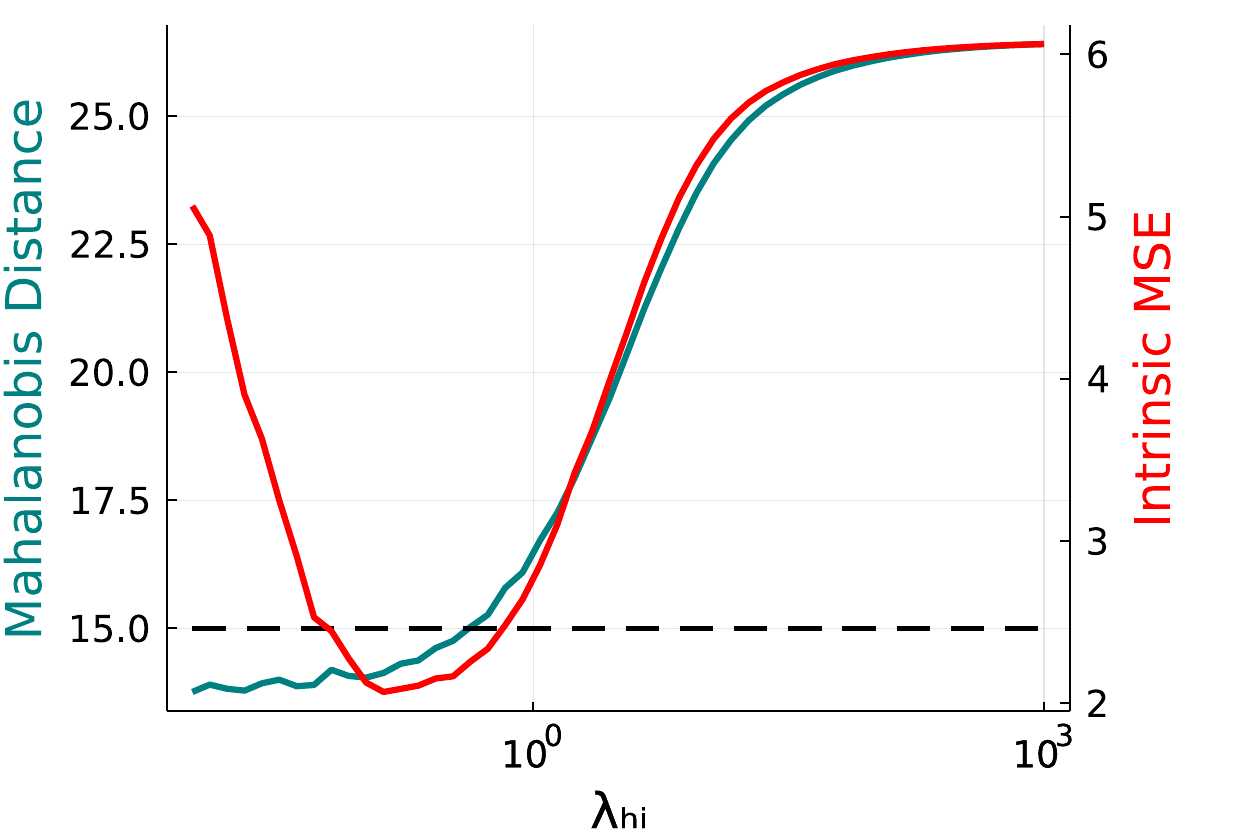}
	\end{subfigure}
\caption{Intrinsic MSE of $\Sigmahihat$ (red) and mean Mahalanobis distance at $\Sigmahihat$ (teal) as a function of regularization parameter $\lambda$ in the fixed-$\Sigmalo$ setting. We vary the dimension $d \in \{3, 4, 5\}$ within a class of simple example problems. The $\lambda$ associated with the minimum of the MSE curves corresponds closely to that associated with mean Mahalanobis distance equal to $\frac{d(d+1)}{2}$, plotted with dashed black lines.} 
\label{fig:regParSelect}
\end{figure}

As with any penalized estimator, the regularization parameters in \cref{eq:mdist_min_reg} should be tuned to balance data fitting, encapsulated in the Mahalanobis distance term, with regularity, accounted for in the intrinsic distance terms. While we leave development of regularization parameter selection methods for the general estimator \cref{eq:mdist_min_reg} to future work, in the specific case of the fixed-$\Sigmalo$ simplification (\Cref{sec:fixSigmalo}) we have found a useful heuristic. The penalized regression problem in this setting reads 
\begin{equation} 
\Sigmahihat = \argmin_{\Sigmahi \in \bbP_d} \langle \log_\bfSigma \bfS, \; \Gamma_\bfS\inv \log_\bfSigma \bfS \rangle + \lambda_{\rm hi} ||\log (\Sigmahi) ||_{\rm F}^2 , \quad \text{s.t. } \bfSigma = \mu_\bfS(\Sigmahi, \Slobar) 
\label{eq:fixlo_penalized}
\end{equation} 
In testing our estimator and sweeping over a wide range of regularization parameters, we found that the choice of $\lambda_{\rm hi}$ minimizing MSE in the intrinsic metric closely corresponded to that yielding a mean squared Mahalanobis distance of $\frac{d(d+1)}{2}$, as demonstrated in \cref{fig:regParSelect}. 
We saw in \Cref{sec:fixSigmalo} that when $\Sigmalo$ is known exactly, the analytical minimum of \eqref{eq:twomat_reg} solved without regularization has expectation $\frac{d(d+1)}{2}$. Thus it appears that the best choice of regularization parameter with $\Sigmalohat$ fixed at $\Slobar \approx \Sigmalo$ is that which ensures that the statistics of our computed solution \cref{eq:fixlo_penalized} match the statistics of the theoretical solution \eqref{eq:regest_nleq}. 

\subsection{Square root parameterization} 
Solving \cref{eq:mdist_min_I} directly requires optimization over the manifold-valued variables $\Sigma_0, \dots, \Sigma_L \in \bbP_d$,
to which standard gradient-based methods are not directly applicable. Although there do exist methods and software packages for manifold optimization, e.g., \cite{absil2009optimization, boumal2014manopt, sra2015conic}, we choose to circumvent their machinery by reformulating the problem in terms of matrix square roots. 
Instead of solving \cref{eq:mdist_min_I}, we solve 
\begin{equation}
	(\hat B_0, \dots, \hat B_L) \in \argmin_{B_0, \dots, B_L \in \bbH_d} \left\langle\log_{\bfSigma} \bfS,\, \Gamma\inv_{\bfS, \bfI}\log_\bfSigma \bfS \right\rangle \quad \text{s.t. } \bfSigma = \mu_\bfS(B_0^2, \dots, B_L^2)
	\label{eq:sqrtOpt},
\end{equation}
optimizing over the {matrix square roots} $(B_0, \dots, B_L) \equiv \left(\Sigma_0^{1/2}, \dots, \Sigma_L^{1/2} \right)$ which inhabit the Euclidean vector space $\bbH_d$. The formulation \eqref{eq:sqrtOpt} lends itself to unconstrained gradient descent methods because, given a starting point consisting of $L+1$ symmetric matrices, as long as the descent directions are computed such that they lie in $\bbH_d^L$ (see, e.g., \cite{matrixcookbook, minka2000old}), the result of optimization will also be in $\bbH_d^L$. 

A slight subtlety of the square root formulation \cref{eq:sqrtOpt} is that it does not guarantee that the resulting $(\hat\Sigma_0, \dots, \hat\Sigma_L) = (\hat B_0^2, \dots, \hat B_L^2)$ will be strictly positive definite; rather it is only necessary that they be positive semidefinite. \rev{We enforce strict positive definiteness in \cref{eq:sqrtOpt} by adding regularization as in \Cref{sec:regularization}}. 

\section{Numerical results}
\label{sec:numerics}
In this section we demonstrate the performance of our multifidelity covariance estimator \cref{eq:mdist_min} in a forward uncertainty quantification setting (\Cref{sec:simpleGaussian}) and in a downstream machine-learning task known as metric learning (\Cref{sec:metricLearning}).

\subsection{Simple Gaussian example}
\label{sec:simpleGaussian}
The first test problem we consider is that of estimating the covariance of a high-fidelity four-dimensional Gaussian random variable $\Xhi \sim \calN(\mathbf{0}, \Sigmahi)$ by incorporating samples of a low-fidelity random variable related to $\Xhi$ by
\[
\Xlo = \Xhi + \varepsilon,
\]
where $\varepsilon \sim \calN(\mathbf{0}, \sigma^2 I)$ is independent of $\Xhi$. $\Xhi$ and $\Xlo$ are jointly Gaussian with 
\[
\begin{bmatrix}
    \Xhi \\ \Xlo 
\end{bmatrix} 
\sim \calN \left(\mathbf{0}, \; \begin{bmatrix}\Sigmahi & \Sigmahi \\ \Sigmahi & \Sigmahi + \sigma^2 I\end{bmatrix} \right). 
\]
We set $\sigma^2 = 0.7$ and choose $\Sigmahi$ from the Wishart ensemble in $d = 4$ dimensions, i.e., $\Sigmahi = A\t A$ where the entries of $A \in \R^{4 \times 4}$ were sampled i.i.d.\ from the standard normal distribution. 

We (artificially) impose costs $c_{\rm hi} = 1$ to sample $\Xhi$ and $c_{\rm lo} = 10^{-2}$ to sample $X_{\rm lo}$ and vary the total sampling budget $B$ in the interval $[6, 206]$. For each budget value we compute a regularized fixed-$\Sigmalo$ multifidelity regression estimator \cref{eq:fixlo_three} and EMF and LEMF control variate estimators using the optimal sample-allocation corresponding to the Euclidean estimator; see \cite{maurais2023logEuclidean} for details. We additionally compute equivalent-cost single-fidelity estimators using high-fidelity samples alone and low-fidelity samples alone for comparison.
\begin{figure}[H] 
\centering 
    \begin{subfigure}{0.32\linewidth}
    \includegraphics[width=\linewidth]{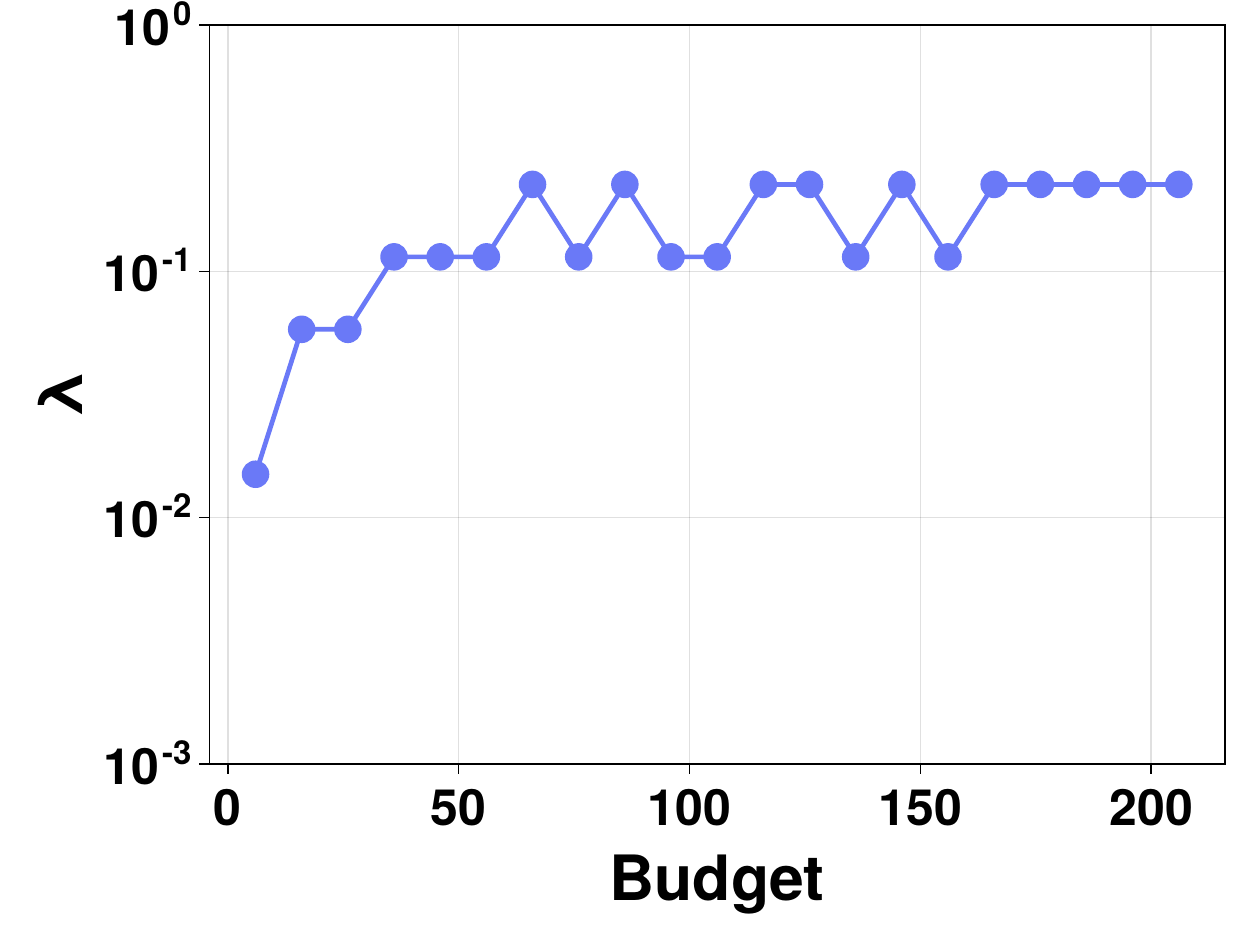}        
    \end{subfigure}
    \begin{subfigure}{0.32\linewidth}
    \includegraphics[width=\linewidth]{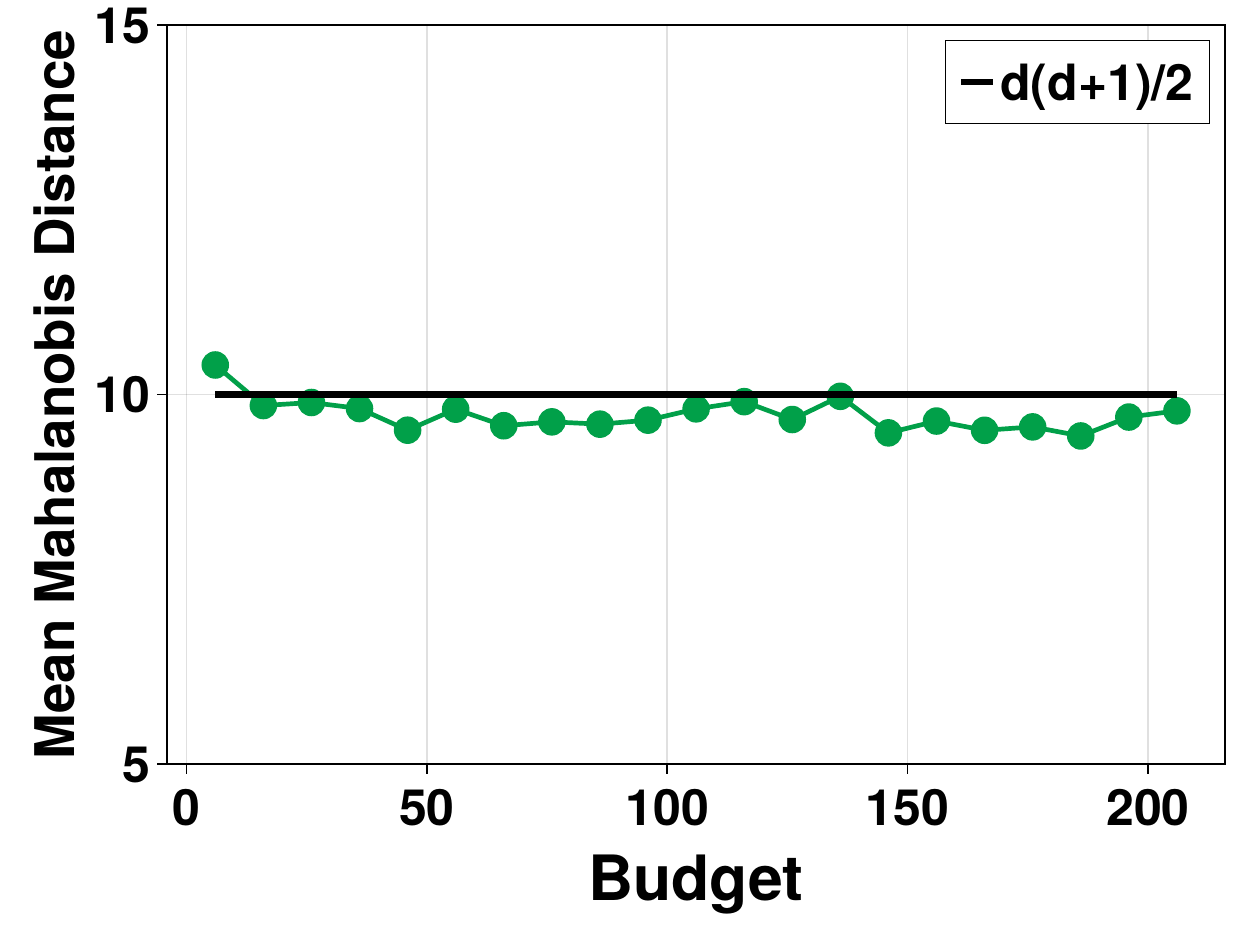}
    \end{subfigure}
    \begin{subfigure}{0.32\linewidth}
    \includegraphics[width=\linewidth]{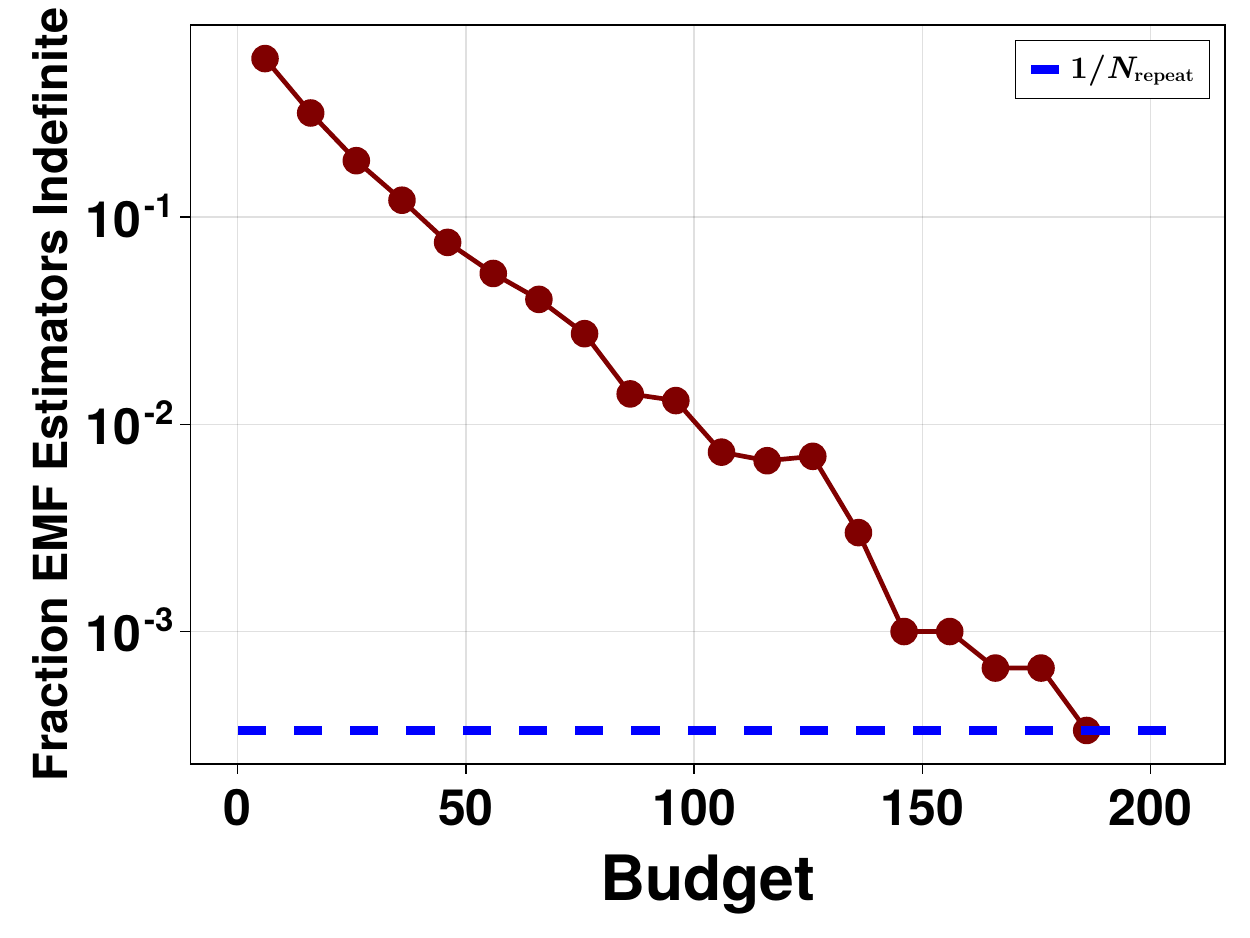}        
    \end{subfigure}
    \caption{Simple Gaussian example: Regularization parameters selected by matching mean minimum Mahalanobis distance over 32 pilot trials to $\frac{d(d+1)}{2}$ (left), resulting mean minimum Mahalanobis distance over 3000 trials using the selected regularization parameters (middle), and fraction of EMF estimators which were indefinite over 3000 repeated trials (right). All budgets except $B = 196$ resulted in at least one indefinite EMF estimator.}
    \label{fig:diagnostics_gaussian}
\end{figure}
For each value of the budget $B$ we pre-compute the covariance operator $\Gamma_{\bf S, I}$ using 1000 pilot samples. We additionally pre-compute the regularization parameter $\lambda_{\rm hi}$ in \cref{eq:fixlo_penalized} \textit{admissibly} by testing 18 values of $\lambda_{\rm hi}$ logarithmically spaced over $[10^{-3}, 10^2]$ and choosing the one corresponding most closely to a sample average minimum Mahalanobis distance of $\frac{d(d+1)}{2} = 10$ (computed over 32 trials), \rev{as in \cref{sec:regparselect}}. A plot of the selected regularization parameters and resulting mean Mahalanobis distance in the ensuing trials for each value of $B$ can be seen in \cref{fig:diagnostics_gaussian}. \rev{Due to the Gaussian structure in this example, the coefficients needed to construct the EMF and LEMF estimators are available and closed form (see \cite{maurais2023logEuclidean}), \textit{but} these closed forms depend on the values of $\Sigmahi$ and $\Sigmalo$. Though we use the closed forms of the LEMF and EMF parameters in this example in order to provide a higher standard of comparison for the MRMF estimator, in practice pilot samples would be needed to estimate these parameters as well. Hence, we do not include the cost of computing the pilot samples in the budgets in \cref{fig:diagnostics_gaussian,fig:se_gaussian,fig:SE_hists,fig:SE_hists_intrinsic}, as these costs are incurred equally by each multifidelity estimator in question.} 
\begin{figure}[h]
    \centering
      \begin{subfigure}{0.49\linewidth}
        \includegraphics[width=\linewidth]{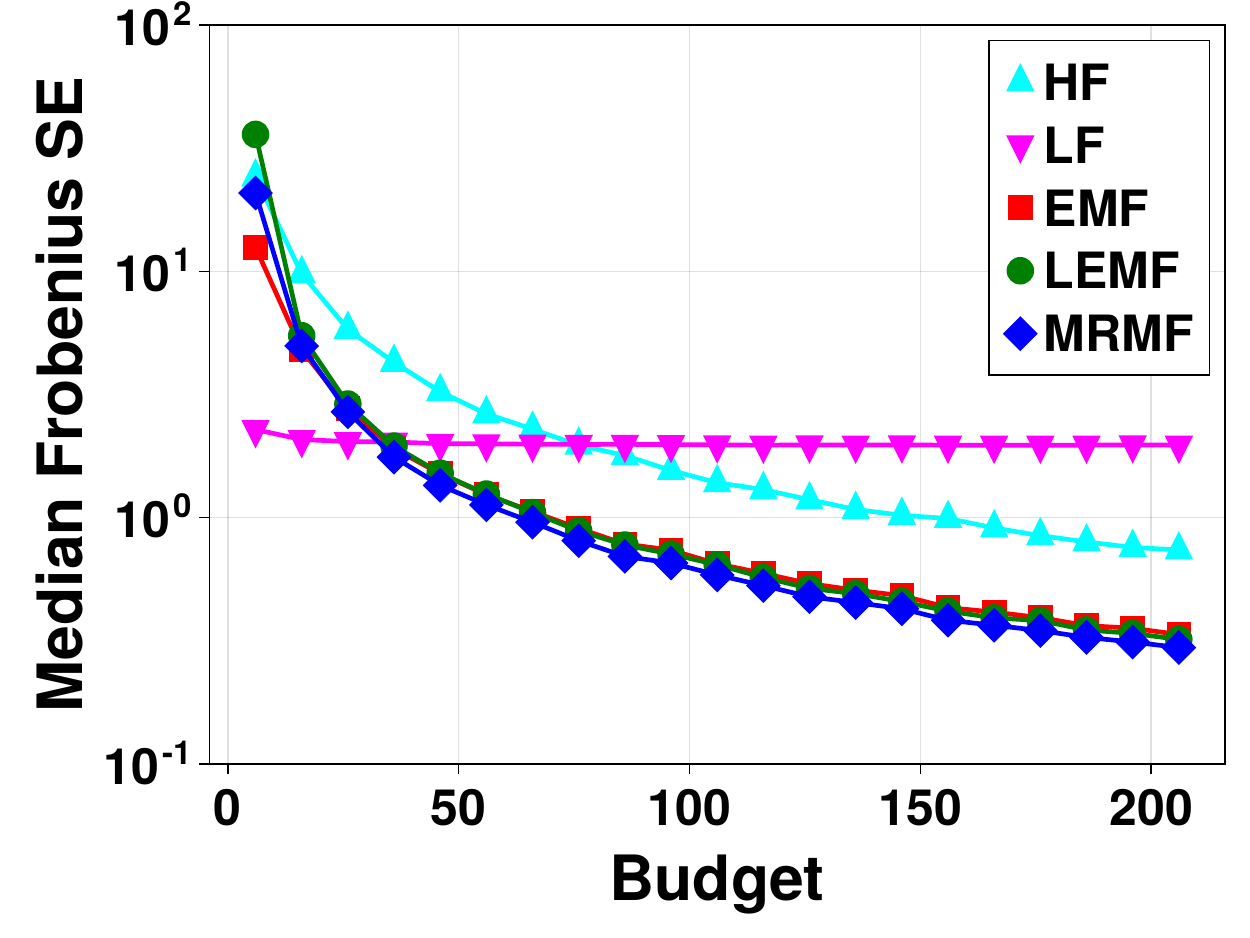}
    \end{subfigure}
    \begin{subfigure}{0.49\linewidth}
        \includegraphics[width=\linewidth]{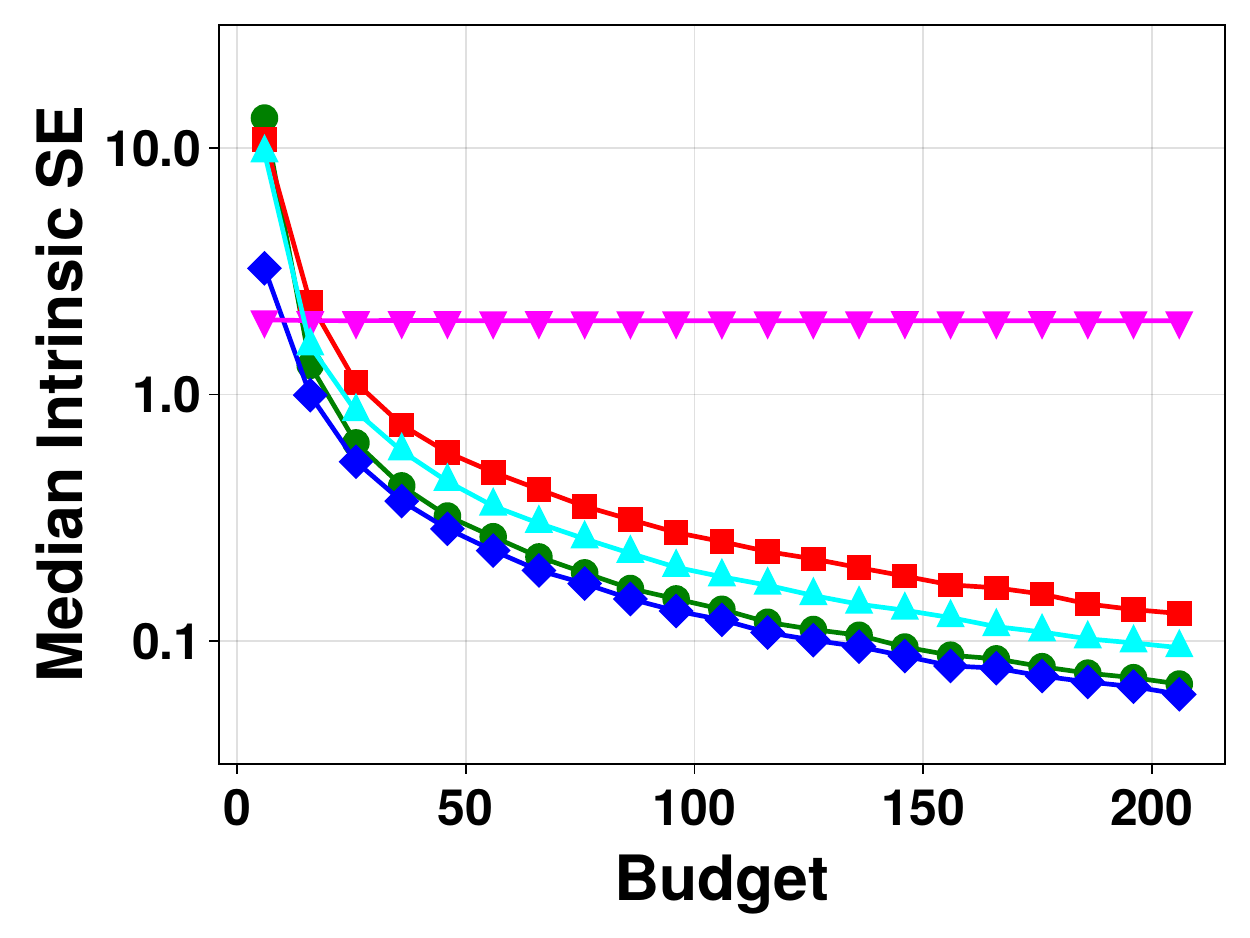}
    \end{subfigure}
    \caption{Simple Gaussian example: Median squared error in the Frobenius norm (left) and intrinsic metric (right).}
    \label{fig:se_gaussian}
\end{figure}

In \cref{fig:se_gaussian,fig:SE_hists,fig:SE_hists_intrinsic} we examine the squared error of our regression estimator $\Sigmahihat^{\rm MRMF}$ \cref{eq:fixlo_three}, the LEMF and EMF estimators $\Sigmahihat^{\rm LEMF}$ \cref{eq:LEMF} and $\Sigmahihat^{\rm EMF}$ \cref{eq:sigmaLCV} of \cite{maurais2023logEuclidean}, and equivalent-cost low-fidelity and high-fidelity estimators $\Sigmahihat^{\rm LF}$ and $\Sigmahihat^{\rm HF}$ 
over 2000 repeated trials. 
We see that $\Sigmahihat^{\rm MRMF}$ outperforms $\Sigmahihat^{\rm HF}$ at all budgets and often has an edge on $\Sigmahihat^{\rm LEMF}$ and $\Sigmahihat^{\rm EMF}$ as well.
While the performances of $\Sigmahihat^{\rm MRMF}$ and $\Sigmahihat^{\rm EMF}$ are similar in the Frobenius metric, $\Sigmahihat^{\rm EMF}$ does poorly in the intrinsic metric because it is frequently indefinite, and indefinite matrices are infinitely far away from $\Sigmahi$ in the intrinsic metric. The rates at which $\Sigmahihat^{\rm EMF}$ loses definiteness are shown in the right panel of \cref{fig:diagnostics_gaussian}; at the four lowest budgets they exceed 10\%. 
\begin{figure}[h]
\includegraphics[width=\linewidth]{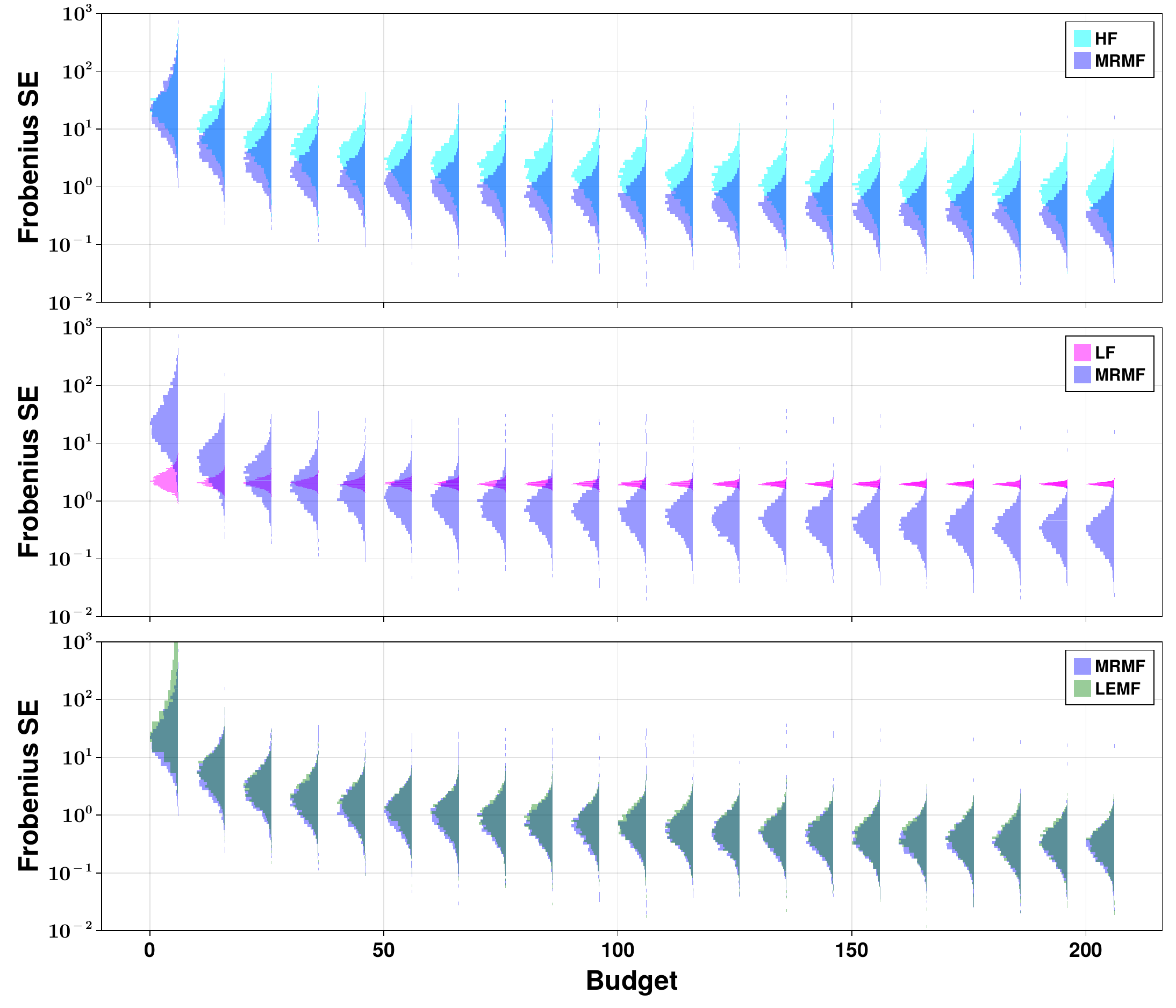}
\caption{Simple Gaussian example: Frobenius squared error histograms of $\Sigmahihat^{\rm MRMF}$ compared to %
$\Sigmahihat^{\rm HF}$ (top),  $\Sigmahihat^{\rm LF}$ (middle), and $\Sigmahihat^{\rm LEMF}$ (bottom). $\Sigmahihat^{\rm MRMF}$ attains significantly lower error than $\Sigmahihat^{\rm HF}$ at all budgets, intuitively because it obtains more information, via recourse to correlated low-fidelity samples, at the same cost. For small budgets $\Sigmahihat^{\rm LF}$ has lower squared error than $\Sigmahihat^{\rm MRMF}$ because its variability is small due to the large number of samples comprising it, but as the budget increases its bias becomes apparent and $\Sigmahihat^{\rm MRMF}$ yields estimates with lower error.}
\vskip -0.4cm
\label{fig:SE_hists}
\end{figure} 
Performances of $\Sigmahihat^{\rm MRMF}$ and $\Sigmahihat^{\rm LEMF}$ are similar in both metrics except for lower values of budget $B$, at which we have noticed that $\Sigmahihat^{\rm LEMF}$ can become unstable. Indeed, while the median squared Frobenius error of $\Sigmahihat^{\rm LEMF}$ looks reasonable in \cref{fig:se_gaussian}, the \textit{mean} squared error in the Frobenius metric at the lowest budget was quite high, on the order of $10^8$, due to a few extreme outliers. Likewise, in \cref{fig:SE_hists_intrinsic} we see that at the lowest budget the squared error distribution of $\Sigmahihat^{\rm LEMF}$ is shifted significantly upward from that of $\Sigmahihat^{\rm MRMF}$.

While the low-fidelity estimator $\Sigmahihat^{\rm LF}$ does out-perform the other estimators at the lowest budgets, its error stagnates as the budget is increased due to the presence of bias. By contrast, the squared errors of the multifidelity estimators decrease with increasing budget and quickly fall below that of $\Sigmahihat^{\rm LF}$ to such an extent that their histograms have almost no common support. 

\begin{figure}[h]
\includegraphics[width=\linewidth]{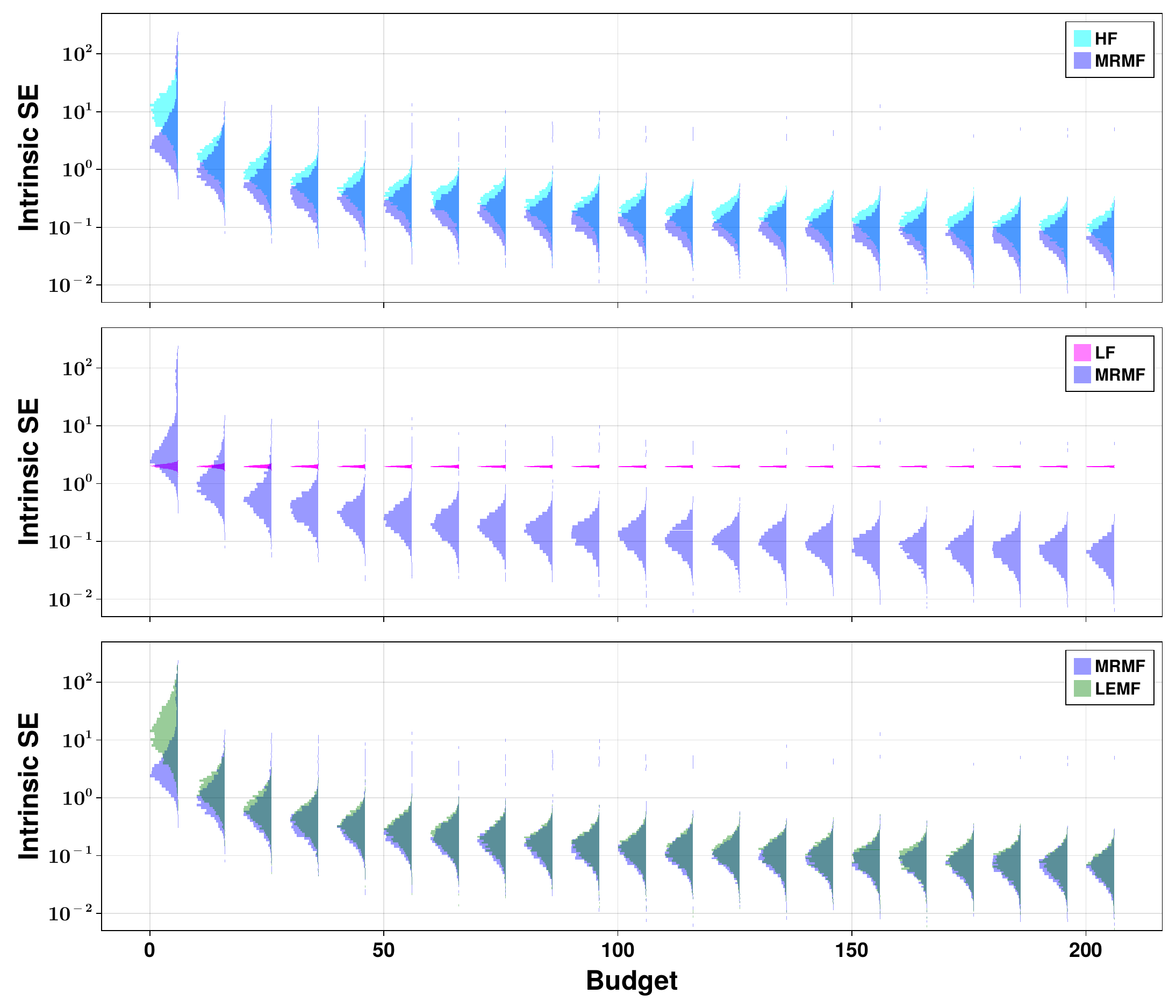}
\caption{Simple Gaussian example: Intrinsic squared error distributions of $\Sigmahihat^{\rm MRMF}$ as compared to $\Sigmahihat^{\rm HF}$ (top),  $\Sigmahihat^{\rm LF}$ (middle), and $\Sigmahihat^{\rm LEMF}$ (bottom). The advantages of $\Sigmahihat^{\rm MRMF}$ relative to $\Sigmahihat^{\rm HF}$ and $\Sigmahihat^{\rm LF}$ are more pronounced in the intrinsic metric, which compares matrices as operators by examining their generalized eigenvalues, than in the Frobenius metric, which compares matrices as vectors. The intrinsic metric also reveals the poor performance of $\Sigmahihat^{\rm LEMF}$ at low budgets, though at higher budgets the performances of $\Sigmahihat^{\rm MRMF}$ and $\Sigmahihat^{\rm LEMF}$ are comparable.}
\label{fig:SE_hists_intrinsic}
\end{figure}

\subsection{Metric learning with the surface quasi-geostrophic equation}
\label{sec:metricLearning}
In this second example we demonstrate the utility of our multifidelity covariance regression estimator applied within the geometric mean metric learning framework of \cite{zadeh2016geometric}. 

\subsubsection{Geometric mean metric learning}
Broadly speaking, the goal of metric learning is to obtain a distance measure over Euclidean space such that machine-learning tasks, including clustering and classification, are easier in the new metric for a given dataset \cite{xing2002distance, weinberger2009distance, kulis2013metric, bellet2013survey}. Supposing, for example, that we have a dataset containing points belonging to $K \geq 2$ distinct classes, an effective learned metric $d(\cdot, \cdot)$ on this space should place points from the same class close together while placing those from different classes far apart. If we consider only metrics which take the form of a Mahalanobis distance,
\[
d_A(\bfy, \bfy') = \sqrt{(\bfy - \bfy')\t A(\bfy - \bfy')}, 
\]
where $\bfy, \bfy' \in \R^d$ and $A \in \bbP_d$, the task of metric learning reduces to the task of obtaining a suitable symmetric positive definite ``metric matrix'' $A$. 
In \cite{zadeh2016geometric}, the authors propose a novel family of objective functions for the semi-supervised Mahalanobis metric learning problem. The optimal metric matrices admit closed form expressions as points on geodesics of $\bbP_d$ and are, up to scaling by constant factors, 
\begin{equation}
A_{\rm GMML} = T^{-\frac{1}{2}}(T^{\frac{1}{2}} D T^{\frac{1}{2}})^t T^{-\frac{1}{2}},
\label{eq:GMML}
\end{equation} 
where $t\in [0,1]$ and $T$ and $D$ are the \textit{similarity matrix} and \textit{dissimilarity matrix,}
\[
T = \E_{\mathrm{class}(\bfy) = \mathrm{class}(\bfy')}\left[\left(\bfy - \bfy'\right)\left(\bfy - \bfy'\right)\t \right], \quad D = \E_{\mathrm{class}(\bfy) \neq \mathrm{class}(\bfy')}\left[\left(\bfy - \bfy'\right)\left(\bfy - \bfy'\right)\t \right].
\]
When the dataset is drawn from an equal mixture of $K = 2$ classes, $T$ and $D$ can be written 
\begin{equation}
T = \Gamma_0 + \Gamma_1, \quad D = T + (\bfm_0 - \bfm_1)(\bfm_0 - \bfm_1)\t,
\label{eq:TandD_covs}
\end{equation} 
with $\Gamma_i = \Cov[\bfy \mid \mathrm{class}(\bfy) = i]$ and $\bfm_i = \E[\bfy \mid \mathrm{class}(\bfy) = i]$, $i \in \{0, 1\}$. We see from the formulations in \cref{eq:TandD_covs} that our ability to learn the metric matrix \eqref{eq:GMML} is strongly dependent on our ability to learn the covariance matrices $\Gamma_0$ and $\Gamma_1 \in \bbP_d$. 

\subsubsection{Surface quasi-geostrophic equation}
In this example we consider a metric learning problem in which our data are observations of solutions to a surface quasi-geostrophic (SQG) equation \cite{HeldEtAl1985} with parameters drawn according to a two-class mixture distribution. 
The SQG equation describes the evolution of the buoyancy $b(\bfx, t)$ over a periodic spatial domain $\calX = [-\pi, \pi] \times [-\pi, \pi]$ and is given by 
\begin{equation}
    \partial_t b(\bfx, t; \bftheta) + J(\psi, b) = 0 \, ,
\label{eq:SQG}
\end{equation}
where $\bfx$ is the spatial coordinate, $\psi$ is the streamfunction, $J(\psi, b)$ is the Jacobian determinant
\begin{equation*}
    J(\psi, b) = \left( \frac{\partial \psi}{\partial x_1} \right)\left(\frac{\partial b}{\partial x_2}  \right) - \left( \frac{\partial b}{\partial x_1}  \right)\left(\frac{\partial \psi}{\partial x_2}  \right) 
\end{equation*}
and $\bftheta \in \R^5$ are parameters. We specify the initial condition as
\[
    b_0(\bfx;\bftheta) = -\frac{1}{(2\pi/ |\theta_5|)^2} \exp\left( -x_1^2 - \exp(2\theta_1) x_2^2 \right),
\]
the contours of which form ellipses parameterized by $\theta_1$ and $\theta_5$. The remaining parameters $\theta_2$, $\theta_3$, and $\theta_4$ govern the dynamics \cref{eq:SQG}. The parameters $\bftheta$ are drawn from an equal mixture of $\pi_0\sim \calN(\bfmu_0, C)$ and $\pi_1 \sim (\bfmu_1, C)$, which differ only in the mean of the log aspect-ratio $\theta_1$; see \cref{app:sqg} for details. We sample the solution to \eqref{eq:SQG} at nine equally spaced points in the domain $\calX$ to obtain observations $\bfy \in \R^9$. Our goal in the metric learning setting is to be able to distinguish samples of $\bfy \mid \bftheta \sim \pi_0$ from samples of $\bfy \mid \bftheta \sim \pi_1$. 

\subsubsection{Multifidelity metric learning}
The metric \cref{eq:GMML} can be learned by estimating $\Gamma_i = \Cov[\bfy \mid \bftheta \sim \pi_i]$ and $m_i = \Cov[\bfy \mid \bftheta \sim \pi_i]$, $i \in \{0,1\}$ from samples of $\bfy \mid \bftheta \sim \pi_0$ and $\bfy \mid \bftheta \sim \pi_1$. We cast this metric learning problem in the multifidelity setting as follows: Let $\bfy_{\rm hi}$ correspond to realizations of the observable when the SQG equation \cref{eq:SQG} is solved numerically over 256 grid points in each coordinate direction, and $\bfy_{\rm lo}$ correspond to observations when the SQG equation \cref{eq:SQG} is solved numerically over just 16 grid points in each direction. In both cases we compute the solution to time $T = 24$ with a time step of $\Delta t = 0.05$, so we associate the costs of sampling $\bfy_{\rm hi}$ and $\bfy_{\rm lo}$ with the number of grid points in the solver; $c_{\rm hi} = 256^2 = 65,536$ and $c_{\rm lo} = 16^2 = 256$. $\bfy_{\rm hi}$ is thus 256 times more expensive to sample than $\bfy_{\rm lo}$. 

Our goal is to learn the covariance matrices $\Gamma_0$ and $\Gamma_1$, and subsequently the metric matrix $A_{\rm GMML}$, by taking advantage of the multifidelity structure in this problem. We allocate a computational budget of $B = 17c_{\rm hi}$ to learning each of $\Gamma_0$ and $\Gamma_1$ and apply a manifold regression multifidelity (MRMF) estimator to each, with the numbers of high- and low-fidelity samples involved determined according the optimal allocation given in \cite{maurais2023logEuclidean}. 
For comparison we also consider the log-Euclidean multifidelity (LEMF) and Euclidean multifidelity (EMF) estimators of \cite{maurais2023logEuclidean} with the same sample allocation, and equivalent cost estimators using only high-fidelity samples $\bfy_{\rm hi} \mid \bftheta \sim \pi_i$ or only low-fidelity samples $\bfy_{\rm lo} \mid \bftheta \sim \pi_i$. The specific values of the sample allocations for each class $i \in \{0,1\}$ and each type of estimator can be seen in \cref{tab:samp_alloc}. We use the estimates we obtain of $\Gamma_0$ and $\Gamma_1$ to construct an estimate of $A_{\rm GMML}$. 
\begin{table}[H]
    \centering
    \begin{tabular}{lcccc}
    \toprule  
     &  \multicolumn{2}{c}{Class 0} & \multicolumn{2}{c}{Class 1} \\ \cmidrule(lr){2-3}\cmidrule(lr){4-5}
     & $N_{\rm hi}$ & $N_{\rm lo}$ & $N_{\rm hi}$ & $N_{\rm lo}$ \\ 
     \midrule 
    High-fidelity alone & 17 & 0 & 17 & 0 \\ 
    Low-fidelity alone & 0 & 4352 & 0 & 4352\\ 
    Multifidelity & 15 & 512 & 14 & 768 \\ 
    \bottomrule  
    \end{tabular}
    \caption{SQG metric learning: Sample allocations for single- and multi-fidelity estimators of $\Gamma_0$ and $\Gamma_1$. Each allocation requires the same computational budget, and the multifidelity allocations differ between classes due to differing values of generalized correlation determining the allocations according to \cite{maurais2023logEuclidean}.}
    \label{tab:samp_alloc}
\end{table}
\subsubsection{Results}
\begin{figure}[h]
\captionsetup[subfigure]{justification=centering}
\begin{subfigure}[t]{0.32\linewidth}
\centering
\textbf{\small Low-fidelity only} \\
\includegraphics[width=\linewidth]{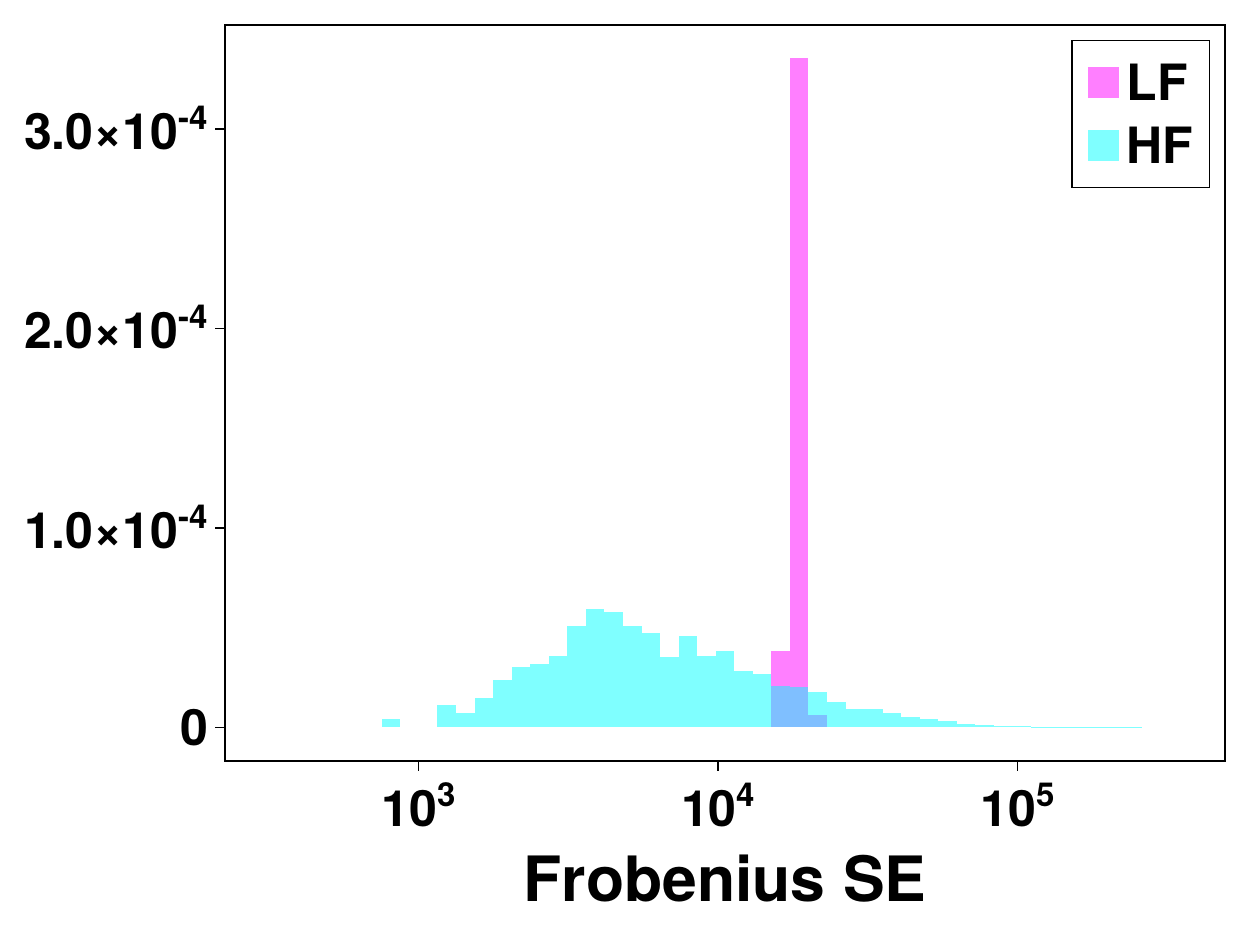}    
\subcaption*{23\% decrease in Frobenius MSE}
\end{subfigure}
\begin{subfigure}[t]{0.32\linewidth}
\centering 
\textbf{\small Log-Euclidean multifidelity} \\ 
\includegraphics[width=\linewidth]{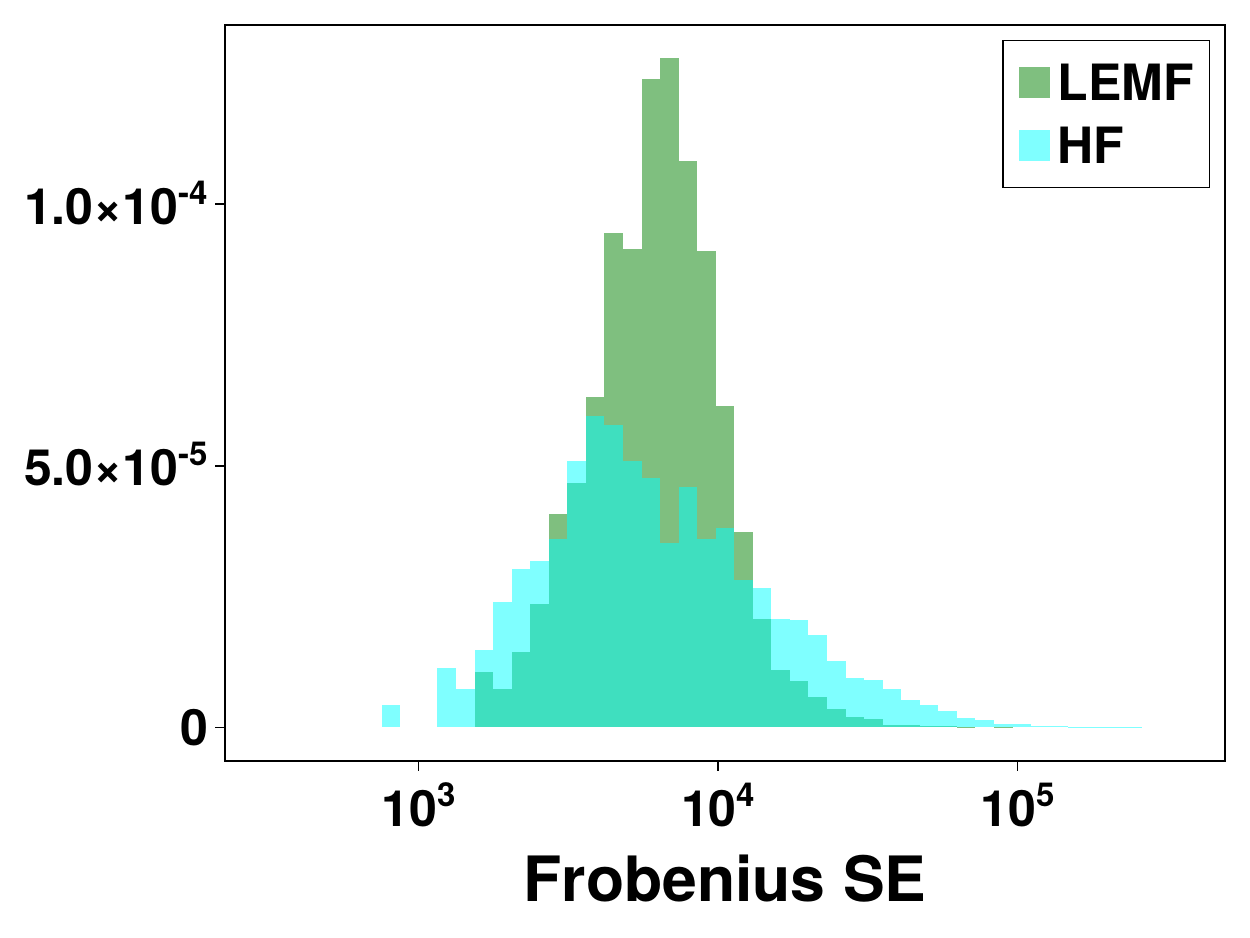}    
\subcaption*{60\% decrease in Frobenius MSE}
\end{subfigure}
\begin{subfigure}[t]{0.32\linewidth}
\centering 
\textbf{\small Multifidelity regression} \\
\includegraphics[width=\linewidth]{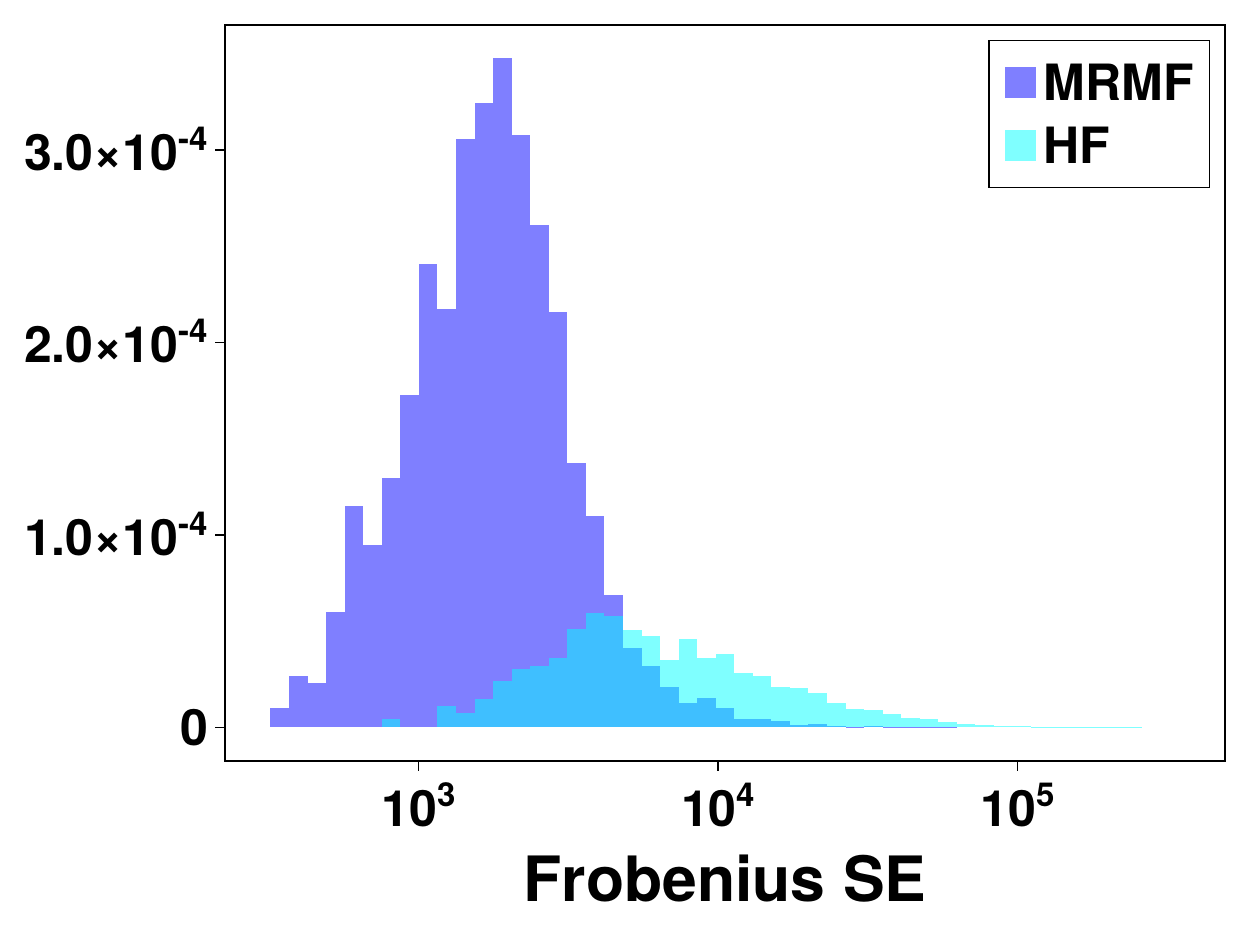}    
\subcaption*{\textbf{84\% decrease} in Frobenius MSE }
\end{subfigure}
\\
\begin{subfigure}[t]{0.32\linewidth}
\includegraphics[width=\linewidth]{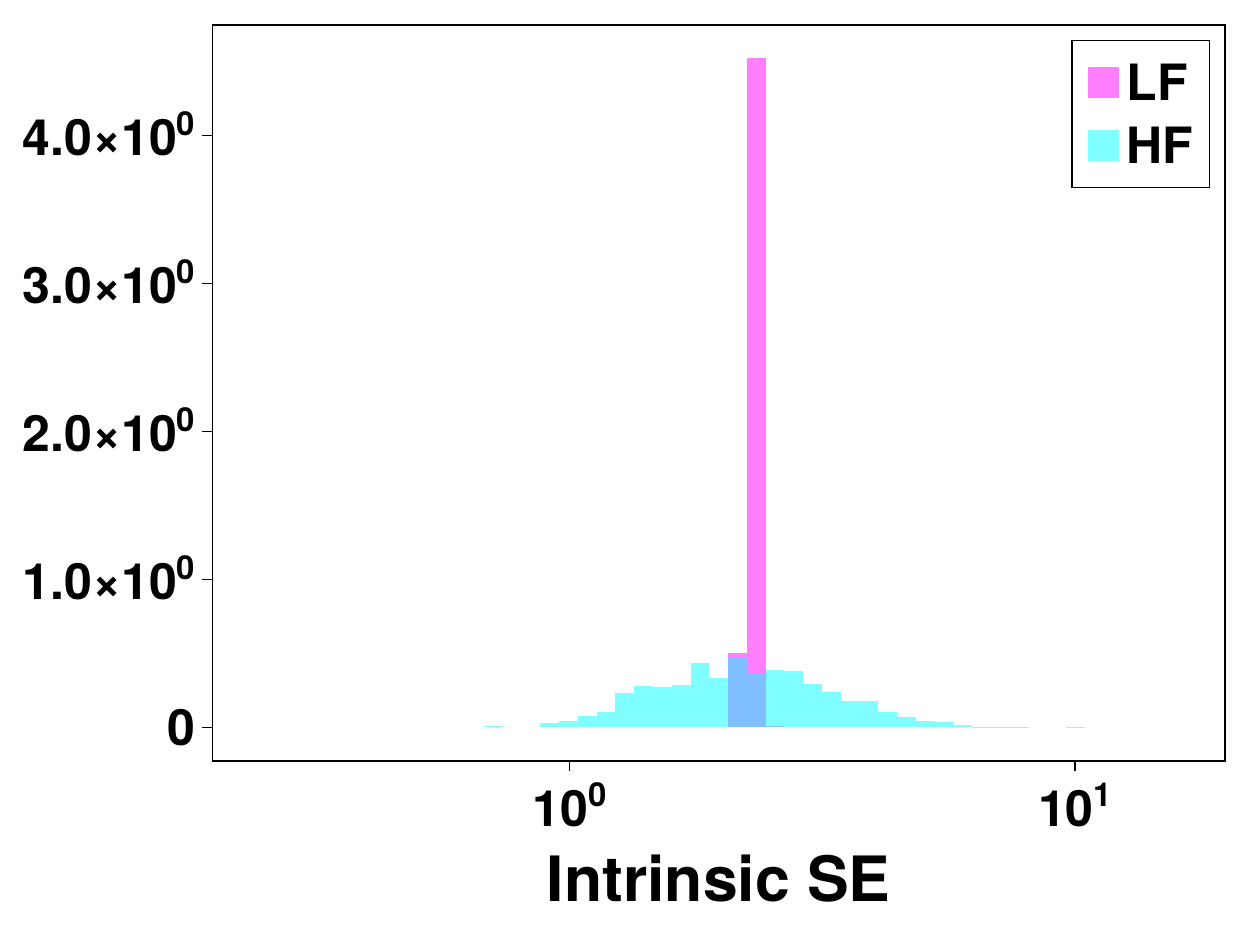}    
\subcaption*{17\% decrease in intrinsic MSE}
\end{subfigure}
\begin{subfigure}[t]{0.32\linewidth}
\includegraphics[width=\linewidth]{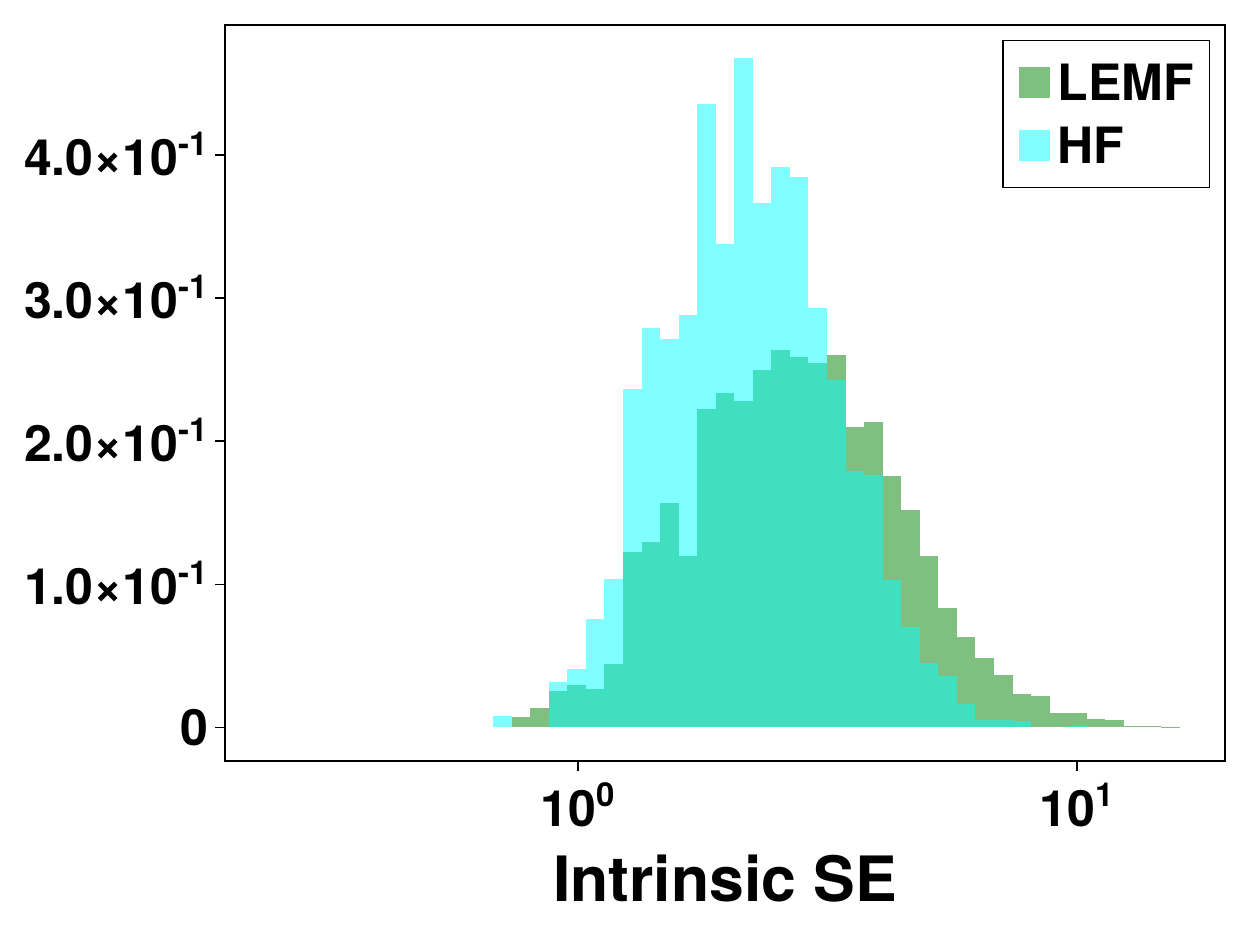}    
\subcaption*{39\% \textit{increase} in intrinsic MSE}
\end{subfigure}
\begin{subfigure}[t]{0.32\linewidth}
\includegraphics[width=\linewidth]{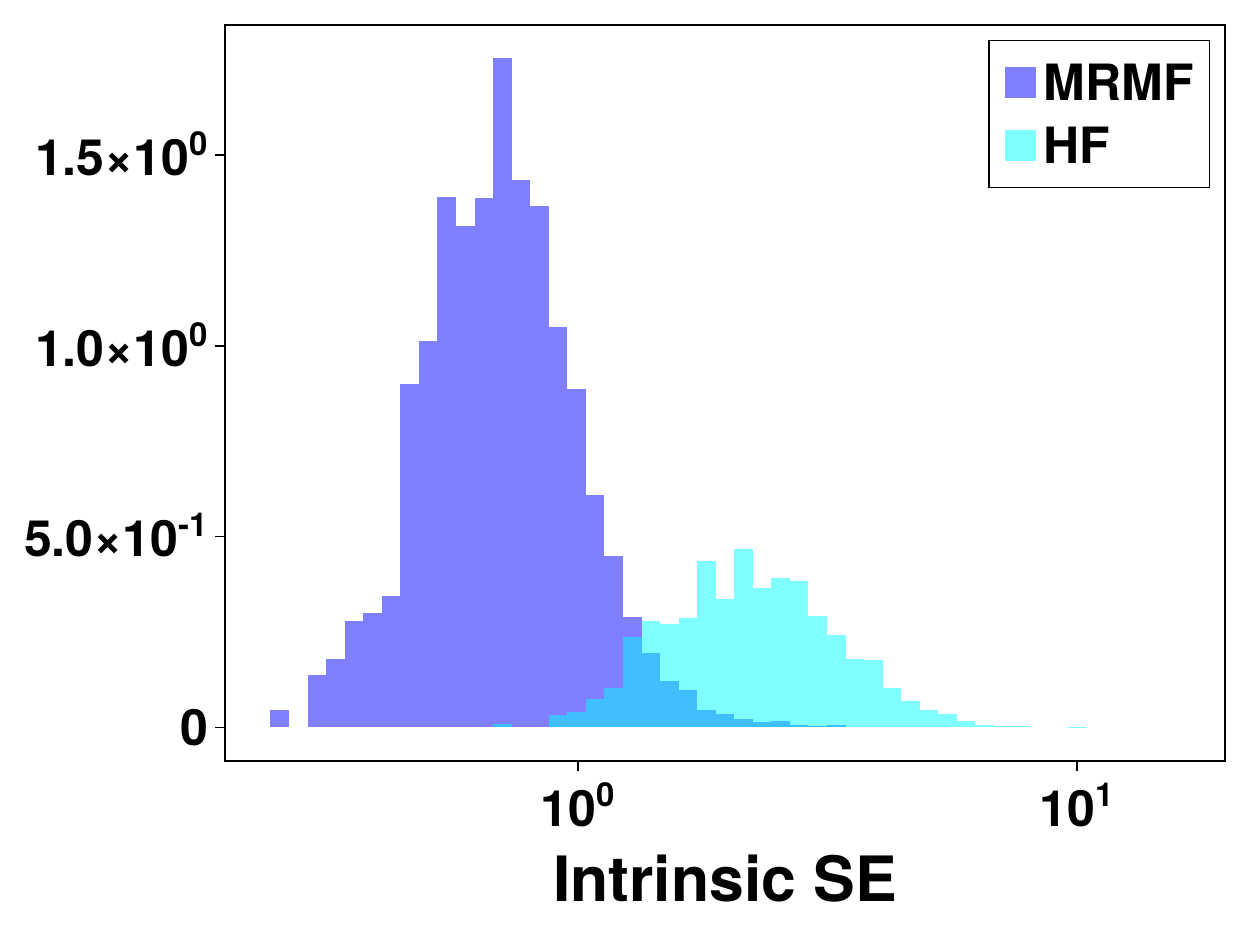}    
\subcaption*{\textbf{69\% decrease} in intrinsic MSE}
\end{subfigure}
\caption{SQG metric learning: Squared error of $\hat A^{\rm LF}$ (left), $\hat A^{\rm LEMF}$, (center), and $\hat A^{\rm MRMF}$ (right), computed in the Frobenius norm (top) and intrinsic metric \cref{eq:intrinsicdist} (bottom). The squared-error histograms of $\hat A^{\rm HF}$ are overlaid in cyan for comparison, and underneath each plot we note the percentage change in mean squared error (MSE) of each estimator relative to $\hat A^{\rm HF}$. Note that histogram horizontal axes are log-scaled.}  
\label{fig:sqg_mse_HF}
\end{figure}

\begin{figure}[h]
\centering 
\begin{subfigure}{0.35\linewidth}
\includegraphics[width=\linewidth]{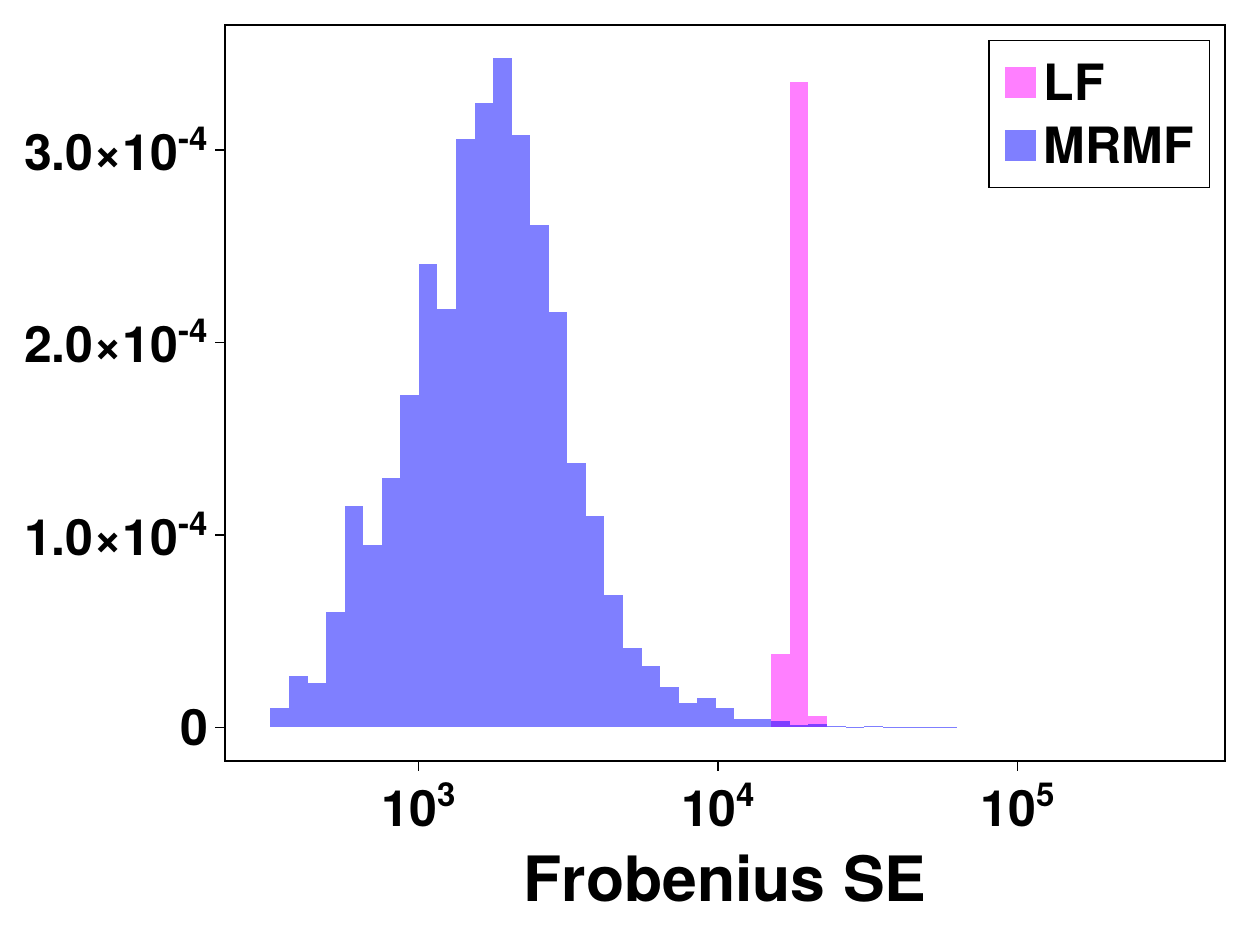}    
\end{subfigure}
\begin{subfigure}{0.35\linewidth}
\includegraphics[width=\linewidth]{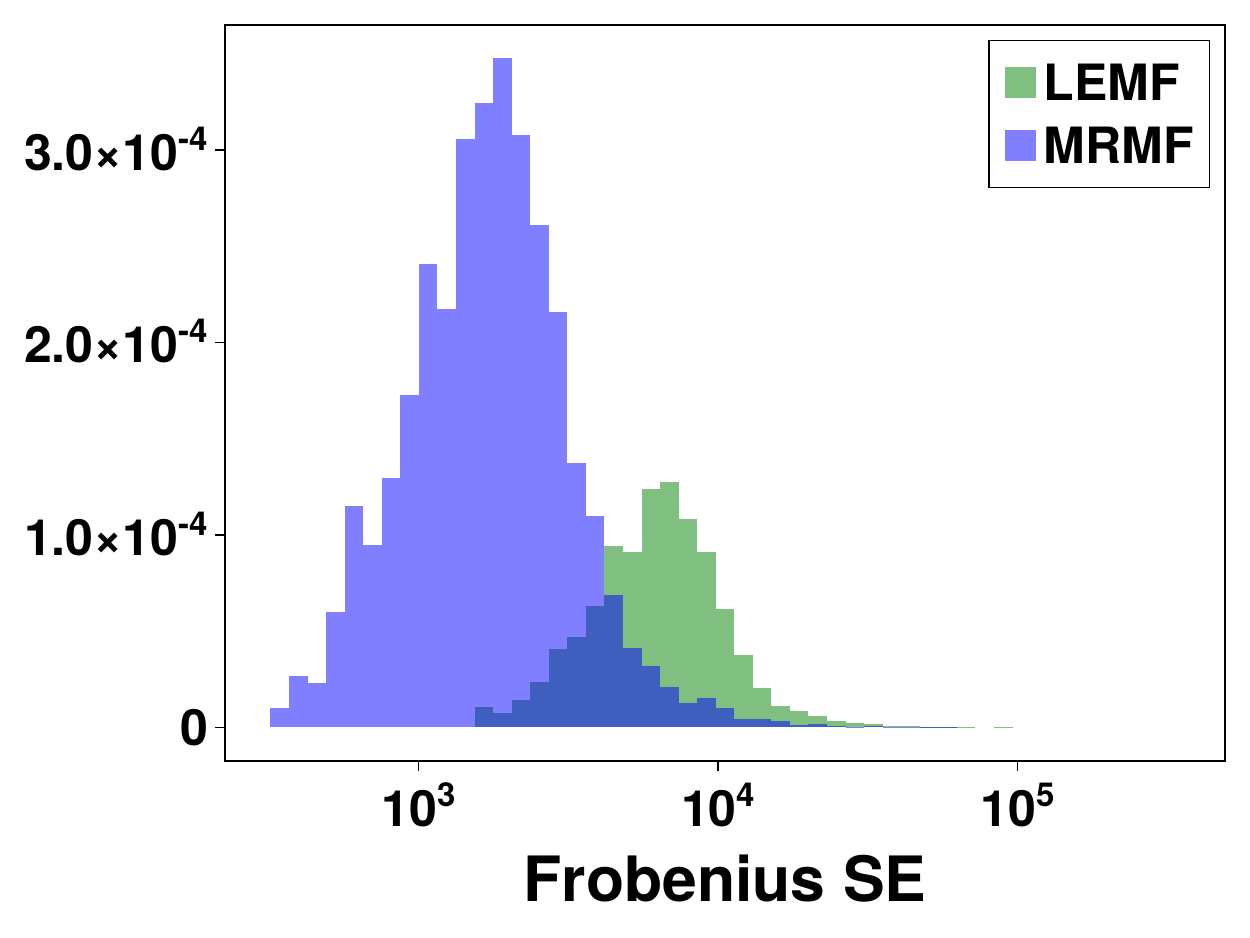}    
\end{subfigure}
\\
\begin{subfigure}{0.35\linewidth}
\includegraphics[width=\linewidth]{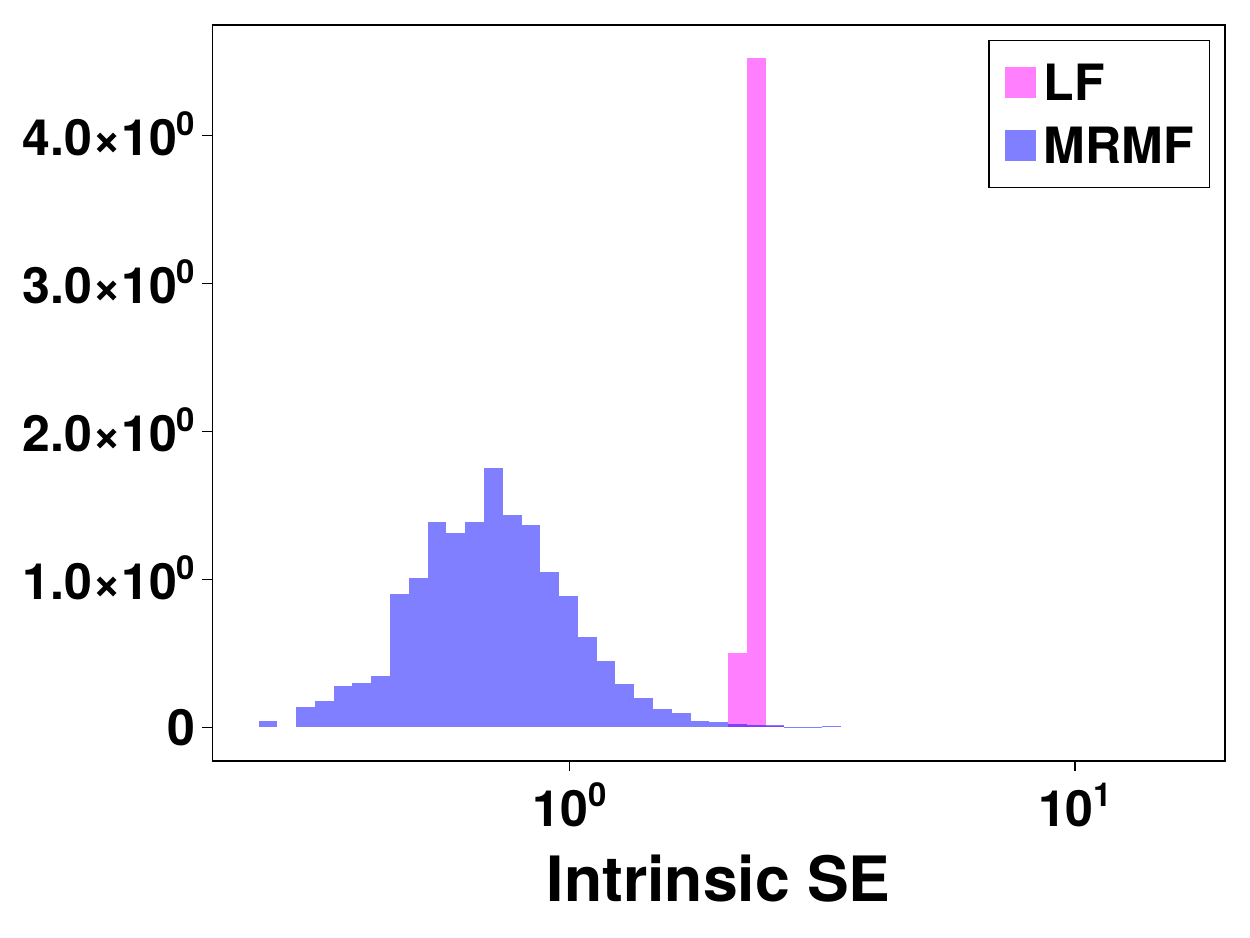}    
\end{subfigure}
\begin{subfigure}{0.35\linewidth}
\includegraphics[width=\linewidth]{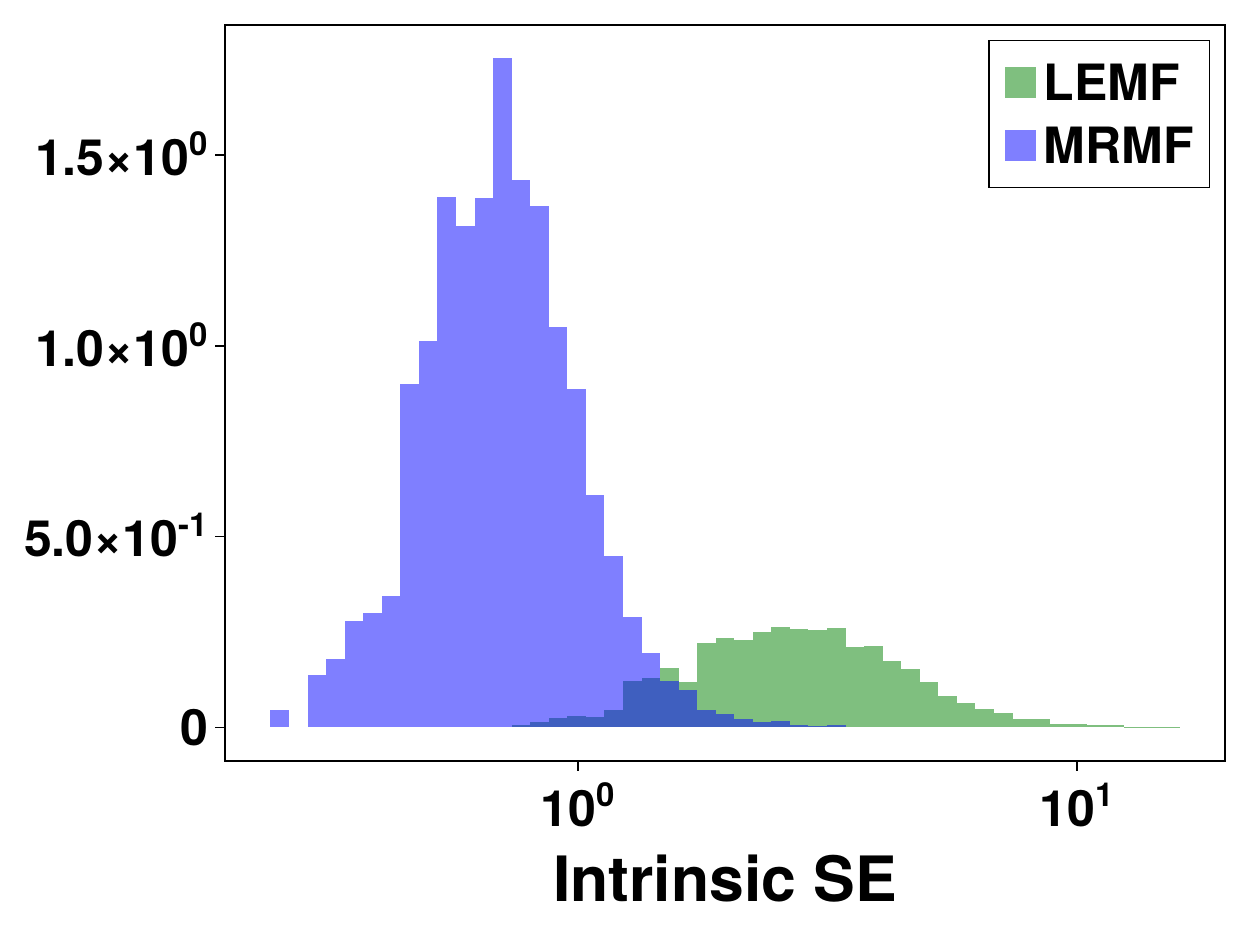}    
\end{subfigure}
\caption{SQG metric learning: Squared error distributions of $\hat A^{\rm LF}$ (left) and $\hat A^{\rm LEMF}$ compared to those of $\hat A^{\rm MRMF}$, overlaid in blue, with squared error computed in the Frobenius norm (top) and intrinsic distance (bottom). $\hat A^{\rm MRMF}$ achieves 80.\% lower Frobenius MSE than $\hat A^{\rm LF}$ and 60.\% lower Frobenius MSE than $\hat A^{\rm LEMF}$, and achieves 63\% lower intrinsic MSE than $\hat A^{\rm LF}$ and 78\% lower intrinsic MSE than $\hat A^{\rm LEMF}$. Note that histogram horizontal axes are log-scaled.} 
\label{fig:sqg_mse_reg}
\end{figure}

Prior to applying our MRMF estimator and the LEMF and EMF estimators of \cite{maurais2023logEuclidean} to the tasks of estimating $\Gamma_0$ and $\Gamma_1$, we simulate 12,000 high-fidelity and 12,000 low-fidelity pilot samples of each of $\bfy \mid \bftheta \sim \pi_0$ and $\bfy \mid \bftheta \sim \pi_1$ in order to estimate the required hyperparameters: generalized correlations and variances for the LEMF and EMF estimators, and $\Gamma_\bfS$ for the MRMF estimator. \rev{Because pilot samples are required to estimate parameters for each of the MRMF, LEMF, and EMF estimators, we keep the cost of cost of obtaining the pilot samples separate from the budgets in \cref{fig:sqg_mse_HF,fig:sqg_mse_reg}.} We additionally compute a reference estimate of $A_{\rm GMML}$ with these samples, which we use to approximate the error in the estimators we consider.

Using the sample-allocations in \cref{tab:samp_alloc}, we compute estimates $\hat \Gamma_i^{\rm HF}$, $\hat \Gamma_i^{\rm LF}$, $\hat \Gamma_i^{\rm EMF}$, $\hat \Gamma_i^{\rm LEMF}$, and $\hat \Gamma_i^{\rm MRMF}$ using high-fidelity samples alone, low-fidelity samples alone, the LEMF estimator \cref{eq:LEMF}, the EMF estimator \cref{eq:sigmaLCV}, and the MRMF estimator \cref{eq:fixlo_penalized}, respectively, $i \in \{0, 1\}$. We specifically employ the regularized, fixed-$\Sigmalo$ version of the regression estimator \cref{eq:fixlo_penalized} and in this example choose optimal values of the regularization parameters via a coarse direct search. We combine these estimates of $\Gamma_0$ and $\Gamma_1$ with estimates of $\bfm_0$ and $\bfm_1$, obtained using multifidelity Monte Carlo \cite{bpkwmg} for the multifidelity covariance estimators and standard (single-fidelity) Monte Carlo for the single-fidelity covariance estimators, to obtain ensuing estimates of $A$, denoted $\hat A^{\rm HF}$, $\hat A^{\rm LF}$, $\hat A^{\rm EMF}$, $\hat A^{\rm LEMF}$, and $\hat A^{\rm MRMF}$. Motivated by Figure 3 in \cite{zadeh2016geometric}, we set $t = 0.1$ in \cref{eq:GMML} for our experiments. The results presented in this section reflect performance of each estimator of $A$ over 2000 repeated trials; for results detailing performance in estimating each of $\Gamma_0$ and $\Gamma_1$ individually see \cref{app:sqg:addResults}. 
\paragraph{Efficacy in Estimating $A_{\rm GMML}$}
In \cref{fig:sqg_mse_HF} we show distributions of the squared error of each of $\hat A^{\rm LF}$, $\hat A^{\rm LEMF}$, and $\hat A^{\rm MRMF}$ in comparison to the squared error distribution of $\hat A^{\rm HF}$. We do not show results for $\hat A^{\rm EMF}$ because one of $\hat\Gamma_0^{\rm EMF}$ or $\hat\Gamma_1^{\rm EMF}$ was \textit{indefinite in 94\% of trials} (a known liability of the EMF estimator; see \cite{maurais2023logEuclidean} for further discussion), precluding meaningful computation of the geodesic \cref{eq:GMML} defining $A_{\rm GMML}$. We see from \cref{fig:sqg_mse_HF} that $\hat A^{\rm LF}$, $\hat A^{\rm LEMF}$, and $\hat A^{\rm MRMF}$ all provide significant decreases in Frobenius MSE relative to the baseline approach of estimating $A_{\rm GMML}$ with high-fidelity samples alone and that $\hat A^{\rm MRMF}$ enjoys the best performance by a sizeable margin. Interestingly, use of $\hat A^{\rm LEMF}$ causes an \textit{increase} in intrinsic MSE relative to $\hat A^{\rm HF}$. We notice a similar phenomenon in the results of \Cref{sec:simpleGaussian} for low budgets, and posit that it may result from amplification of error by the matrix exponential. In \cref{fig:sqg_mse_reg} we further compare the squared error distributions of $\hat A^{\rm LF}$ and $\hat A^{\rm LEMF}$ to that of $\hat A^{\rm MRMF}$ and demonstrate that use of the regression estimator \cref{eq:fixlo_penalized} results in substantial decreases in squared error even in comparison to use of low-fidelity samples alone or the LEMF estimator of \cite{maurais2023logEuclidean}. In particular, there is very little overlap between the supports of the squared error distributions of $\hat A^{\rm LF}$ and those of $\hat A^{\rm MRMF}$ in \cref{fig:sqg_mse_reg}. 

In light of this observation, we note that although $\hat A^{\rm LF}$ features low variance (in a generalized sense) due to the large numbers of samples involved in computing $\hat\Gamma_0^{\rm LF}$ and $\hat\Gamma_1^{\rm LF}$, it has high bias due to the coarseness of the discretization generating $\bfy_{\rm lo}$. The multifidelity methods, by contrast, use low-fidelity samples to reduce variance while retaining a small number of high-fidelity samples to counteract bias, thus achieving an overall lower MSE in this example.

\begin{figure}[h]
    \centering
\begin{subfigure}{0.49\linewidth}
    \includegraphics[width=\linewidth]{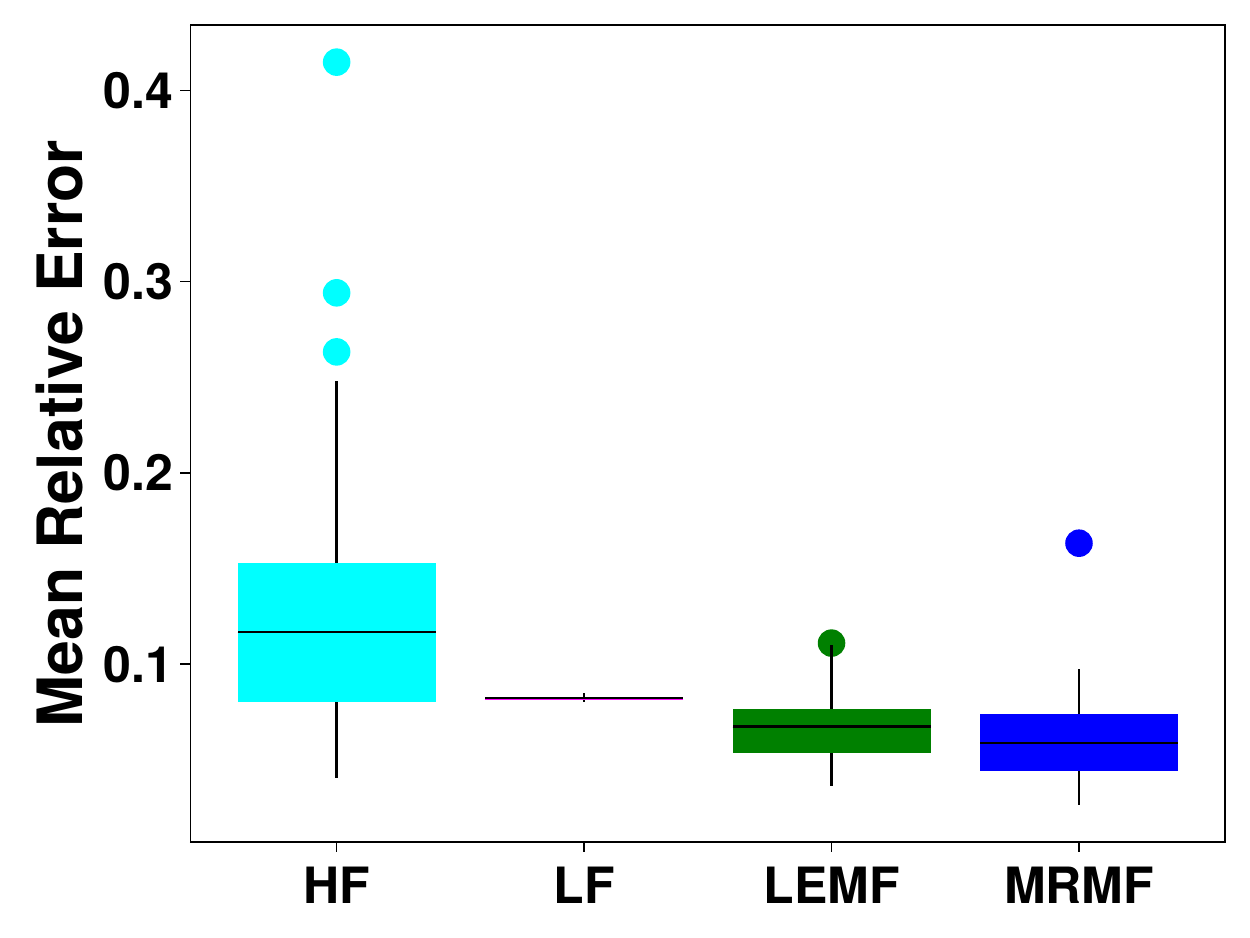}    
\end{subfigure}
\begin{subfigure}{0.49\linewidth}
    \includegraphics[width=\linewidth]{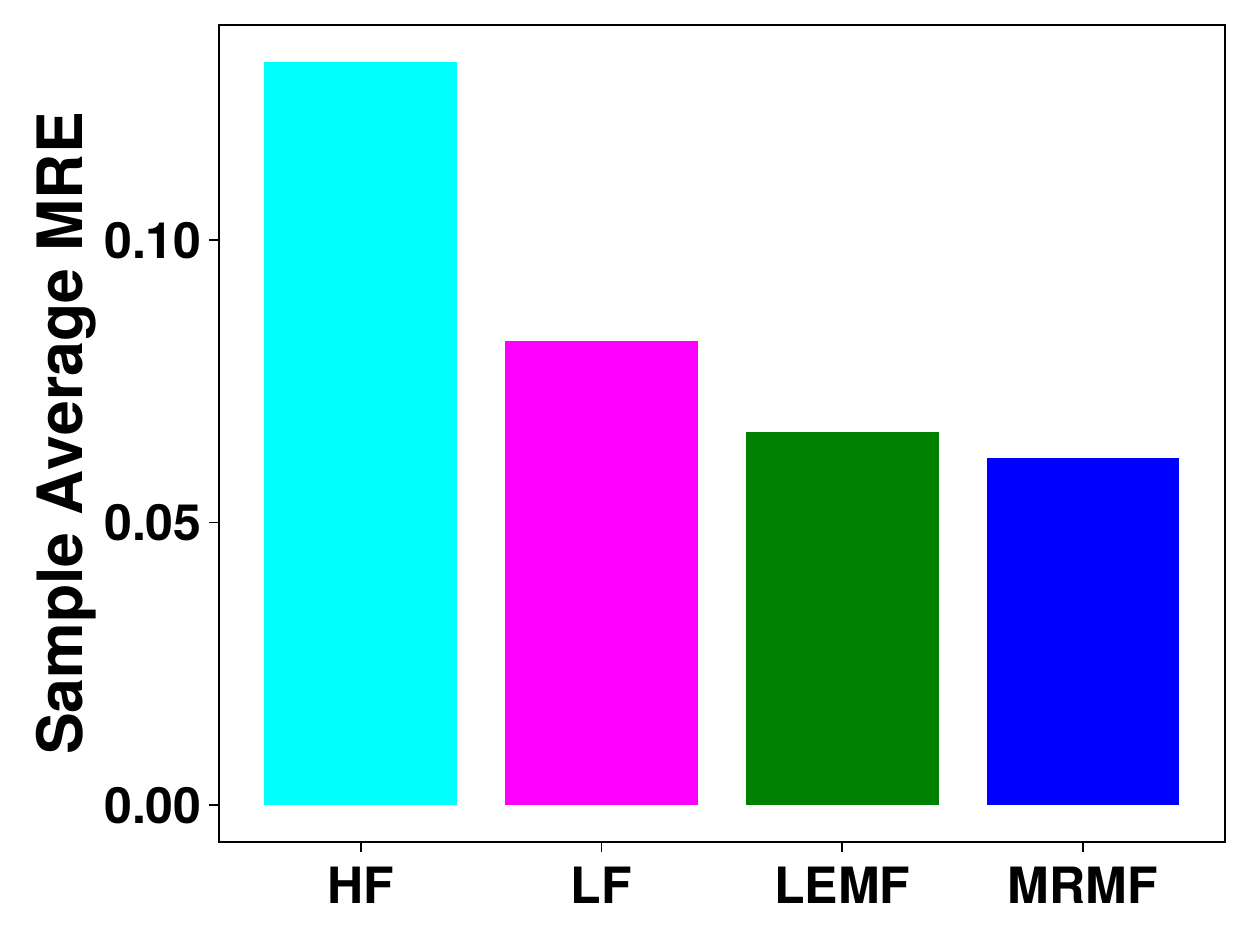}    
\end{subfigure}
    \caption{Boxplot (left) and bar graph of $\mathrm{MRE}(\hat A)$ for $\hat A \in \{\hat A^{\rm HF}, \hat A^{\rm LF}, \hat A^{\rm LEMF}, \hat A^{\rm MRMF}\}$. Empirical MRE was computed over 5000 test samples of $\bfy$ \cref{eq:MRE_mc} for 50 realizations of each $\hat A$. Use of multifidelity regression to estimate $\Gamma_0$ and $\Gamma_1$ decreases average MRE by 53\% relative to when $\Gamma_0$ and $\Gamma_1$ are estimated from high-fidelity samples alone. Average MRE corresponding to the regression estimator is additionally 25\% lower than that corresponding to the low-fidelity-only estimator and 6.9\% lower than that corresponding to the LEMF estimator.}
    \label{fig:mre}
\end{figure}

\paragraph{Downstream performance quantified by mean relative error}
One way of quantifying the goodness of an estimate of $A_{\rm GMML}$ is to examine the mean relative error between the norms induced by $A_{\rm GMML}$ and those by the estimate \cite{zadeh2016geometric}. For $A \in \bbP_d$ we denote the norm corresponding to the distance $d_{A}(\cdot, \cdot)$ by $\| \cdot \|_A = d(\cdot, 0)$, i.e., for $\bfy \in \R^d$ we have $\| \bfy \|_A = \sqrt{\bfy\t A \bfy}$. The mean relative error (MRE) associated with an estimate $\hat A$ of $A_{\rm GMML}$ is given by 
\begin{equation} 
\mathrm{MRE}(\hat A) = \E_\bfy\left[ \frac{\left| \| \bfy\|_{\hat A} - \| \bfy \|_{A_{\rm GMML}} \right|}{\| \bfy \|_{A_{\rm GMML}}} \right].
\label{eq:MRE_def}
\end{equation} 
For each estimator $\hat A \in \{\hat A^{\rm HF}, \hat A^{\rm LF}, \hat A^{\rm LEMF},\hat A^{\rm MRMF} \}$ we estimate $\mathrm{MRE}(\hat A)$ by approximating \cref{eq:MRE_def} with a Monte Carlo estimate over a test set of observations $\{\bfy^{(i)}\}_{i=1}^{5000}$,
\begin{equation}
\mathrm{MRE}(\hat A) \approx \frac{1}{5000}\sum_{i=1}^{5000} \frac{\left| \| \bfy^{(i)}\|_{\hat A} - \| \bfy^{(i)} \|_{A_{\rm GMML}} \right|}{\| \bfy^{(i)} \|_{A_{\rm GMML}}},
\label{eq:MRE_mc}
\end{equation}
where the parameters generating the test observations are sampled from an equal mixture of $\pi_0$ and $\pi_1$. We compute \cref{eq:MRE_mc} for 50 realizations of each $\hat A \in \{\hat A^{\rm HF}, \hat A^{\rm LF}, \hat A^{\rm LEMF},\hat A^{\rm MRMF} \}$ and visualize the resulting values of empirical MRE in \cref{fig:mre}.
As can be seen, use of any of $\hat A^{\rm LF}$, $\hat A^{\rm LEMF}$, or $\hat A^{\rm MRMF}$ results in a substantial decrease in MRE over $\hat A^{\rm HF}$, with $\hat A^{\rm MRMF}$ providing the steepest decrease: the average MRE of $\hat A^{\rm MRMF}$ is 53\% lower than that of $\hat A^{\rm HF}$ and additionally 25\% lower than that of $\hat A^{\rm LF}$ and 6.9\% lower than that of $\hat A^{\rm LEMF}$.
\section{Conclusions}
\label{sec:conclusion}
We have introduced a manifold regression multifidelity (MRMF) estimator of covariance matrices, formulated as the solution to a regression problem on the manifold of SPD matrices. The estimator maintains positive definiteness by construction, 
 provides significant decreases in squared estimation error relative to single-fidelity and other multifidelity covariance estimators, and can benefit downstream tasks such as metric learning. Furthermore, our multifidelity regression framework \textit{encompasses} existing multifidelity covariance estimators based on control variates \cite{maurais2023logEuclidean}, and suggests a general approach to multifidelity estimation of objects residing on Riemannian manifolds. 

Herein we specifically focused on estimation of covariance matrices, and in doing so employed the affine-invariant geometry for SPD matrices; using this geometry enabled us to exploit appealing theoretical properties of the resulting Mahalanobis distance and demonstrate the viability of our multifidelity regression approach in the absence of vector space structure. More broadly, however, the Riemannian multifidelity regression framework we lay out in \Cref{sec:formulation} is applicable to estimation of \textit{any} object residing on a nonlinear manifold, with covariance matrices being just one use case. To generalize our estimator in this way, one would adapt the definitions of mean and covariance from \cite{pennec2006intrinsic} to the particular manifold of interest and substitute them into the random variable formulation \cref{eq:S_rv} and Mahalanobis distance minimization problem \cref{eq:mdist_min} given here. Objects which reside on Riemannian manifolds and may be good candidates for multifidelity estimation include rotation matrices \cite{moakher2002means}, elementwise positive matrices \cite{gillis2014NMF}, and probability measures \cite{amari2016information, villani2009optimal, hanApproximateControlVariates2023}. 

\appendix 
\section{Proofs}
\label{app:proofs}

\subsection{Proof of \cref{thm:ts_agnostic}}
We begin with a lemma establishing \rev{self-adjointness} of the operators weighing the inner- and outer-products of tangent spaces to $\bbP_d^{N}$.
\begin{lemma}
    Let $\bfY = (Y_1, \dots, Y_N) \in \bbP_d^N$, where $N \in \Z^+$, and define the linear operator $G_\bfY: \bbH_d^N \to \bbH_d^N$ by 
\[
\bfA = (A_1, \dots, A_N) \mapsto  (Y_1 \inv A_1 Y_1 \inv, \dots, Y_N \inv A_{N} Y_{N} \inv) = G_\bfY\bfA. 
\]
$G_\bfY$ is \rev{self-adjoint with respect to the inner product on $\bbH_d^N$.}
\begin{proof}
    Let $\bfA, \bfB \in \bbH_d^N$. Using the definition of the inner product for \rev{elements of $\bbH_d$ (symmetric matrices)} and the cyclic property of trace, we quickly establish \rev{self-adjointness} of $G_\bfY$,
    \begin{multline*}
    \langle G_\bfY \bfA, \, \bfB \rangle = \sum_{n=1}^N \langle Y_n\inv A_n Y_n\inv,\, B \rangle = \sum_{n=1}^N \trace{ Y_n\inv A_n Y_n\inv B } = 
    \sum_{n=1}^N \trace{  A_n Y_n\inv B Y_n\inv}\\ 
    =  \sum_{n=1}^N \langle  A_n,\, Y_n\inv B Y_n\inv \rangle = \langle \bfA,\, G_\bfY \bfB \rangle. 
    \end{multline*}
\end{proof}
\label{lemma:Gy_symmetric}
\end{lemma}

We next demonstrate that $\Gamma_\bfS$ can be factored in terms of the Riemannian transformation of $\rmT_\bfSigma \bbP_d^N$ in \cref{prop:gammaS_decomp}.
\begin{lemma}
	Consider $\Gamma_\bfS = \E[\log_\bfSigma \bfS \otimes_\bfSigma \log_\bfSigma \bfS]$ for the random variable $\bfS$ with mean $\bfSigma$ in \cref{eq:S_rv}. $\Gamma_\bfS$ can be written 
	\begin{equation*}
	\Gamma_\bfS = \rev{\Gamma_{\bfS, \bfI} \circ G_{\bfSigma}},
	\end{equation*}
	where $\Gamma_{\bfS, \bfI} = \E[\log_\bfSigma \bfS \otimes \log_\bfSigma \bfS]$, and $G_{\bfSigma}$ is the linear operator on $\bbH_d^{N}$ mapping 
 \[
\bfA = (A_1, \dots, A_{N}) \mapsto  \left ((\Sigma^1) \inv A_1 (\Sigma^1) \inv, \dots, (\Sigma^N) \inv A_{N} (\Sigma^N) \inv \right ) = G_{\bfSigma}\bfA,
 \]
where $\Sigma^1, \dots, \Sigma^N$ 
are the $N$ individual $\bbP_d$-valued elements of $\bfSigma = (\bfSigma^{(1)}, \dots, \bfSigma^{(K)})$. 
	\begin{proof} 
The covariance of $\bfS$ is given by
    \begin{equation*}
    \Gamma_\bfS = \E[\log_\bfSigma \bfS \otimes_\bfSigma \log_\bfSigma \bfS] = \E[\log_\bfSigma \bfS \otimes G_\bfSigma \log_\bfSigma \bfS],
    \end{equation*}
    by definition of the outer-product $\otimes_\bfSigma$. We factor the \rev{self-adjoint} linear operator $G_\bfSigma$ out of the outer product and obtain 
    \begin{equation*}
    \Gamma_\bfS =\E[\log_\bfSigma \bfS \otimes G_\bfSigma\log_\bfSigma \bfS] = \rev{\E[\log_\bfSigma \bfS \otimes \log_\bfSigma \bfS] \circ G_\bfSigma^\ast} \equiv \rev{\Gamma_{\bfS, I} \circ G_\bfSigma},
    \end{equation*}
    \rev{where we use $G_\bfSigma^\ast$ to denote the adjoint of $G_\bfSigma$.} 
	\end{proof}
	\label{prop:gammaS_decomp}
\end{lemma}

This decomposition of $\Gamma_\bfS$ allows us to quickly show the \rev{basis independence} in \cref{thm:ts_agnostic}. By \cref{prop:gammaS_decomp}, we can factor the covariance of $\bfS$ as $\Gamma_\bfS = \rev{\Gamma_{\bfS, \bfI} 
\circ G_{\bfSigma}}$, hence its inverse is given by 
\begin{equation*}
\Gamma_\bfS\inv =\rev{G_{\bfSigma}\inv \circ \Gamma_{\bfS, \bfI}\inv}. 
\end{equation*}
The Mahalanobis distance in \cref{eq:mdist_min} is defined in terms of the $\bfSigma$ inner product, which we can write
\begin{equation*}
d^2_S(\bfSigma) = \left\langle \log_\bfSigma \bfS, \; \Gamma_\bfS\inv \log_\bfSigma \bfS \right\rangle_\bfSigma = \left\langle \log_\bfSigma \bfS, \; G_\bfSigma (\Gamma_\bfS\inv \log_\bfSigma \bfS) \right\rangle,
\end{equation*}
with $G_\bfSigma$ as in \cref{prop:gammaS_decomp}. Substituting $\Gamma_\bfS\inv = \rev{G_\bfSigma\inv \circ \Gamma_{\bfS, \bfI}\inv}$ above, we obtain the desired result, 
\begin{equation*}
d^2_S(\bfSigma) = \left\langle \log_\bfSigma \bfS, \; G_\bfSigma (G_\bfSigma\inv(\Gamma_{\bfS, \bfI}\inv \log_\bfSigma \bfS)) \right\rangle = \left\langle \log_\bfSigma \bfS, \; \Gamma_{\bfS, \bfI} \inv \log_\bfSigma \bfS \right\rangle.
\end{equation*}

\subsection{Proof of \cref{thm:affineinvariance}}
\subsubsection{Preliminaries}
In order to show affine-invariance of the Mahalanobis distance, we require a result relating $\log_{\tilde A} \tilde B$ to $\log_A B$, where $\tilde A = Y\inv A Y\inv$ and $\tilde B = Y\inv B Y\inv$ for some $Y \in \bbP_d$, which we will then extend to $\bbP_d^N$. 

\begin{lemma}
	Let $A, B, Y \in \bbP_d$, and define $\tilde A = Y\inv A Y\inv$ and $\tilde B = Y\inv B Y\inv$. It holds that 
	\begin{equation}
	\log_{\tilde A} \tilde B = Y\inv (\log_A B) Y\inv,
	\label{eq:logequiv}
	\end{equation}
	i.e., affine transformations on $\bbP_d$ correspond to affine transformations on tangent spaces to $\bbP_d$. 
	
\begin{proof}
Recall the definition of $\log_A B$ for $A, B \in \bbP_d$, 
\begin{equation*}
\log_A B = A^{\frac{1}{2}}\log(A^{-\frac{1}{2}}B A^{-\frac{1}{2}})A^{\frac{1}{2}} = A\log(A\inv B). 
\end{equation*}	
$\log_{\tilde A} \tilde B$ is given by 
\begin{equation}
	\begin{aligned}
		\log_{\tilde A}\tilde B = \tilde A\log(\tilde A\inv \tilde B) &= Y\inv AY\inv\log(Y A\inv Y Y\inv BY\inv )  \\
		&= Y\inv AY\inv \log(Y A\inv B Y\inv ).
	\end{aligned}
	\label{eq:logtilde}
\end{equation} 
In \cite{bhatiaposdef} \rev{(page 219)} we find mentioned that for $T \in \bbM_d$ with no eigenvalues on $(\infty, 0]$ and $S\in\bbM_d$ invertible, with $\bbM_d$ denoting the set of real $d \times d$ matrices, $S\log(T)S\inv = \log(STS\inv)$. Applying this fact to the last line of \eqref{eq:logtilde}, we see that
\begin{equation*}
\log_{\tilde A}\tilde B = Y\inv A Y\inv\log(Y A\inv B Y\inv) = Y\inv A\log(A\inv B)Y\inv \equiv Y\inv \log_A(B) Y \inv,
\end{equation*}
yielding the desired equivalence \cref{eq:logequiv}. 
\end{proof}
\label{lemma:logequiv}
\end{lemma}

The extension of this relationship to an analogous one $\bbP_d^N$ is immediate:

\begin{corollary}
	Let $\bfA = (A_1, \dots, A_N) \in \bbP_d^N$, $\bfB = (B_1, \dots, B_N) \in \bbP_d^N$ and define 
 \[
 \tilde\bfA = (Y_1\inv A_1 Y_1\inv, \dots, Y_N\inv A_N Y_N\inv) = G_\bfY\bfA, \; \tilde\bfB = (Y_1\inv B_1 Y_1\inv, \dots, Y_N\inv B_N Y_N\inv) = G_\bfY\bfB
 \]
 where $\bfY \in \bbP_d^N$ and $G_\bfY: \bbH_d^N \to \bbH_d^N$ is the \rev{self-adjoint} linear operator mapping 
 \[
\bfC = (C_1, \dots, C_N) \mapsto  (Y_1 \inv C_1 Y_1 \inv, \dots, Y_N \inv C_{N} Y_{N} \inv) = G_\bfY\bfC
 \]
 with $\bfC \in \bbH_d^N$. Then 
	\begin{equation*}
		\log_{\tilde \bfA}\tilde \bfB = \log_{G_\bfY\bfA} (G_\bfY\bfB) = G_\bfY \log_\bfA \bfB.
	\end{equation*}
	\begin{proof}
	Applying \cref{lemma:logequiv} elementwise to $\log_{\tilde \bfA} \tilde \bfB$, we see
	\begin{equation*}
		\log_{\tilde\bfA} \tilde \bfB = \begin{bmatrix} \log_{\tilde A_1}\tilde B_1 \\ \vdots \\ \log_{\tilde A_N} \tilde B_N \end{bmatrix} = \begin{bmatrix} Y_1\inv(\log_{A_1}B_1)Y_1\inv \\ \vdots \\ Y_N \inv(\log_{A_N}B_N)Y_N\inv \end{bmatrix} = G_\bfY \log_\bfA \bfB.
	\end{equation*}	
	\end{proof}
\label{corr:logequiv_K}
\end{corollary}

\subsubsection{Main Result}
With \cref{lemma:logequiv} and \cref{corr:logequiv_K}, we can now show that the Mahalanobis distance objective \cref{eq:mdist_min} is affine-invariant. 

By \cref{thm:ts_agnostic}, the Mahalanobis distance in \cref{eq:mdist_thm} is equal to 
		\begin{equation*}
		d^2_\bfS(\bfSigma) = \langle \log_\bfSigma \bfS, \; \Gamma_\bfS\inv \log_\bfSigma \bfS \rangle_\bfSigma = \langle \log_\bfSigma \bfS, \; \Gamma_{\bfS, \bfI}\inv \log_\bfSigma \bfS \rangle 
		\end{equation*}
		where $\Gamma_{\bfS, \bfI} =\E[\log_\bfSigma \bfS \otimes \log_\bfSigma \bfS]$.  In computing the Mahalanobis distance for $d^2_{\tilde \bfS}(\tilde \bfSigma)$ we will similarly only concern ourselves with the unweighted inner product $\langle \cdot, \cdot \rangle$ and $\Gamma_{\tilde\bfS, \bfI} = \E \left[\log_{\tilde \bfSigma}\tilde\bfS \otimes \log_{\tilde \bfSigma}\tilde\bfS \right]$. Using \cref{corr:logequiv_K}, factoring linear operators out of the outer product, and noting that $G_\bfY$ is \rev{self-adjoint}, we write $\Gamma_{\tilde\bfS, I}$ as 
  \begin{equation*}
		\begin{aligned}
			\Gamma_{\tilde\bfS, I} &= \E \left[\log_{\tilde \bfSigma}\tilde\bfS \otimes \log_{\tilde \bfSigma}\tilde\bfS \right] \\
			&= \E \left[G_\bfY\log_{\bfSigma}\bfS \otimes G_\bfY\log_{\bfSigma} \bfS \right] \\
			&= \rev{G_\bfY \circ \E \left[ \log_{\bfSigma}\bfS \otimes \log_{\bfSigma} \bfS  \right] \circ G_\bfY^\ast} \\
			&= \rev{G_\bfY \circ \Gamma_{\bfS, I} \circ G_\bfY},
		\end{aligned}
  \end{equation*}
	 \rev{where we use $G_\bfY^\ast$ to denote the adjoint of $G_\bfY$.} Hence, $\Gamma_{\tilde\bfS, \bfI}\inv = \rev{G_\bfY\inv \circ \Gamma_{\bfS, \bfI}\inv \circ G_\bfY\inv}$. The Mahalanobis distance $d^2_{\tilde \bfS}(\tilde \bfSigma)$ is thus 
  \begin{equation*}
	 \begin{aligned}
	 	d^2_{\tilde \bfS}(\tilde \bfSigma) &= \left\langle \log_{\tilde \bfSigma}\tilde\bfS,\; \Gamma_{\tilde\bfS, \bfI}\inv \log_{\tilde\bfSigma}\tilde\bfS \right\rangle \\
	 	&= \left\langle G_\bfY\log_{ \bfSigma}\bfS,\; \rev{(G_\bfY\inv \circ \Gamma_{\bfS, \bfI}\inv \circ G_\bfY\inv)} G_\bfY\log_{ \bfSigma} \bfS \right\rangle \\
	 	&= \left\langle \log_{ \bfSigma}\bfS,\; \Gamma_{\bfS, \bfI}\inv \log_{ \bfSigma} \bfS \right\rangle \\
	 	&= d^2_\bfS(\bfSigma).
	 \end{aligned}
\end{equation*}

\subsection{Proof of \cref{prop:mle}}
\cref{thm:ts_agnostic} shows that instead of directly minimizing the Mahalanobis distance objective in \cref{eq:mdist_min}, which is rigorously defined using the inner- and outer-products specific to $\rmT_\bfSigma \bbP_d^N$, we can equivalently solve 
\begin{equation}
	(\hat\Sigma_0,\dots, \hat\Sigma_L) = \argmin_{\Sigma_0, \dots,\Sigma_L \in \bbP_d} \langle \log_{\bfSigma} \bfS,\; \Gamma_{\bfS, \bfI}\inv \log_{\bfSigma} \bfS \rangle \quad \text{s.t. } \bfSigma = \mu_\bfS(\Sigma_0, \dots, \Sigma_L),
	\label{eq:mdist_min_tsa}
\end{equation}
defined with the standard Euclidean products $\langle \cdot, \cdot \rangle$ and $\otimes$. 
   Because $\bbH_d^N$ is a Euclidean vector space, the density of $\bm\calE \sim \calN_{\bbH_d^N}(\mathbf{0}, \Gamma_{\bm \calE})$ can be written 
   \[
   p(\bm{\calE}) \propto \exp\left(-\ltfrac{1}{2}\langle \bm{\calE},\; \Gamma_{\bm{\calE}}\inv \bm{\calE} \rangle \right) = \exp \left(-\ltfrac{1}{2}\langle \log_\bfSigma \bfS,\; \Gamma_{\bfS, \bfI}\inv \log_\bfSigma \bfS \rangle \right),
   \]
   where we have used the fact that $\Gamma_{\bm{\calE}} = \Gamma_{\bfS, \bfI}$. Maximizing the above is equivalent to minimizing its logarithm, which is what occurs in \cref{eq:mdist_min_tsa} and equivalently \cref{eq:mdist_min}. 

\subsection{Proof of \cref{prop:nleq}}
\label{app:proof_nleq}
\rev{Throughout this proof, for conciseness, we use juxtaposition to denote composition of linear operators, e.g., if $A, B: \bbP_d \to \bbP_d$ are linear, then $AB$ denotes $A \circ B$}. The inner product in \cref{eq:twomat_reg} can be decomposed into inner products between the individual components of $\log_\bfSigma \bfS$, 
\begin{equation} 
\begin{aligned}
	\Sigmahihat &= \argmin_{\Sigmahi \in \bbP_d} \left\langle \begin{bmatrix} \log_\Sigmahi(\Shi) \\ \log_\Slobar(\Slo) \end{bmatrix}, \; \Gamma_{S} \inv \begin{bmatrix} \log_\Sigmahi(\Shi) \\ \log_\Slobar(\Slo) \end{bmatrix}  \right\rangle  \\
	&= \argmin_{\Sigmahi \in \bbP_d} \langle \log_\Sigmahi(\Shi), \; C_{\rm hi} \log_\Sigmahi (\Shi) \rangle + 2 \langle \log_\Sigmahi (\Shi), \; C_{\rm lo, hi} \log_\Slobar (\Slo) \rangle \\
 &\qquad\qquad\qquad\qquad\qquad\qquad\qquad\qquad\qquad + \langle \log_\Slobar (\Slo), \; C_{\rm lo} \log_\Slobar (\Slo) \rangle,
\end{aligned}
\label{eq:opt_blocks}
\end{equation} 
where $C_{\rm hi}$, $C_{\rm lo, hi}$, and $C_{\rm lo}: \bbH_d \to \bbH_d$ are blocks of $\Gamma_\bfS\inv$, 
\[
\Gamma_{\bfS}\inv = 
\begin{bmatrix}
C_{\rm hi} & C_{\rm lo, hi} \\
C_{\rm lo, hi}*\ast & C_{\rm lo}
\end{bmatrix} .
\]

Denote $\calShi = \log_\Sigmahi(\Shi)$ and $\calSlo = \log_\Slobar(\Slo)$; $\calShi$ depends on $\Sigmahi$ in \eqref{eq:opt_blocks} whereas $\calSlo$ is fixed. Thus we have 
\begin{equation} 
\begin{aligned}
\Sigmahihat &= \argmin_{\Sigmahi \in \bbP_d}\; \langle \calShi, \; C_{\rm hi} \calShi \rangle + 2 \langle \calShi, \; C_{\rm lo, hi} \calSlo \rangle  + \langle  \calSlo, \; C_{\rm lo} \calSlo \rangle \quad \text{s.t. } \calShi = \log_\Sigmahi \Shi \\
&\equiv \argmin_{\Sigmahi \in \bbP_d}\; f(\calShi) \quad \text{s.t. } \calShi = \log_\Sigmahi \Shi 
\end{aligned}
\label{eq:insert_calS}
\end{equation} 
Since \eqref{eq:insert_calS} depends on $\Sigmahi$ only through $\calShi$, we may optimize the Mahalanobis distance with respect to $\calShi$, 
\[
\calShihat = \argmin_{\calShi \in \bbH_d} f(\calShi) 
\]
which in the end will leave us with a nonlinear equation for $\Sigmahihat$.  
The gradient of $f$ with respect to $\calShi$ is given by
\[
\nabla f(\calShi) = 2 C_{\rm hi} \calShi + 2 C_{\rm lo, hi} \calSlo,
\]
where we recall that $\langle \cdot, \cdot \rangle$ in \eqref{eq:insert_calS} denotes the Frobenius (trace) inner product between symmetric matrices. Setting this gradient equal to the zero matrix and solving the corresponding equation yields an optimum value of $\calShi = \log_\Sigmahi (\Shi)$,
\begin{equation*}
\calShihat = -C_{\rm hi}\inv C_{\rm lo, hi}\calSlo 
\end{equation*}
Because $C_{\rm hi}$ and $C_{\rm lo, hi}$ are linear operators on symmetric matrices, $\calShihat$ is indeed symmetric. Using relevant formulae for blockwise inversion of a square-partitioned \rev{self-adjoint} linear operator \cite{lu2002inverses}, we express $C_{\rm hi}$, $C_{\rm lo}$, and $C_{\rm lo, hi}$ in terms of the blocks of $\Gamma_\bfS$,
\begin{equation}
\begin{gathered}
C_{\rm hi} = (\Gamma_{\rm hi} - \Gamma_{\rm lo, hi}\Gamma_{\rm lo}\inv \Gamma_{\rm lo, hi}^\ast)\inv, \quad C_{\rm lo} = \Gamma_{\rm lo}\inv + \Gamma_{\rm lo}\inv \Gamma_{\rm lo, hi}^\ast(\Gamma_{\rm hi} - \Gamma_{\rm lo, hi} \Gamma_{\rm lo}\inv \Gamma_{\rm lo, hi}^\ast)\inv \Gamma_{\rm lo, hi}\Gamma_{\rm  lo}\inv \\
C_{\rm lo, hi} = -(\Gamma_{\rm hi} - \Gamma_{\rm lo, hi} \Gamma_{\rm lo}\inv \Gamma_{\rm lo, hi}^\ast)\inv \Gamma_{\rm lo, hi}\Gamma_{\rm lo}\inv .
\end{gathered}
\label{eq:Cinvs}
\end{equation} 
Notably, from \eqref{eq:Cinvs} we see 
\[
C_{\rm lo} = \Gamma_{\rm lo}\inv + \Gamma_{\rm lo}\inv  \Gamma_{\rm lo, hi}^\ast C_{\rm hi} \Gamma_{\rm lo, hi}\Gamma_{\rm lo}\inv \quad \text{and} \quad  C_{\rm lo, hi} = - C_{\rm hi} \Gamma_{\rm lo, hi}\Gamma_{\rm lo}\inv. 
\]
Thus the optimum value of $\calShihat$ is given by 
\begin{equation} 
\calShihat = -C_{\rm hi}\inv C_{\rm lo, hi}\calSlo = \Gamma_{\rm lo, hi}\Gamma_{\rm lo}\inv \log_\Slobar(\Slo). 
\label{eq:nleq_calS}
\end{equation}
\Cref{eq:nleq_calS} is a nonlinear equation defining the regression estimate of $\Sigmahi$ when the value of $\Sigmalo$ is fixed at $\Slobar$, 
\begin{equation*}
\begin{aligned} 
\log_\Sigmahi(\Shi) &= \Gamma_{\rm lo, hi} \Gamma_{\rm lo}\inv \log_\Slobar(\Slo) \\
&\Updownarrow \\
\Sigmahi \log(\Sigmahi\inv \Shi) &= \Gamma_{\rm lo, hi} \Gamma_{\rm lo}\inv(\Slobar \log(\Slobar\inv \Slo)).
\end{aligned} 
\end{equation*}

\subsection{Proof of \cref{prop:expectedMdist}}
\rev{Throughout this proof, for conciseness, we use juxtaposition to denote composition of linear operators, e.g., if $A, B: \bbP_d \to \bbP_d$ are linear, then $AB$ denotes $A \circ B$}.
Notice, as before, that the squared Mahalanobis distance in \cref{eq:twomat_reg} depends on $\Sigmahi$ only through $\calShi \equiv \log_\Sigmahi \Shi$. With $\calSlo$ likewise denoting $\log_\Sigmalo \Slo$, we use $f(\cdot)$ to denote the value of the squared Mahalanobis distance objective at a particular $\calShi = \log_\Sigmahi \Shi \in \bbH_d$, 
\[
f(\calShi) = \left\langle \begin{bmatrix} \calShi \\ \calSlo \end{bmatrix}, \; \Gamma_{S} \inv \begin{bmatrix} \calShi \\ \calSlo \end{bmatrix}  \right\rangle.
\]
The value of $f(\cdot)$ at $\calShihat = \Gamma_{\rm lo, hi} \Gamma_{\rm lo}\inv \calS_{\rm lo} = -C_{\rm hi}\inv C_{\rm lo, hi}\calS_{\rm lo}$, is
\begin{equation*}
\begin{aligned}
f(\calShihat) &= \langle C_{\rm hi}\inv C_{\rm lo, hi}\calS_{\rm lo}, \; C_{\rm lo, hi}\calS_{\rm lo} \rangle + -2 \langle C_{\rm hi}\inv C_{\rm lo, hi}\calS_{\rm lo}, \; C_{\rm lo, hi} \calS_{\rm lo} \rangle  + \langle  \calS_{\rm lo}, \; C_{\rm lo} \calS_{\rm lo} \rangle \\
&= -\langle C_{\rm hi}\inv C_{\rm lo, hi}\calS_{\rm lo}, \; C_{\rm lo, hi}\calS_{\rm lo} \rangle + \langle  \calS_{\rm lo}, \; C_{\rm lo} \calS_{\rm lo} \rangle \\
&= -\langle C_{\rm hi}\inv (\shortminus C_{\rm hi} \Gamma_{\rm lo, hi}\Gamma_{\rm lo}\inv)\calS_{\rm lo}, \, \shortminus C_{\rm hi} \Gamma_{\rm lo, hi}\Gamma_{\rm lo}\inv\calS_{\rm lo} \rangle + \langle  \calS_{\rm lo}, \, (\Gamma_{\rm lo}\inv + \Gamma_{\rm lo}\inv  \Gamma_{\rm lo, hi}^\ast C_{\rm hi} \Gamma_{\rm lo, hi}\Gamma_{\rm lo}\inv) \calS_{\rm lo} \rangle \\
&= -\langle \Gamma_{\rm lo, hi}\Gamma_{\rm lo}\inv \calS_{\rm lo}, \; C_{\rm hi} \Gamma_{\rm lo, hi}\Gamma_{\rm lo}\inv\calS_{\rm lo} \rangle  + \langle  \calS_{\rm lo}, \; \Gamma_{\rm lo}\inv  \Gamma_{\rm lo, hi}^\ast C_{\rm hi} \Gamma_{\rm lo, hi}\Gamma_{\rm lo}\inv \calS_{\rm lo} \rangle + \langle  \calS_{\rm lo}, \; \Gamma_{\rm lo}\inv \calS_{\rm lo} \rangle \\
&= -\langle \Gamma_{\rm lo, hi}\Gamma_{\rm lo}\inv \calS_{\rm lo}, \; C_{\rm hi} \Gamma_{\rm lo, hi}\Gamma_{\rm lo}\inv\calS_{\rm lo} \rangle  + \langle  \calS_{\rm lo} \Gamma_{\rm lo}\inv  \Gamma_{\rm lo, hi}^\ast, \;  C_{\rm hi} \Gamma_{\rm lo, hi}\Gamma_{\rm lo}\inv \calS_{\rm lo} \rangle + \langle  \calS_{\rm lo}, \; \Gamma_{\rm lo}\inv \calS_{\rm lo} \rangle \\
&= \langle  \calS_{\rm lo}, \; \Gamma_{\rm lo}\inv \calS_{\rm lo} \rangle \\
&= \langle \log_\Sigmalo \Slo, \; \Gamma_{\rm lo}\inv \log_\Sigmalo \Slo \rangle 
\end{aligned}
\end{equation*}
We see that the value of the squared Mahalanobis distance associated with the minimizer $\Sigmahihat$ satisfying $\log\Sigmahihat \Shi = \Gamma_{\rm lo, hi}\Gamma_{\rm lo}\inv \log_\Sigmalo \Slo$ is exactly {the Mahalanobis distance between $\Slo$ and its marginal distribution}. As noted in \cite{pennec2006intrinsic}, this Mahalanobis distance has expectation $\frac{d(d+1)}{2}$,
\begin{equation} 
\begin{aligned}
	\E[f(\calShihat)] &= \E[\langle \log_\Sigmalo \Slo, \; \Gamma_{\rm lo}\inv \log_\Sigmalo \Slo \rangle] = \E[\trace{ (\log_\Sigmalo \Slo) \Gamma_{\rm lo}\inv (\log_\Sigmalo \Slo) }] \\
	&= \E[\trace{ (\log_\Sigmalo \Slo \otimes \log_\Sigmalo \Slo)\Gamma_{\rm lo}\inv}]] = \trace{\E[\log_\Sigmalo \Slo \otimes \log_\Sigmalo \Slo]\Gamma_{\rm lo}\inv } \\
	&= \trace{\Gamma_{\rm lo }\Gamma_{\rm lo }\inv} = \trace{I_{\bbH_d}}  = \ltfrac{d(d+1)}{2}.
\end{aligned}
\label{eq:mean_mdist_min}
\end{equation}

\subsection{Proof of \cref{prop:BLUEs}}
We first demonstrate that the estimator \cref{eq:cv_euclidean} is a BLUE in the Euclidean geometry for $\bbP_d$, and note that the fact that \cref{eq:cv_logEuclidean} is a BLUE in the log-Euclidean geometry for $\bbP_d$ follows by a directly analogous argument. 

In order to estimate $\Sigmahi = \E[\Shi]$ linearly in the Euclidean geometry from $\Shi$ and $\Slo$ we seek an estimator of the form
\begin{equation}
\Sigmahihat = \bfA\Shi + \bfB\Slo + \tilde C \equiv \bfA\Shi + \bfB(\Slo + C),
\label{eq:euclidean_blue}
\end{equation}
where $\bfA, \bfB: \bbH_d \to \bbH_d$ are linear and $\tilde C \in \bbH_d$ is fixed. For simplicity we assume that $\bfB$ is invertible and employ the change of variables $C = \bfB\inv \tilde C$. We want this estimator \eqref{eq:euclidean_blue} to be unbiased, 
\[
\E[\Sigmahihat] = \bfA\Sigmahi + \bfB(\Sigmalo + C) \equiv \Sigmahi,
\]
which results in the constraint 
\[
(\bfA - \bfI)\Sigmahi = \bfB(\Sigmalo + C) \iff C = \rev{(\bfB\inv \circ (\bfA - \bfI))}\Sigmahi -\Sigmalo. 
\]
We substitute the constraint back into our estimator and obtain 
\begin{equation}
\Sigmahihat = \bfA\Shi + \bfB(\Slo + \rev{(\bfB\inv \circ (\bfA - \bfI))}\Sigmahi -\Sigmalo) = \bfA\Shi + \bfB(\Slo - \Sigmalo) + (\bfA - \bfI)\Sigmahi. 
\label{eq:euclidean_constraint_blue}
\end{equation}
In order for our estimator \cref{eq:euclidean_blue} to be admissible, it cannot depend on $\Sigmahi$. Thus we see from \cref{eq:euclidean_constraint_blue} that we must have $\bfA \equiv \bfI$, simplifying our estimator to 
\begin{equation}
\Sigmahihat = \Shi + \bfB(\Slo - \Sigmalo). 
\label{eq:euclidean_cv_B}
\end{equation}
Now we want to choose the linear operator $\bfB: \bbH_d \to \bbH_d$ such that the Euclidean MSE of \cref{eq:euclidean_cv_B} is minimized. 
\begin{small} 
\begin{equation*}
\begin{aligned}  
\bfB^{\rm opt} &= \argmin_{\bfB: \bbH_d \to \bbH_d} \E[||\Sigmahi - \Shi - \bfB(\Slo - \Sigmalo) ||_F^2] \\
 &= \argmin_{\bfB: \bbH_d \to \bbH_d} \E[\left\langle\Sigmahi - \Shi - \bfB(\Slo - \Sigmalo), \; \Sigmahi - \Shi - \bfB(\Slo - \Sigmalo)\right\rangle ] \\
 &= \argmin_{\bfB: \bbH_d \to \bbH_d} \E [\langle \Sigmahi \!-\! \Shi, \, - \bfB(\Slo \!-\! \Sigmalo) \rangle - \langle \bfB(\Slo \!-\! \Sigmalo), \, \Sigmahi \!-\! \Shi \rangle + \langle \bfB(\Slo \!-\! \Sigmalo), \, \bfB(\Slo \!-\! \Sigmalo) \rangle ] \\
 &= \argmin_{\bfB: \bbH_d \to \bbH_d} \rev{\trace{\Psi_{\rm lo, hi} \circ \bfB^\ast} + \trace{\bfB \circ \Psi_{\rm hi, lo}} + \trace{\bfB \circ \Psi_{\rm lo} \circ \bfB^\ast}},
\end{aligned}
\end{equation*}
\end{small} 
where $\Psi_{\rm lo, hi} = \E[(\Shi - \Sigmahi) \otimes (\Slo - \Sigmalo)]$, $\Psi_{\rm hi, lo} = \E[(\Slo - \Sigmalo) \otimes (\Shi - \Sigmahi)] = \rev{\Psi_{\rm lo, hi}^\ast}$, and $\Psi_{\rm lo} = \E[(\Slo - \Sigmalo) \otimes (\Slo - \Sigmalo)]$ are the two cross-covariances between $\Shi$ and $\Slo$ and the auto-covariance of $\Slo$. \rev{We use $\cdot^\ast$ to denote the adjoint of a linear operator}. We solve for $\bfB^{\rm opt}$ by taking the gradient of the last line of the above with respect to $\bfB$ and setting it equal to zero, obtaining 
\[
\rev{0 = 2\Psi_{\rm lo, hi} + 2\bfB^{\rm opt}\circ \Psi_{\rm lo} \iff \bfB^{\rm opt} = - \Psi_{\rm lo, hi}\circ \Psi_{\rm lo}\inv}.
\]
Substituting this choice of $\bfB$ into \cref{eq:euclidean_cv_B}, we see 
\[
\Sigmahihat = \Shi - \rev{(\Psi_{\rm lo, hi}\circ \Psi_{\rm lo}\inv)}(\Slo - \Sigmalo) = \Shi + \rev{(\Psi_{\rm lo, hi}\circ \Psi_{\rm lo}\inv)}(\Sigmalo - \Slo),
\]
which corresponds to the most general form of the EMF estimator \eqref{eq:cv_euclidean}. Thus, the EMF estimator is a BLUE.  The LEMF estimator \eqref{eq:cv_logEuclidean} can be shown to be a BLUE in the log-Euclidean geometry for $\bbP_d$ by a directly analogous argument.

A slight modification of our argument shows that the fixed-$\Sigmalo$ regression estimator \eqref{eq:cv_affinvar} can be thought of as a BLUE on tangent space. Because we know $\Slo$ and $\Sigmalo$, we can compute the ``difference'' $\log_\Sigmalo \Slo$. Suppose that we want to estimate $\log_\Sigmahi \Shi$ linearly from $\log_\Sigmalo \Slo$, meaning that we seek  
\[
\widehat{\log_\Sigmahi \Shi} = \bfB(\log_\Sigmalo \Slo + C),
\]
where $\bfB: \bbH_d \to \bbH_d$ is linear and $C \in \bbH_d$. Because we know $\Shi$, once we have obtained our estimate of $\log_\Sigmahi \Shi$ we can use it to solve for the corresponding estimate of $\Sigmahi$. We want our estimator to be unbiased, so we require that 
\[
\E[\widehat{\log_\Sigmahi \Shi}] = \E[\bfB(\log_\Sigmalo \Slo + C)] \equiv \E[\log_\Sigmahi \Shi].
\]
By definition, $\E[\log_\Sigmahi \Shi] = \E[\log_\Sigmalo \Slo] = 0$, which gives $C = 0$, yielding 
\begin{equation}
\widehat{\log_\Sigmahi \Shi} = \bfB \log_\Sigmalo \Slo.
\label{eq:affinvar_BLUE}
\end{equation}
We want to choose $\bfB$ such that we minimize the MSE of the estimator on $\bbH_d$, 
\begin{equation*}  
\begin{aligned}
\bfB^{\rm opt} &= \argmin_{\bfB: \bbH_d \to \bbH_d} \E[\langle\bfB\log_\Sigmalo \! \Slo - \log_\Sigmahi \! \Shi, \; \bfB\log_\Sigmalo \! \Slo - \log_\Sigmahi \! \Shi \rangle] \\
&= \argmin_{\bfB: \bbH_d \to \bbH_d} \E[\langle \bfB\log_\Sigmalo \! \Slo, \, \bfB\log_\Sigmalo \! \Slo \rangle \!-\! \langle \log_\Sigmahi \! \Shi,  \bfB \log_\Sigmalo \! \Slo \rangle \!-\! \langle \bfB\log_\Sigmalo \! \Slo, \, \log_\Sigmahi \! \Shi \rangle ] \\
&= \argmin_{\bfB: \bbH_d \to \bbH_d} \rev{\trace{\bfB \circ \Gamma_{\rm lo} \circ \bfB^\ast} - \trace{\Gamma_{\rm lo, hi} \circ \bfB^\ast} - \trace{\bfB \circ \Gamma_{\rm hi, lo}}}.
\end{aligned}
\end{equation*}
The optimization objective on the last line of the above has \rev{similar} form to what we encountered for the Euclidean estimator, resulting in 
\[
\bfB^{\rm opt} = \rev{\Gamma_{\rm lo, hi} \circ \Gamma_{\rm lo}\inv} 
\]
and giving the fixed-$\Sigmalo$ regression estimator 
\[
\log_{\Sigmahihat} \Shi = \rev{(\Gamma_{\rm lo, hi} \circ \Gamma_{\rm lo}\inv)}\log_\Sigmalo \Slo,
\]
which is indeed a type of BLUE on tangent space $\bbH_d$. 

\bibliographystyle{siamplain}
\makeatletter\@input{yy.tex}\makeatother
\bibliography{references}

\end{document}


\maketitle
\section{Supplement to \cref{sec:metricLearning}: metric-learning with the surface quasi--\\geostrophic equation}
\label{app:sqg} 
\begin{figure}[h]  
\centering
\begin{tabular}{ccc}
\includegraphics[width=0.40\textwidth]{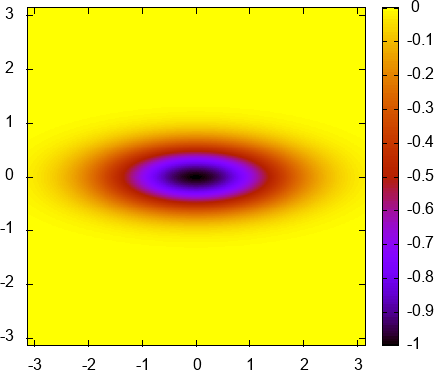}
&
~~~~
&
\includegraphics[width=0.40\textwidth]{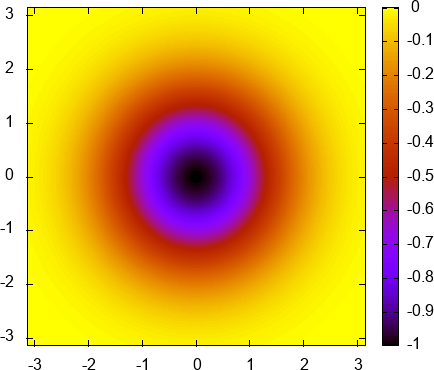}
\\
\small (a) class $i = 0$, initial buoyancy
&&
\small (b) class $i = 1$, initial buoyancy
\\
& \\
\includegraphics[width=0.40\textwidth]{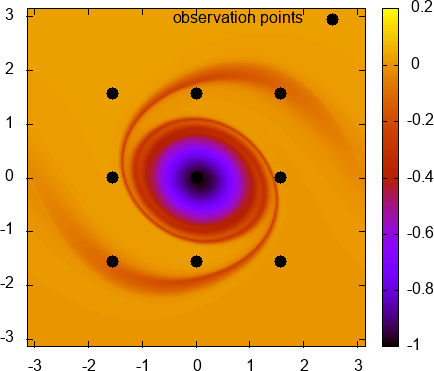}
&&
\includegraphics[width=0.40\textwidth]{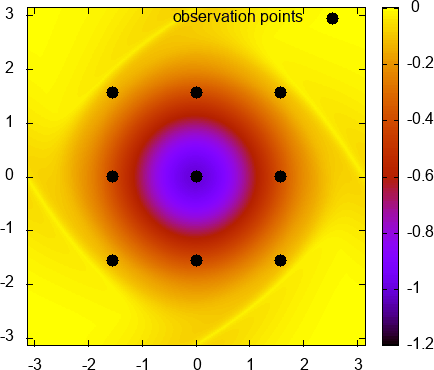}
\\
\small (c) class $i = 0$, final-time buoyancy
&&
\small (d) class $i = 1$, final-time buoyancy
\end{tabular}
\caption{Examples of buoyancy, evolving according to \cref{eq:sqg_system}, at initial (top) and final time (bottom) for $\bftheta$ sampled from class $i = 0$ (left) and $i = 1$ (right). Observations consist of solution values at nine spatial locations in the domain, as depicted in plots (c) and (d). We use the observations to estimate a metric which will distinguish between solutions corresponding to $\bftheta$ sampled from class $i = 0$ and $\bftheta$ sampled from class $i = 1$. }
\label{fig:SQG_solution}
\end{figure}
\subsection{Experimental setup}
The surface quasi-geostrophic equation as presented in \cite{HeldEtAl1985,CapetEtAl2008} describes the evolution of the surface buoyancy $b: \calX \times [0, \infty) \to \R$ on the periodic spatial domain $\calX = [-\pi, \pi] \times [-\pi, \pi]$ via 
\begin{equation} 
\begin{aligned}
\frac{\partial }{\partial t} b(\bfx, t; \bftheta) + J(\psi, b) = 0, \quad z = 0 \\ 
b = \frac{\partial}{\partial z} \psi \\ 
\Delta \psi = 0,\quad z < 0 \\ 
\psi \to 0, \quad z \to -\infty, 
\end{aligned}
\label{eq:sqg_system}
\end{equation} 
where $\bfx = (x_1,x_2)$ is the surface spatial coordinate, $\psi: \calX \times (-\infty, 0] \to \R $ is the stream-function, and $J(\psi, b)$ denotes the Jacobian determinant 
\[
J(\psi, b) = \left( \frac{\partial \psi}{\partial x_1} \right) \left( \frac{\partial b}{\partial x_2} \right) - \left(\frac{\partial b}{\partial x_1} \right)\left( \frac{\partial \psi}{\partial x_2} \right).  
\]
The parameters $\bftheta \in \R^5$ determine the initial condition $b_0$ and some aspects of the dynamics \cref{eq:sqg_system}; we set 
\[
    b_0(\bfx;\bftheta) = -\frac{1}{(2\pi/ |\theta_5|)^2} \exp\left( -x_1^2 - \exp(2\theta_1) x_2^2 \right),
\]
the contours of which form ellipses parametrized by the log aspect-ratio $\theta_1$ and the amplitude $\theta_5$. The gradient Coriolis parameter $\theta_2$, log buoyancy frequency $\theta_3$, and background zonal flow $\theta_4$ all determine aspects of the dynamics.  

In our metric learning experiment in \cref{sec:metricLearning} we draw the parameters $\bftheta$ from an equal two-component Gaussian mixture, i.e., 
\[
p(\bftheta \mid i) = \calN(\bfmu_i, C) = \pi_i, \quad i \sim \mathrm{Ber}(1/2),
\]
where $\bfmu_0$ and $\bfmu_1$ differ only in their first components,
\[
\bfmu_0 = \begin{bmatrix}
1 & 0 & 0 & 0 & 4 
\end{bmatrix}\t, \quad \bfmu_1 = \begin{bmatrix}
0.1 & 0 & 0 & 0 & 4 
\end{bmatrix}\t
\]
and the parameter covariance $C$ is given by 
\[
C = \begin{bmatrix}
    0.3^2 \\ 
    & 0.003^2 \\ 
    & & 0 \\ 
    & & & 0.08^2 \\ 
    & & & & 0.3^2
\end{bmatrix}.
\]
Note that this choice of $C$ indicates that the log buoyancy frequency $\theta_3 = 0$ is deterministic, but the observational covariances $\Gamma_0$ and $\Gamma_1$ which we learn in \cref{sec:metricLearning} are still full-rank. In \cref{fig:SQG_solution} we show examples of the initial buoyancy $b_0$ and final buoyancy $b$ at time $T = 24$ for samples of $\bftheta$ from both mixture components.  

\subsection{Additional results}
\label{app:sqg:addResults}
In the following subsections we display results pertaining to estimation of $\Gamma_0 = \Cov[\bfy \mid \bftheta \sim \pi_0]$ and $\Gamma_1 = \Cov[\bfy \mid \bftheta \sim \pi_1]$. We see in \cref{sec:metricLearning} that the best estimates of $A_{\rm GMML}$ in both the Frobenius and intrinsic metrics are obtained with $\Gamma_0$ and $\Gamma_1$ estimated via multifidelity regression, even though, as we show below, multifidelity regression is generally not the best-performing estimator for $\Gamma_0$ and $\Gamma_1$ in the Frobenius metric. This behavior is sensible when one considers that (a) $A_{\rm GMML}$ is defined as a point on a geodesic between two SPD matrices in the affine-invariant geometry, and the regression estimator, being constructed using the affine-invariant geometry, is thus the ``natural'' choice in this application, (b) while the LEMF estimator out-performs the regression estimator in the Frobenius metric for estimation of $\Gamma_1$, it does quite poorly in estimating $\Gamma_0$ and thus yields relatively poor estimates of $A_{\rm GMML}$, and (c) we are generally unable to construct $A_{\rm GMML}$ from estimates of $\Gamma_0$ and $\Gamma_1$ computed with the EMF estimator due to a high frequency (94\%) of indefiniteness of $\hat\Gamma_{0}^{\rm EMF}$ or $\hat\Gamma_1^{\rm EMF}$. 
\begin{figure}[h]
    \centering
    \captionsetup[subfigure]{justification=centering}
    \begin{subfigure}[t]{0.24\linewidth}
        \includegraphics[width=\linewidth]{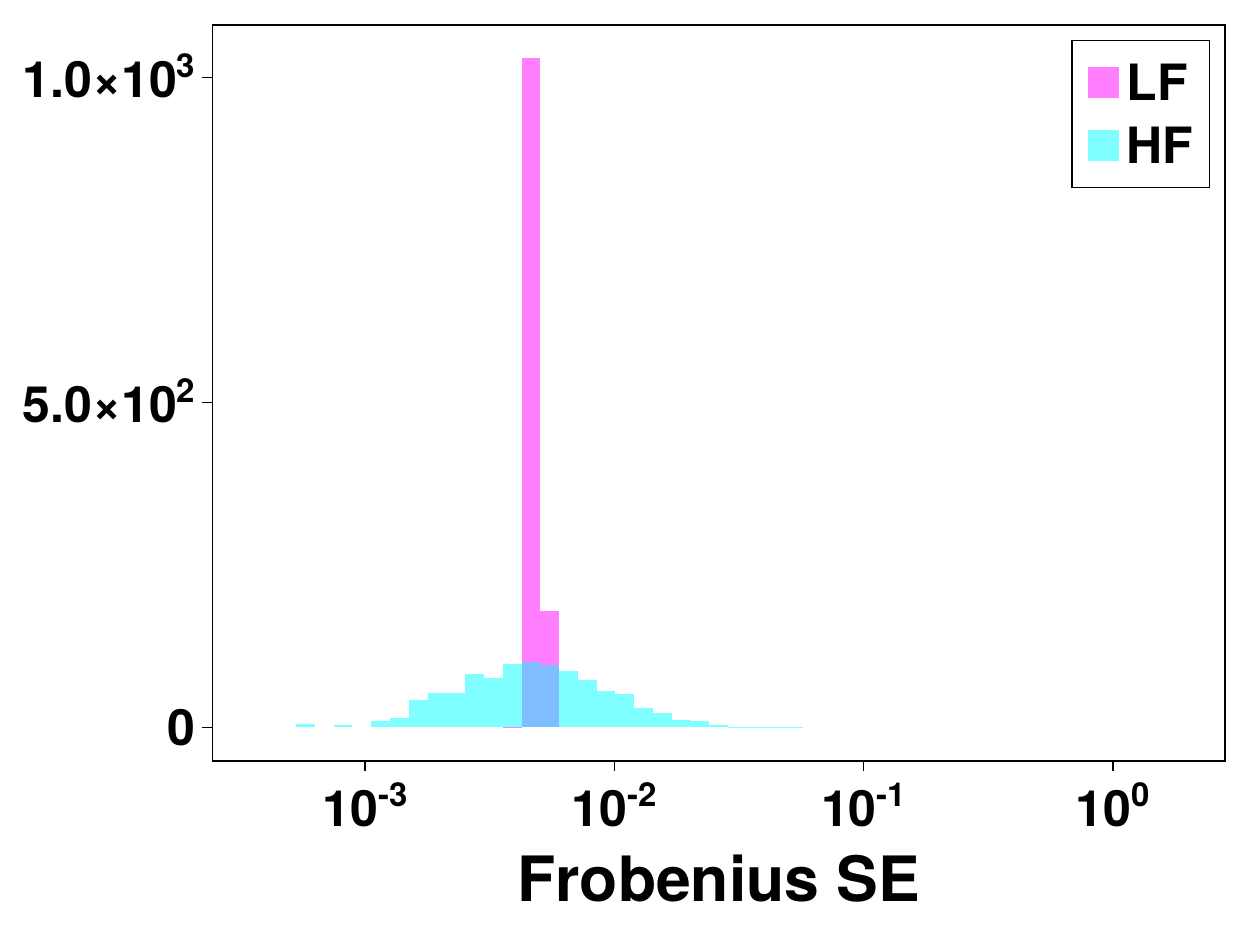}
        \subcaption*{Low-fidelity: 46\% decrease in Frobenius MSE}
    \end{subfigure}
    \begin{subfigure}[t]{0.24\linewidth}
        \includegraphics[width=\linewidth]{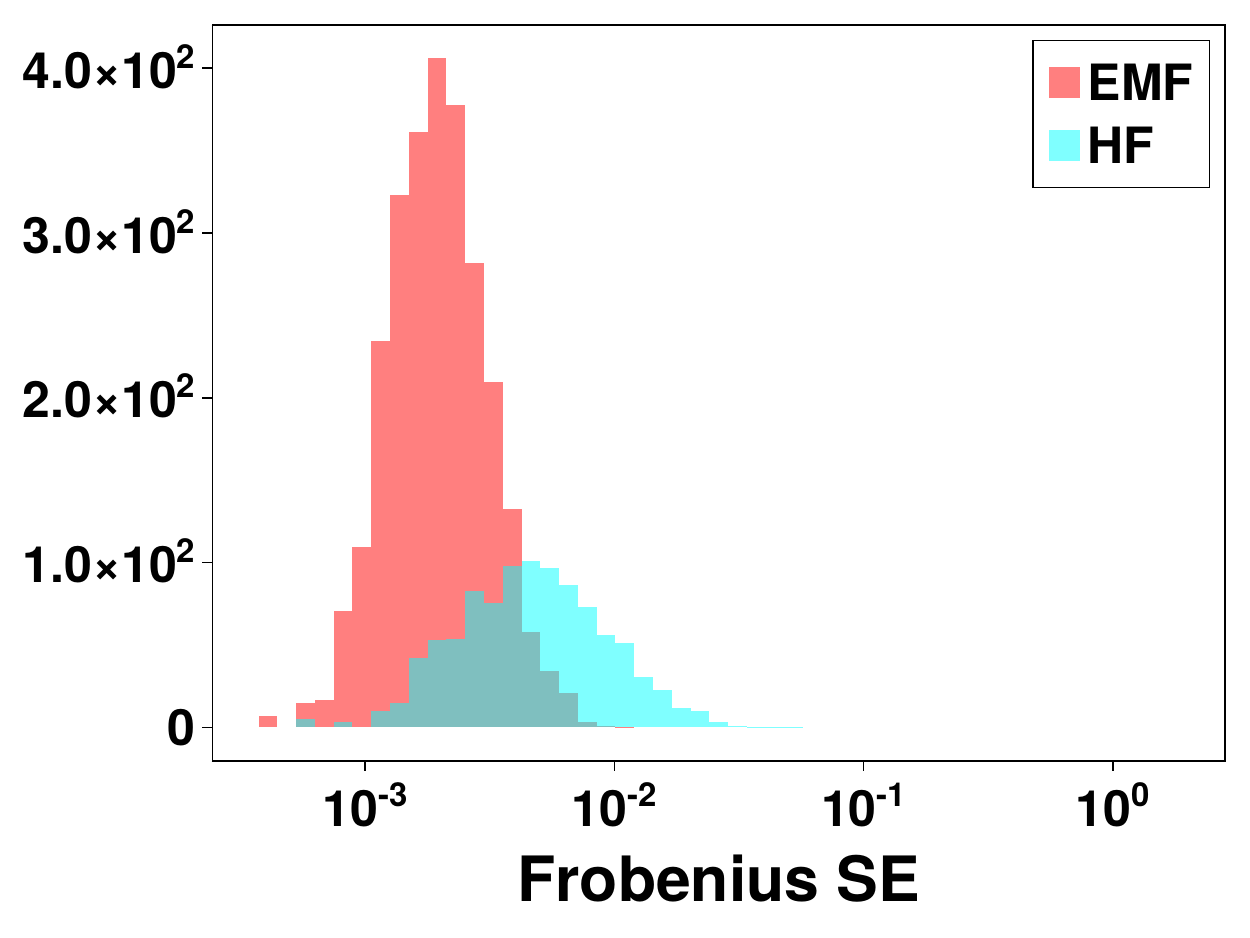}
        \subcaption*{EMF: 70\% decrease in Frobenius MSE}
    \end{subfigure}
    \begin{subfigure}[t]{0.24\linewidth}
        \includegraphics[width=\linewidth]{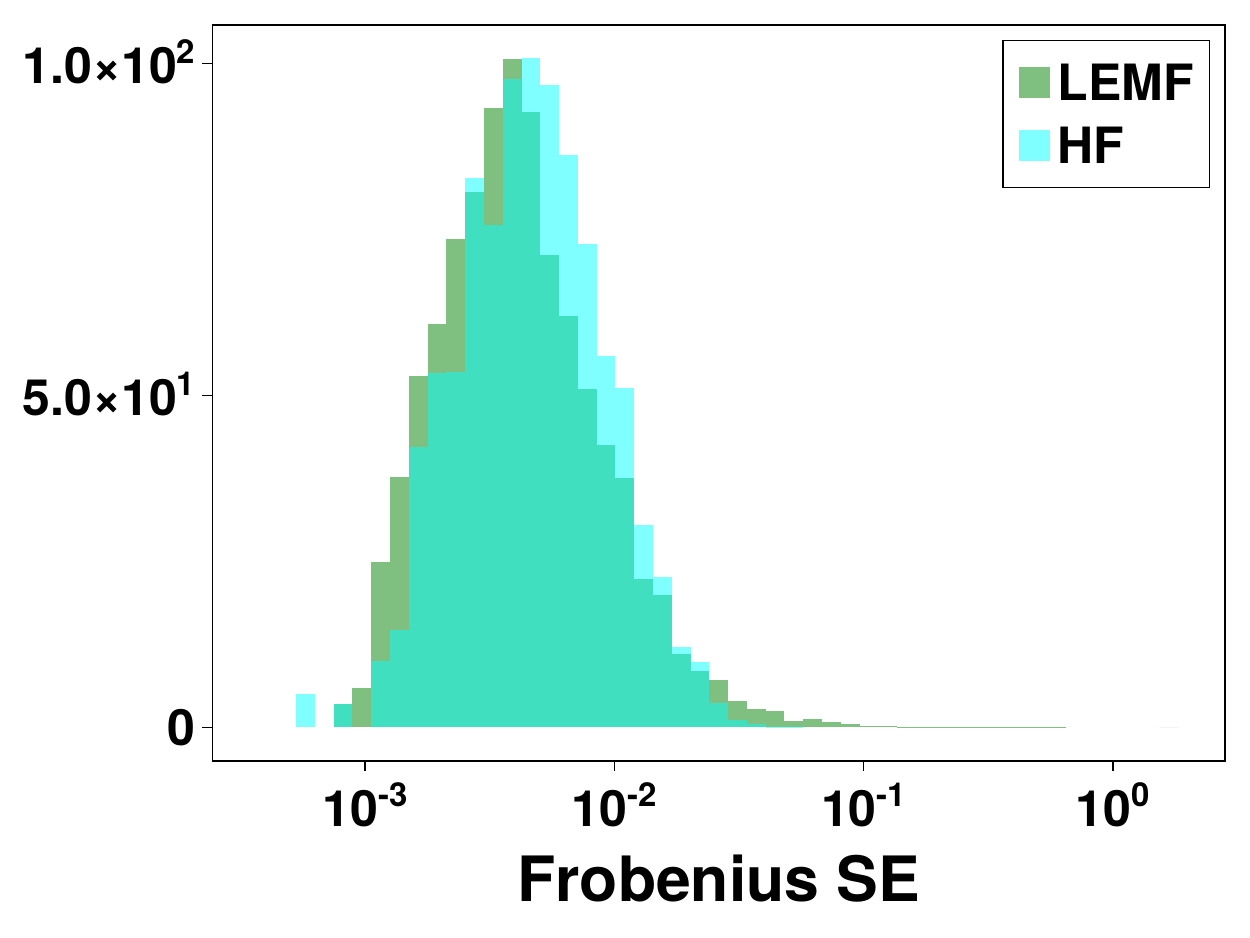}
        \subcaption*{LEMF: 135\% \textit{increase} in Frobenius MSE}
    \end{subfigure}
    \begin{subfigure}[t]{0.24\linewidth}
        \includegraphics[width=\linewidth]{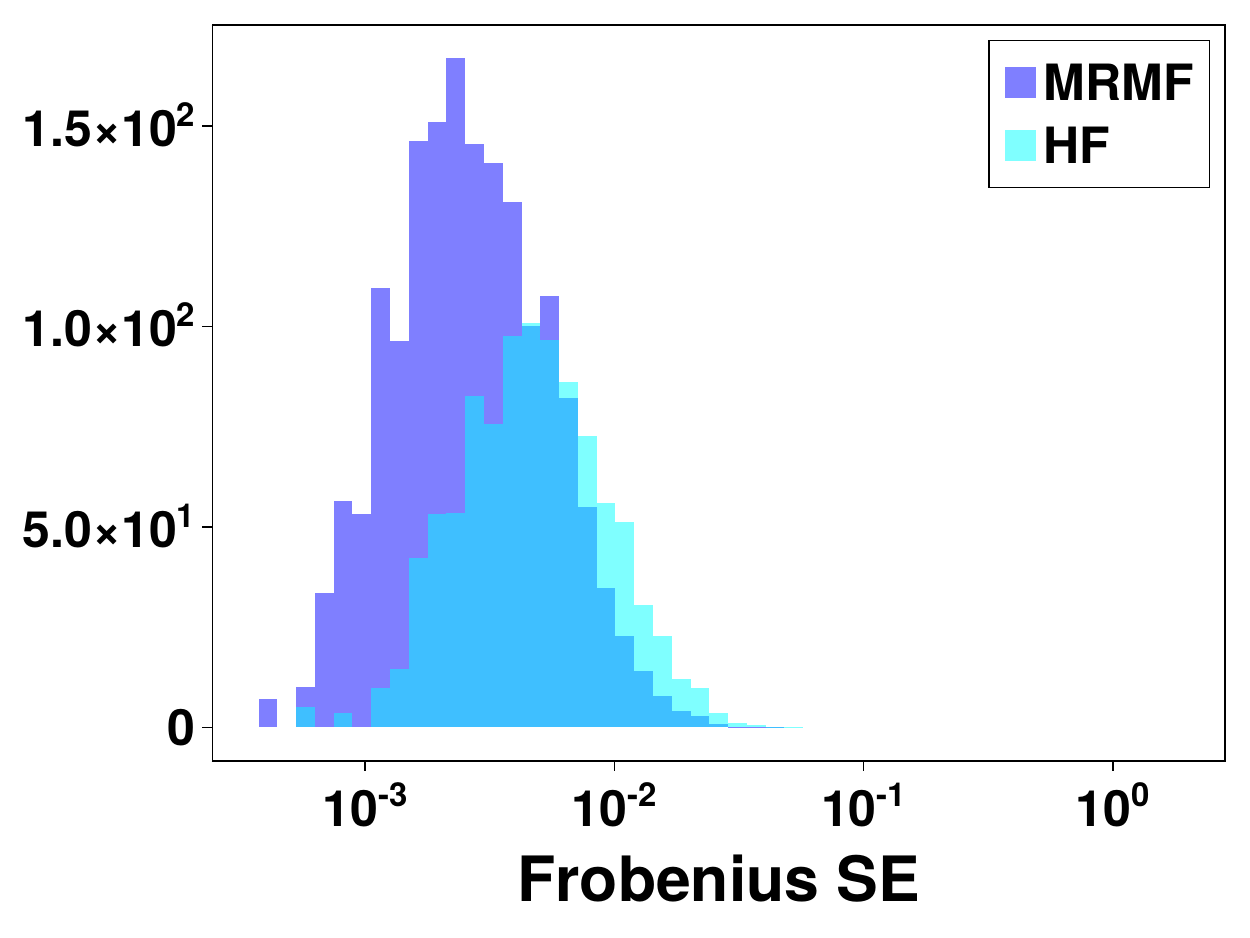}
        \subcaption*{Regression: 37\% decrease in Frobenius MSE}
    \end{subfigure}
    \\
    \begin{subfigure}[t]{0.32\linewidth}
        \includegraphics[width=\linewidth]{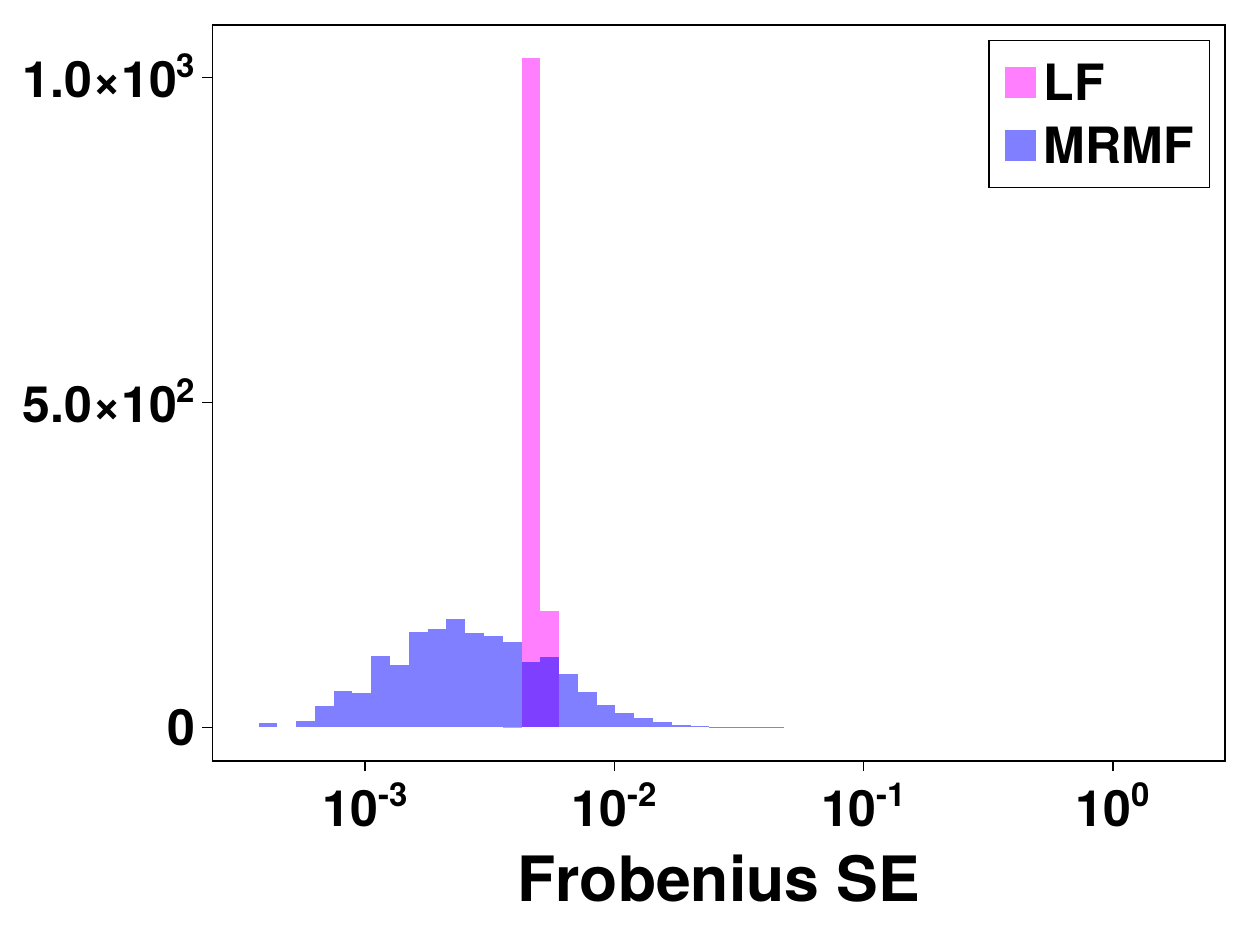}
    \end{subfigure}
    \begin{subfigure}[t]{0.32\linewidth}
        \includegraphics[width=\linewidth]{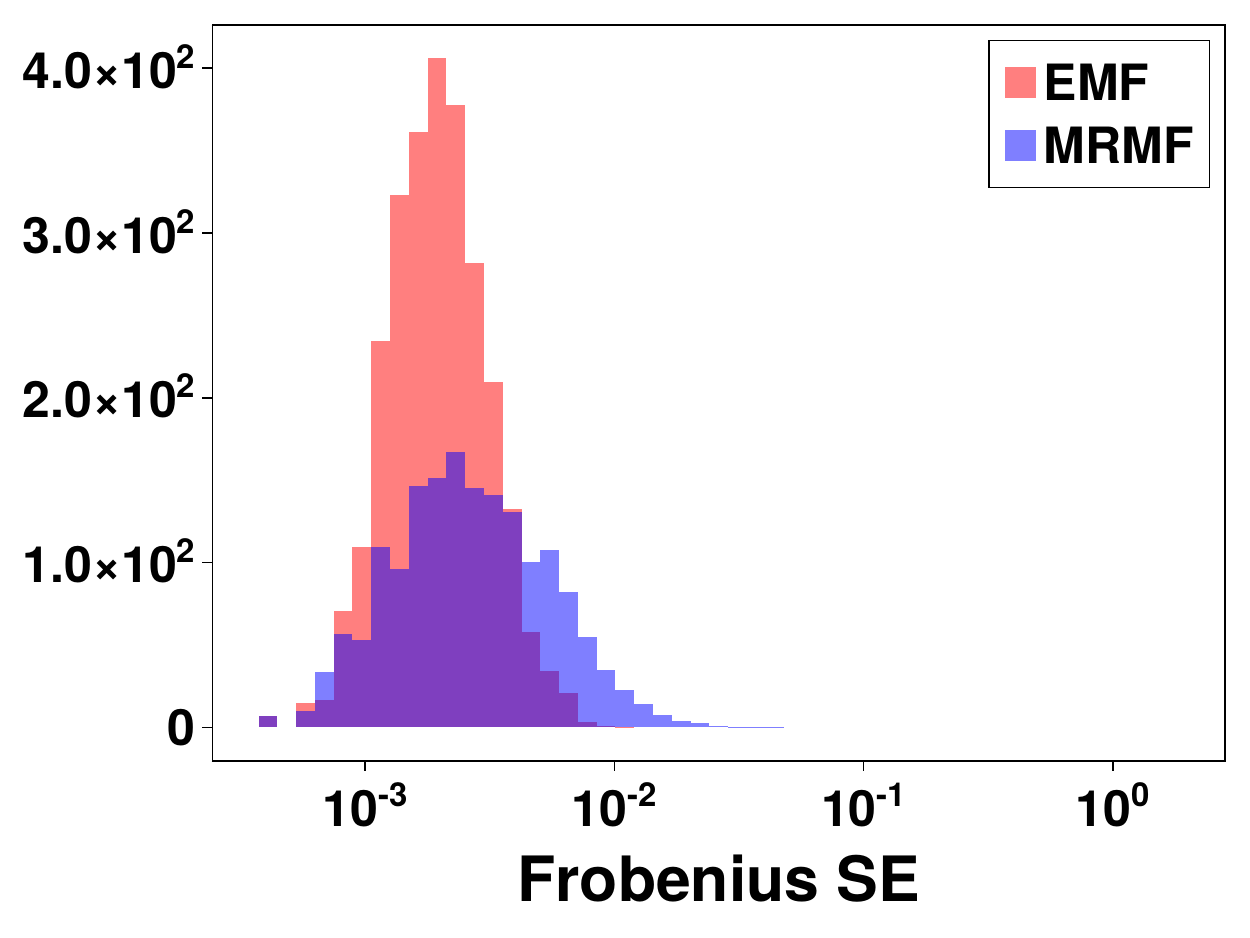}
    \end{subfigure}
    \begin{subfigure}[t]{0.32\linewidth}
        \includegraphics[width=\linewidth]{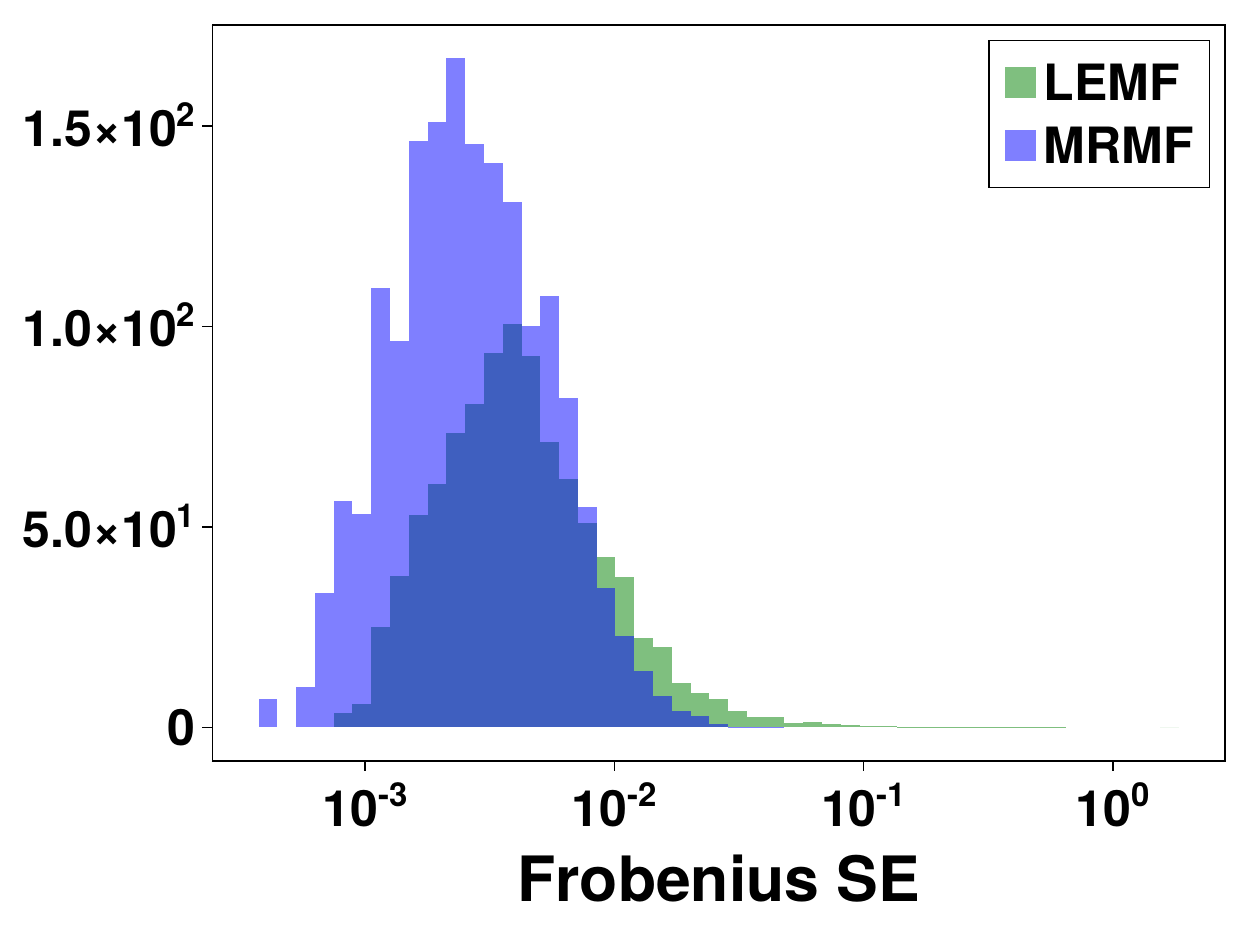}
    \end{subfigure}
    \caption{\textbf{Top:} Frobenius squared error histograms corresponding to (from left to right) $\hat\Gamma_0^{\rm LF}$, $\hat\Gamma_0^{\rm EMF}$, $\hat\Gamma_0^{\rm LEMF}$, and $\hat\Gamma_0^{\rm MRMF}$ compared to that of $\hat\Gamma_0^{\rm HF}$, overlaid in cyan. Reported changes in MSE are relative to $\hat\Gamma_0^{\rm HF}$. \textbf{Bottom:}  Frobenius squared error histograms of $\hat\Gamma_0^{\rm LF}$ (left), $\hat\Gamma_0^{\rm EMF}$ (center), and $\hat\Gamma_0^{\rm LEMF}$ (right) compared to that of $\hat\Gamma_0^{\rm MRMF}$, overlaid in blue. 
    In the Frobenius metric $\hat\Gamma_{\rm 0}^{\rm EMF}$ is the best-performing estimator, which is to be expected, as $\hat\Gamma_{\rm 0}^{\rm EMF}$ is optimized to minimize Frobenius MSE.
.    }
    \label{fig:Gamma0_fro}
\end{figure}
\subsubsection{Estimation of $\Gamma_0$}
\begin{figure}[h]
    \centering
    \captionsetup[subfigure]{justification=centering}
    \begin{subfigure}[t]{0.32\linewidth}
        \includegraphics[width=\linewidth]{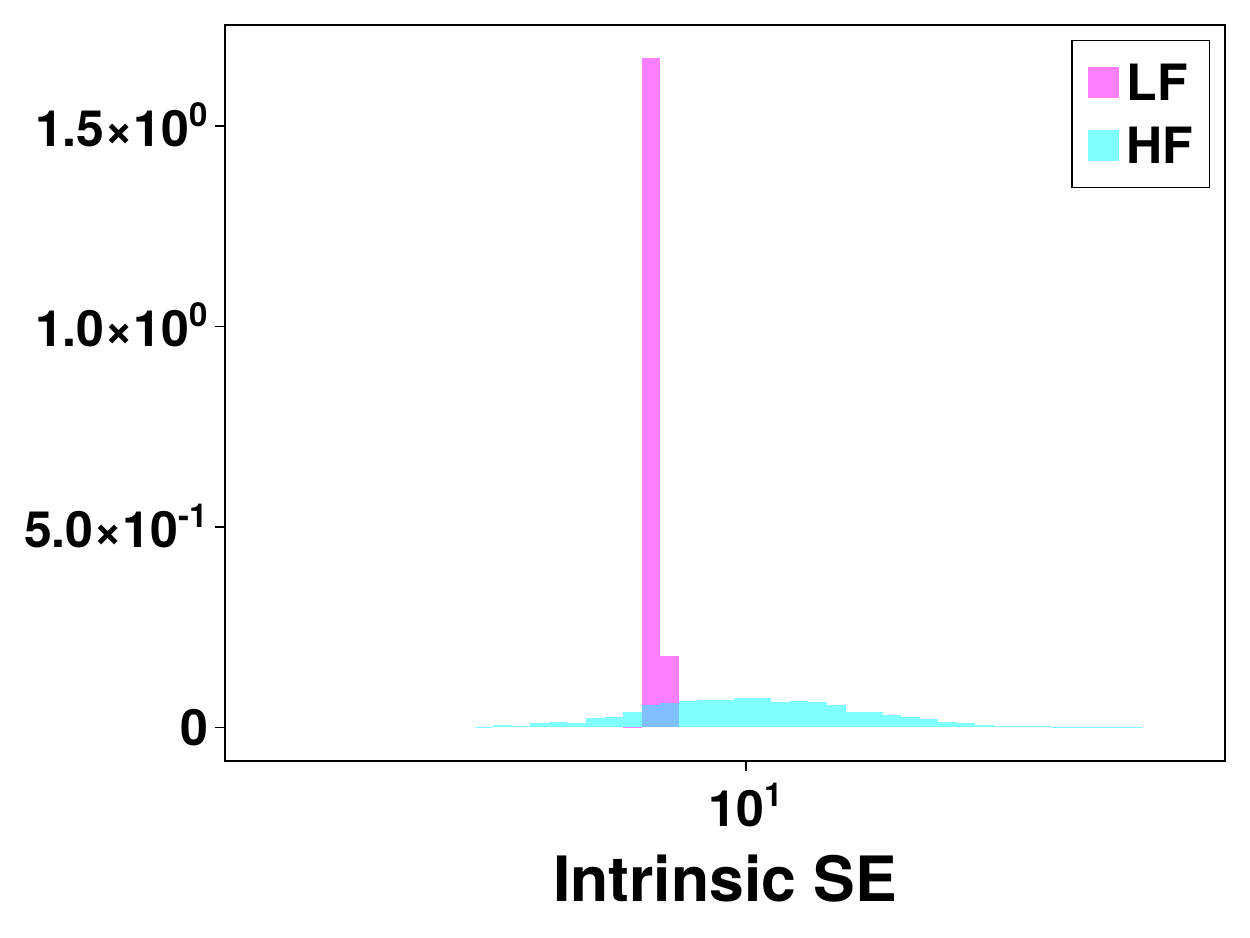}
        \subcaption*{Low-fidelity: 54\% decrease in intrinsic MSE}
    \end{subfigure}
    \begin{subfigure}[t]{0.32\linewidth}
        \includegraphics[width=\linewidth]{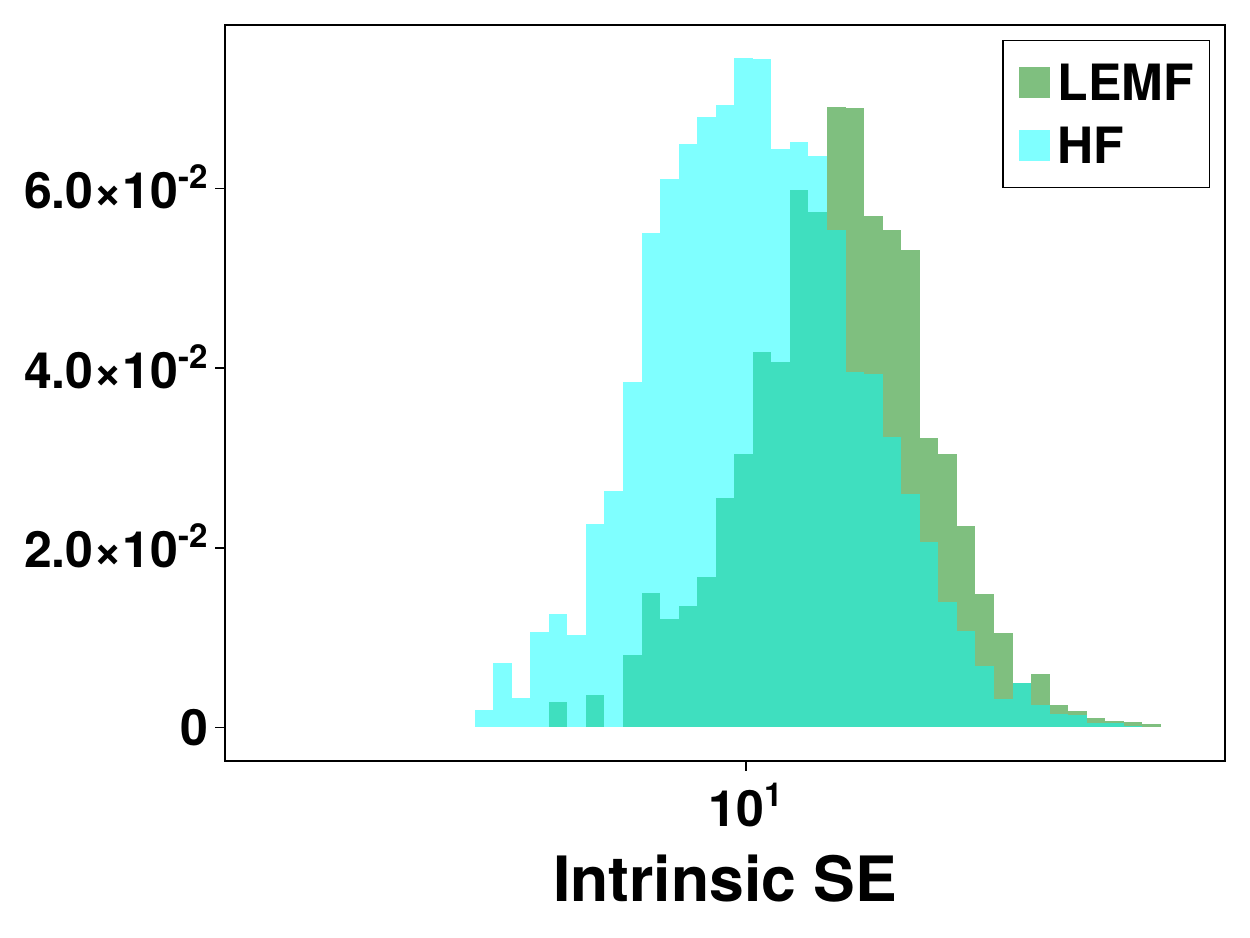}
        \subcaption*{LEMF: 30.\% \textit{increase} in intrinsic MSE}
    \end{subfigure}
    \begin{subfigure}[t]{0.32\linewidth}
        \includegraphics[width=\linewidth]{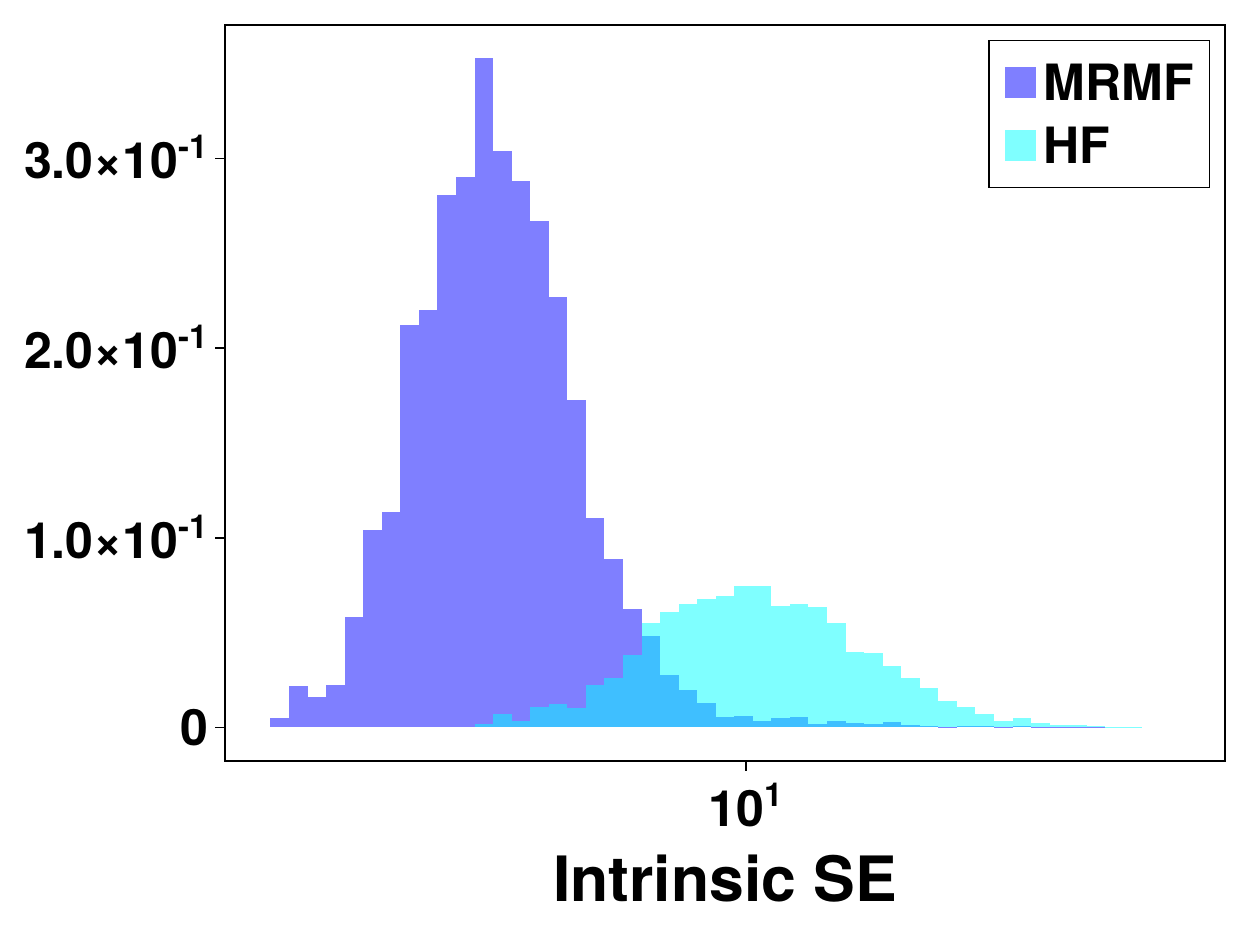}
        \caption*{Regression: 68\% decrease in intrinsic MSE}
    \end{subfigure}
    \\
    \begin{subfigure}[t]{0.32\linewidth}
        \includegraphics[width=\linewidth]{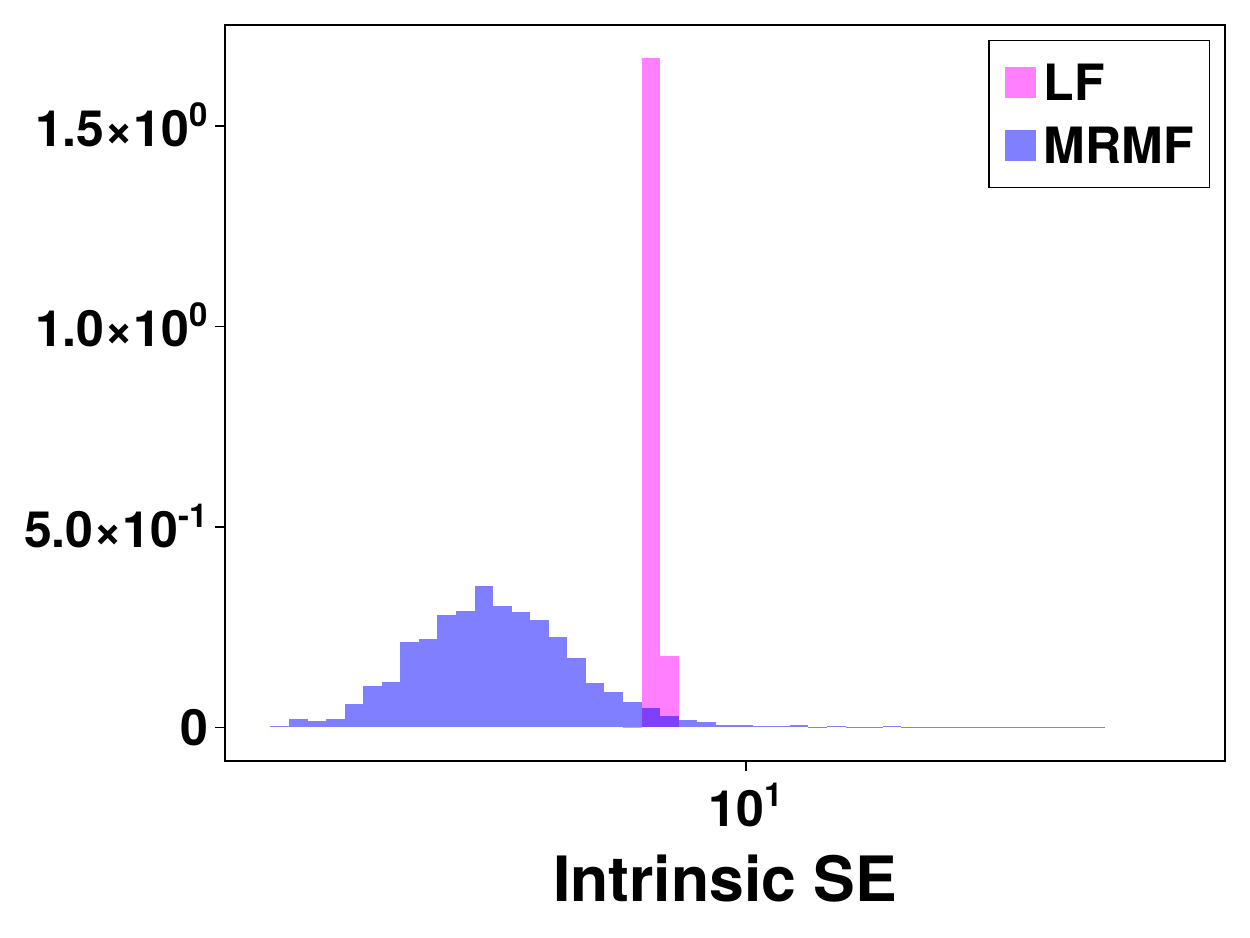}
    \end{subfigure}
    \begin{subfigure}[t]{0.32\linewidth}
        \includegraphics[width=\linewidth]{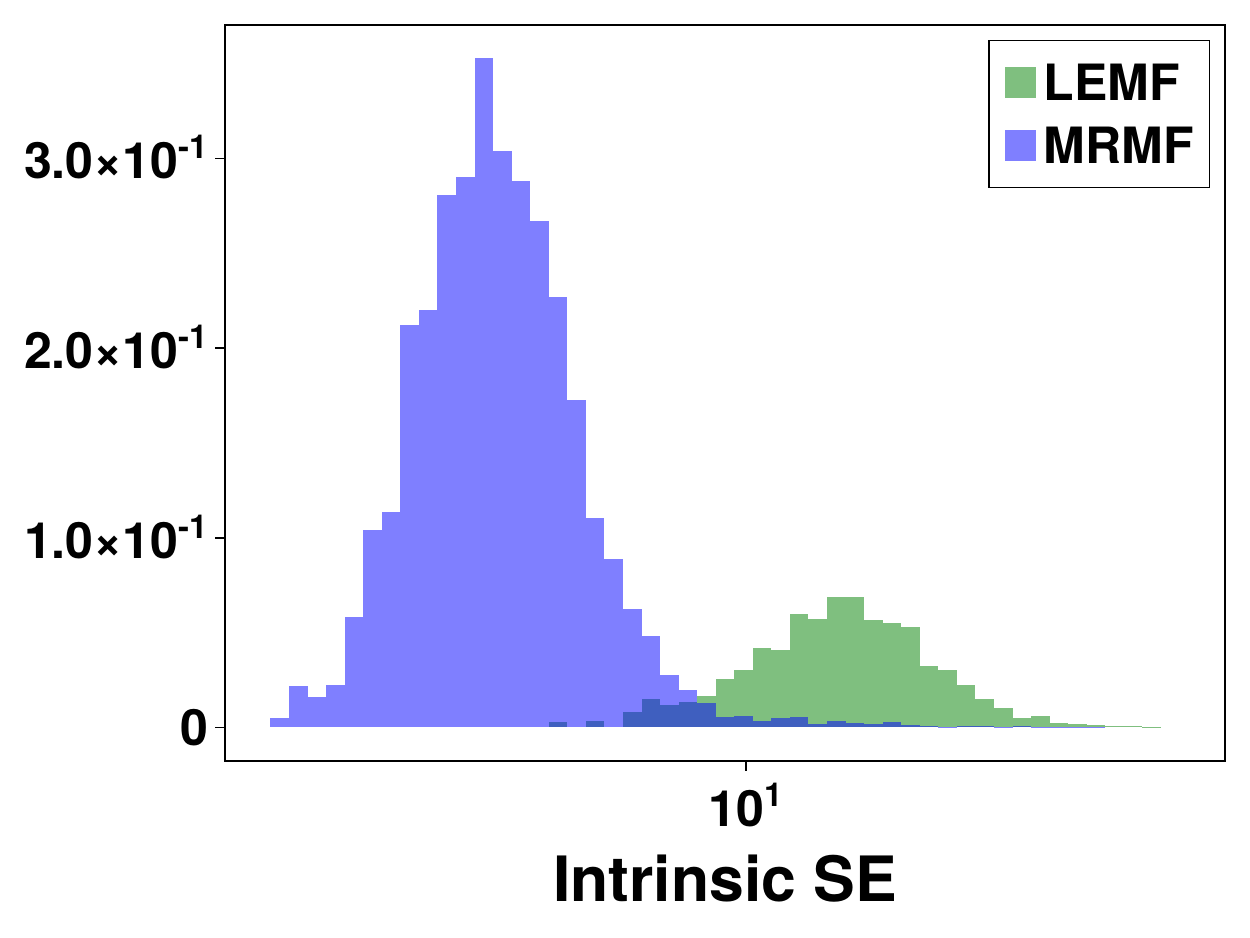}
    \end{subfigure}
    \caption{\textbf{Top:} Intrinsic squared error histograms corresponding to $\hat\Gamma_0^{\rm LF}$ (left), $\hat\Gamma_0^{\rm LEMF}$ (center), and $\hat\Gamma_0^{\rm MRMF}$ (right). Squared error histograms for $\hat\Gamma_0^{\rm HF}$ are overlaid in cyan for comparison. Reported changes in MSE are relative to $\hat\Gamma_0^{\rm HF}$. \textbf{Bottom:} Intrinsic squared error histograms of $\hat\Gamma_0^{\rm LF}$ (left) and $\hat\Gamma_0^{\rm LEMF}$ (right) compared to that of $\hat\Gamma_0^{\rm MRMF}$, overlaid in blue. Use of $\hat\Gamma_0^{\rm MRMF}$ results in an 31\% decrease in intrinsic MSE relative to that of $\hat\Gamma_0^{\rm LF}$ and a 76\% decrease relative to that of $\hat\Gamma_0^{\rm LEMF}$. We do not report results for $\hat\Gamma_{0}^{\rm EMF}$ in the intrinsic metric, which is only defined for SPD arguments, because 82.4\% of its  realizations are indefinite. }
    \label{fig:Gamma0_AI_LE}
\end{figure}

\vskip -0.25cm
In \cref{fig:Gamma0_fro,fig:Gamma0_AI_LE} we show squared error histograms corresponding to $\hat\Gamma_0^{\rm HF}$, $\hat\Gamma_0^{\rm LF}$, $\hat\Gamma_0^{\rm EMF}$, $\hat\Gamma_0^{\rm LEMF}$, and $\hat\Gamma_0^{\rm MRMF}$. In general $\hat\Gamma_0^{\rm LF}$, $\hat\Gamma_0^{\rm EMF}$, and $\hat\Gamma_0^{\rm MRMF}$ all yield substantial decreases in squared error relative to $\hat\Gamma_0^{\rm HF}$, while interestingly $\hat\Gamma_0^{\rm LEMF}$ results in an \textit{increase} squared error relative to $\hat\Gamma_0^{\rm HF}$, perhaps due to amplification of error by the matrix exponential. As one might expect, $\hat\Gamma_0^{\rm MRMF}$ achieves the lowest MSE in the intrinsic metric, while $\hat\Gamma_0^{\rm EMF}$ achieves lowest MSE in the Frobenius metric. At the same time, 82.4\% of realizations of $\hat\Gamma_0^{\rm EMF}$ are indefinite and thus useless for construction of $\hat A_{\rm GMML}$. 

\subsubsection{Estimation of $\Gamma_1$}
\begin{figure}[h]
    \centering
    \captionsetup[subfigure]{justification=centering}
    \begin{subfigure}[t]{0.24\linewidth}
        \includegraphics[width=\linewidth]{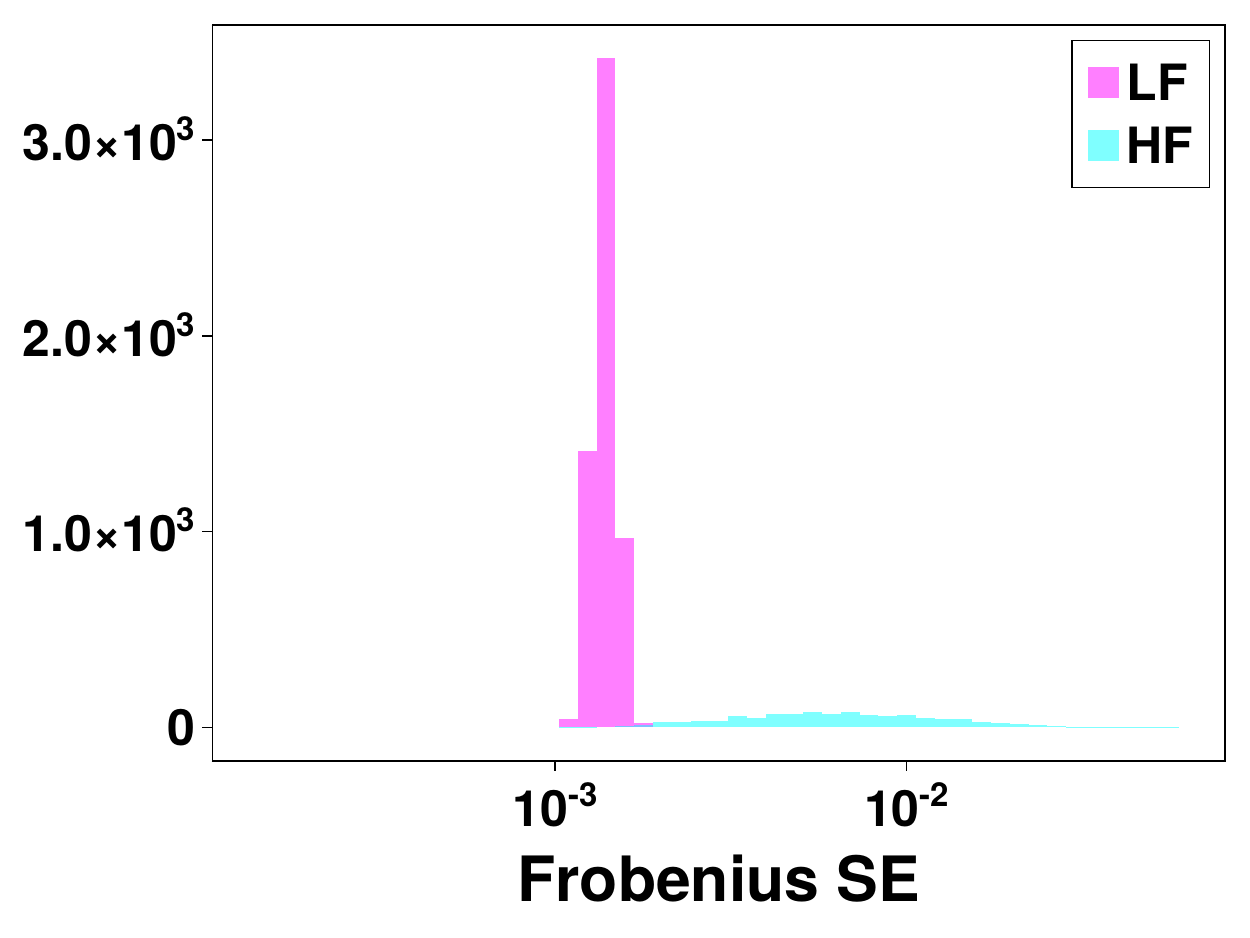}
        \subcaption*{Low-fidelity: 88\% decrease in Frobenius MSE}
    \end{subfigure}
    \begin{subfigure}[t]{0.24\linewidth}
        \includegraphics[width=\linewidth]{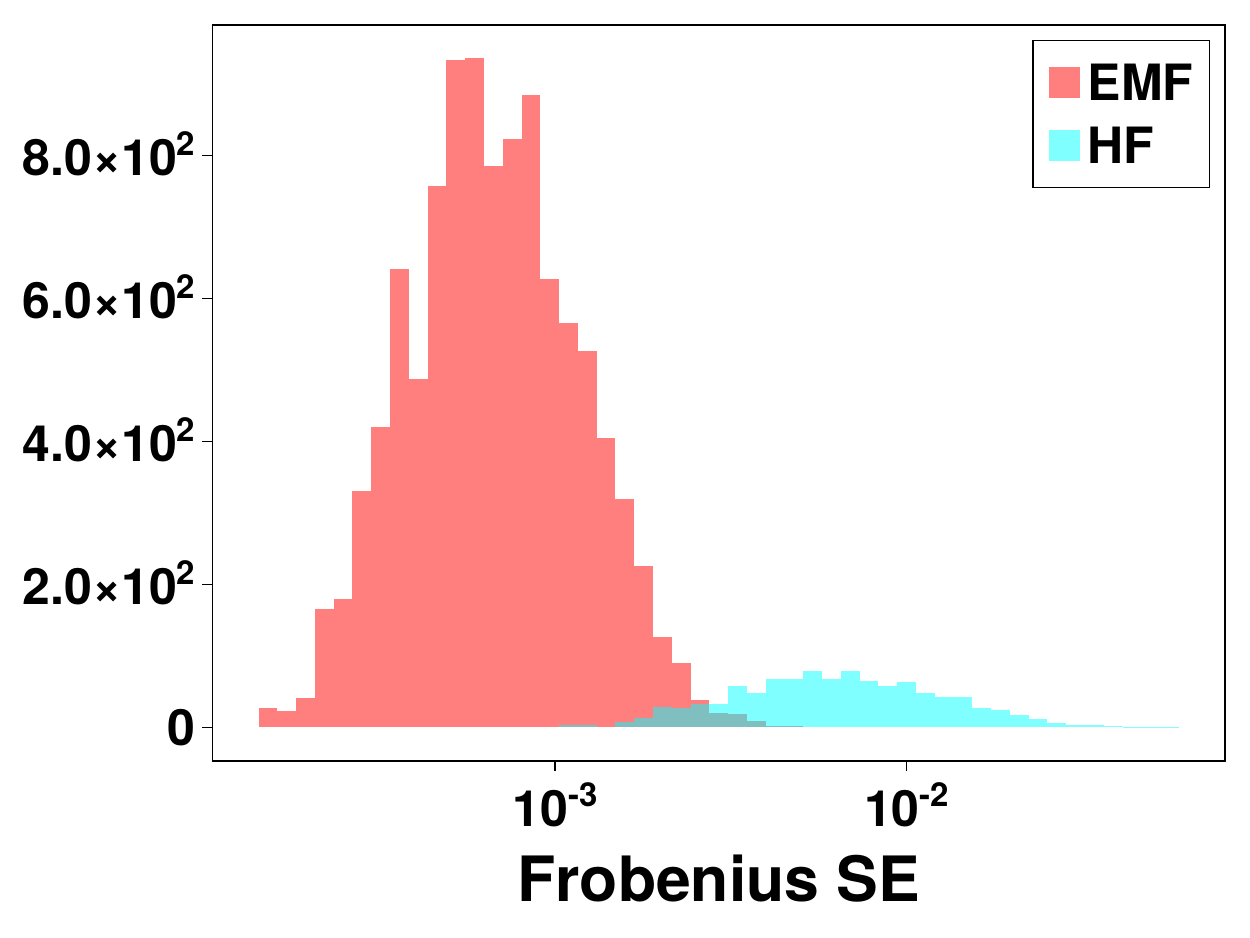}
        \subcaption*{EMF: 91\% decrease in Frobenius MSE}
    \end{subfigure}
    \begin{subfigure}[t]{0.24\linewidth}
        \includegraphics[width=\linewidth]{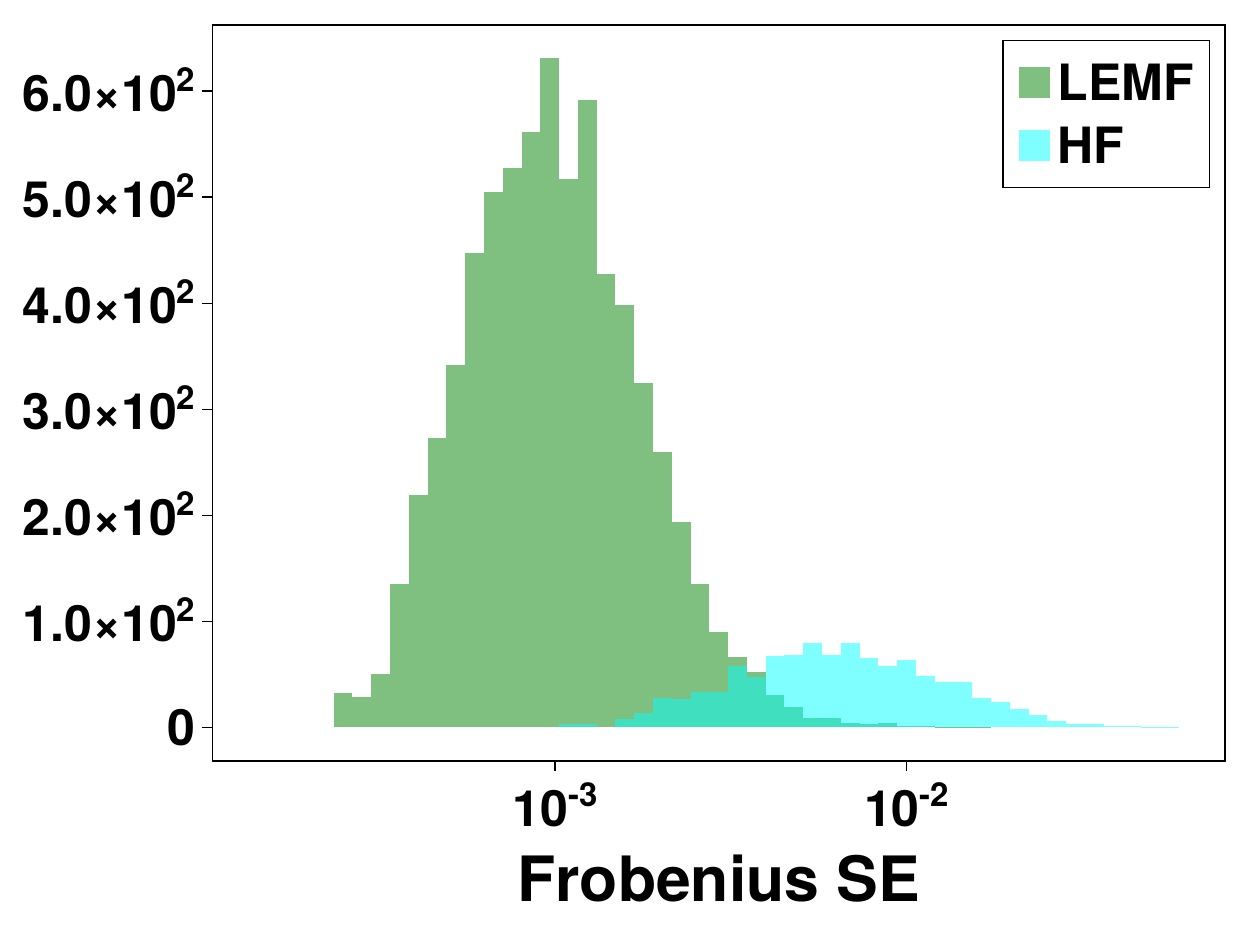}
        \subcaption*{LEMF: 85\% {decrease} in Frobenius MSE}
    \end{subfigure}
    \begin{subfigure}[t]{0.24\linewidth}
        \includegraphics[width=\linewidth]{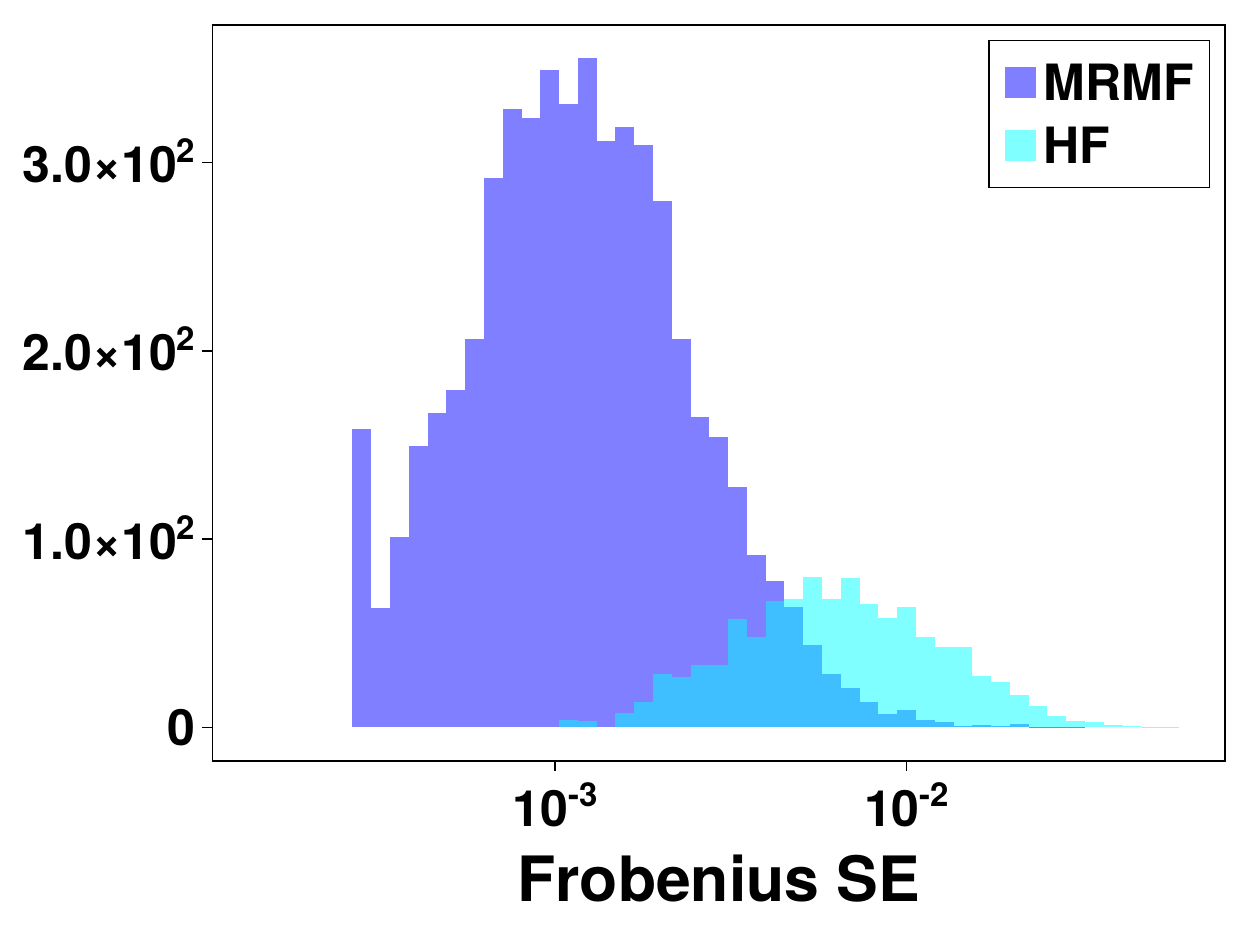}
        \subcaption*{Regression: 76\% decrease in Frobenius MSE}
    \end{subfigure}
    \\
    \begin{subfigure}[t]{0.32\linewidth}
        \includegraphics[width=\linewidth]{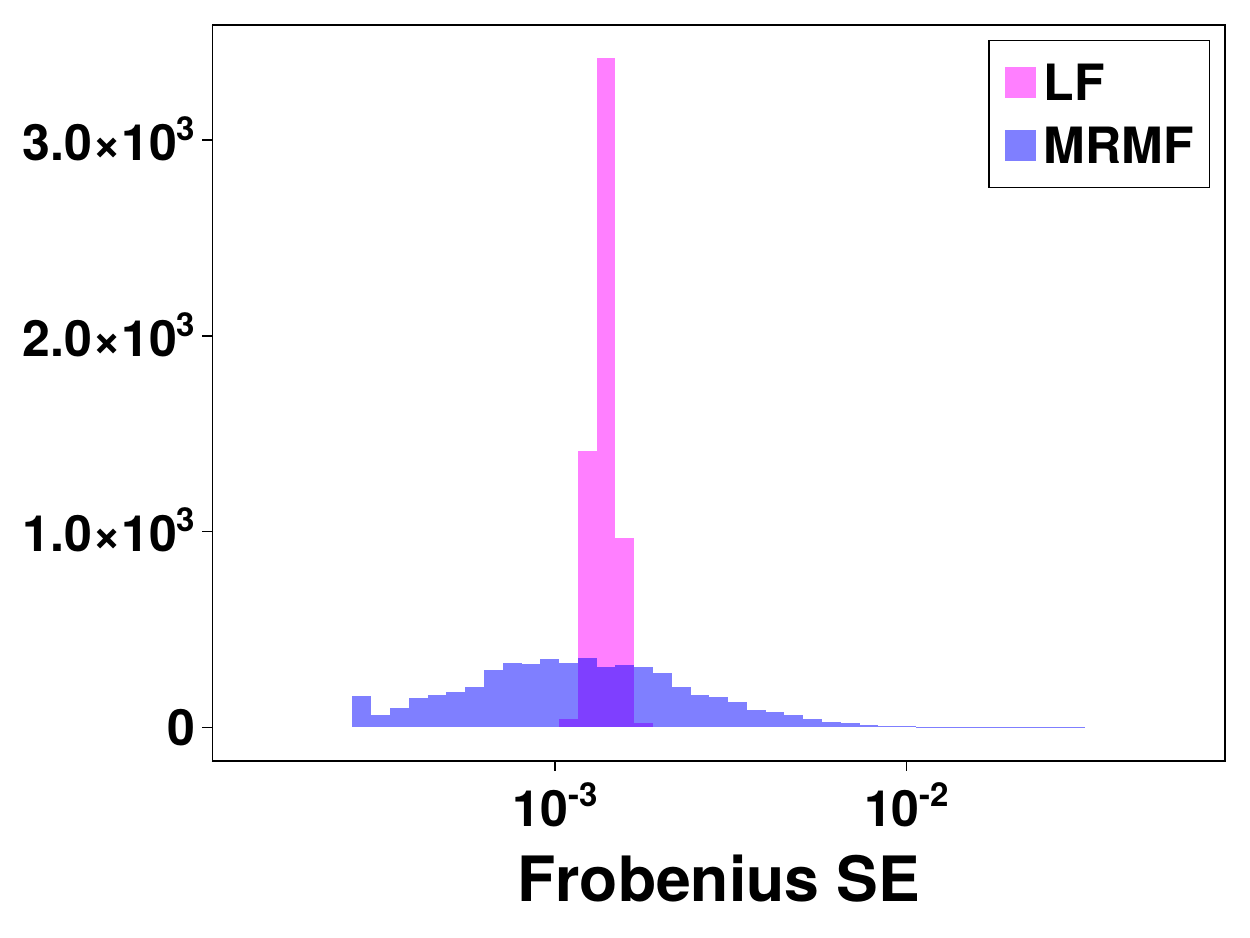}
    \end{subfigure}
    \begin{subfigure}[t]{0.32\linewidth}
        \includegraphics[width=\linewidth]{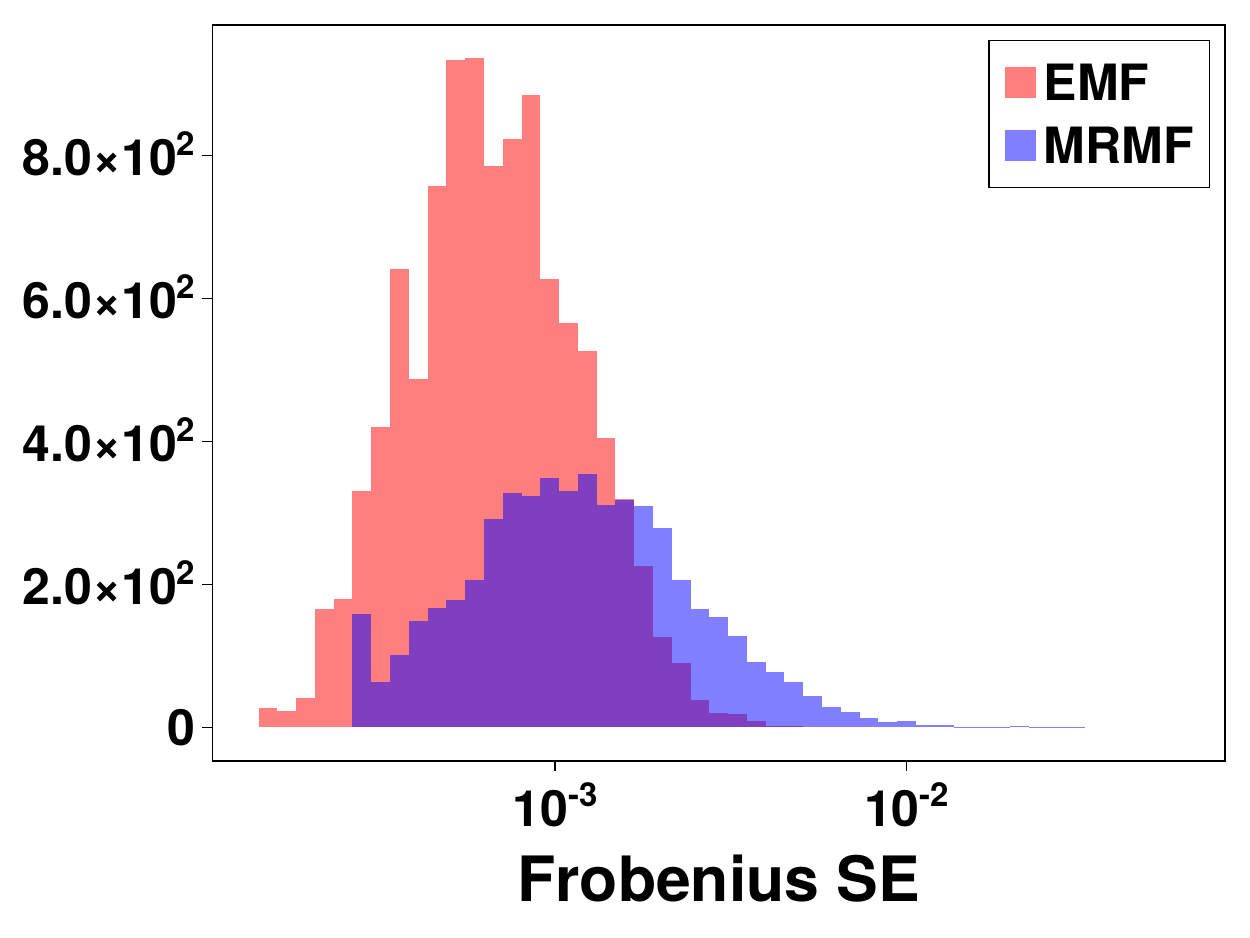}
    \end{subfigure}
    \begin{subfigure}[t]{0.32\linewidth}
        \includegraphics[width=\linewidth]{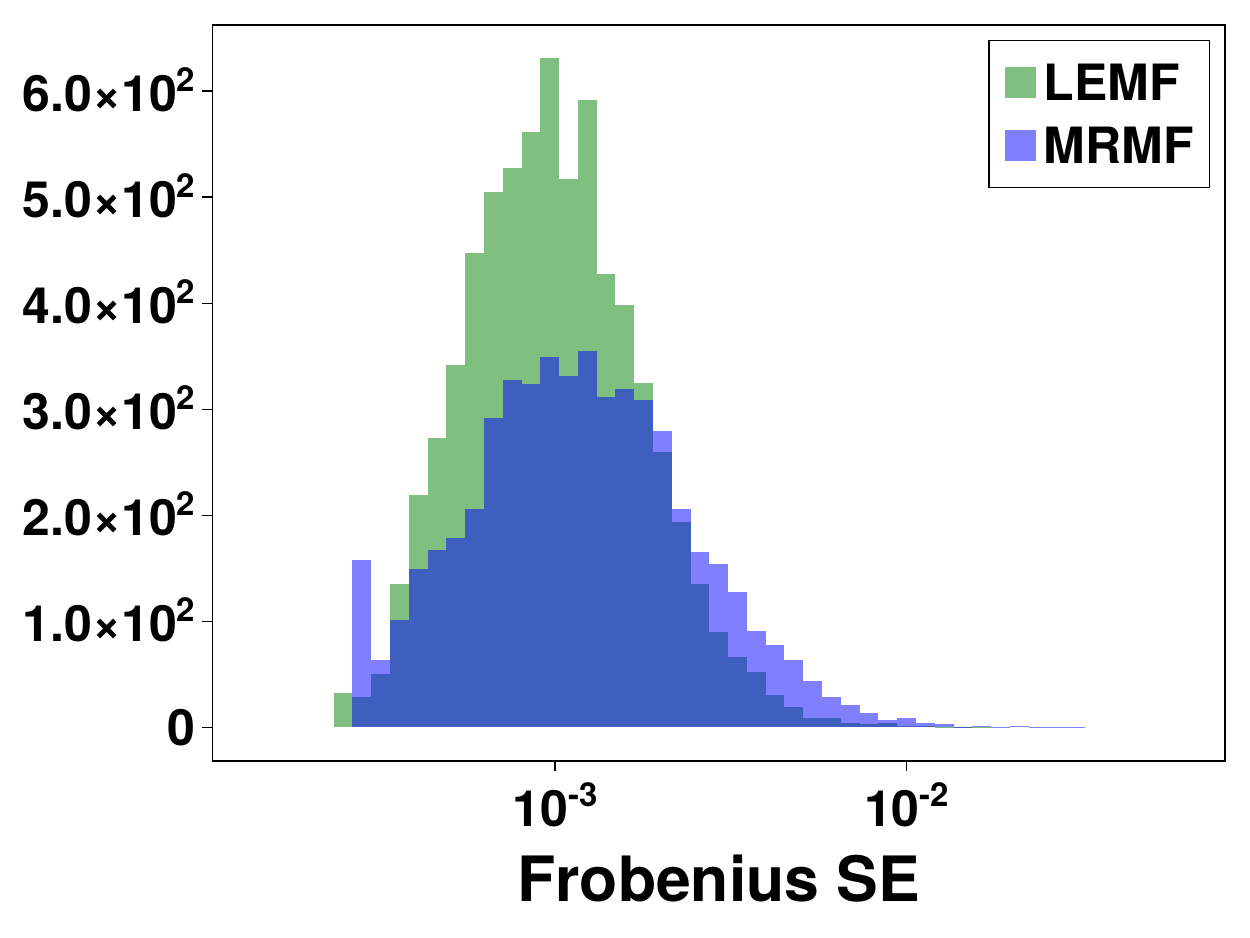}
    \end{subfigure}
    \caption{\textbf{Top:} Frobenius squared error histograms corresponding to (from left to right) $\hat\Gamma_1^{\rm LF}$, $\hat\Gamma_1^{\rm EMF}$, $\hat\Gamma_1^{\rm LEMF}$, and $\hat\Gamma_1^{\rm MRMF}$ compared to that of $\hat\Gamma_1^{\rm HF}$, overlaid in cyan. Reported changes in MSE are relative to $\hat\Gamma_1^{\rm HF}$. \textbf{Bottom:}  Frobenius squared error histograms of $\hat\Gamma_1^{\rm LF}$ (left), $\hat\Gamma_1^{\rm EMF}$ (center), and $\hat\Gamma_1^{\rm LEMF}$ (right) compared to that of $\hat\Gamma_1^{\rm MRMF}$, overlaid in blue. 
    All four estimators obtain substantial decreases in MSE relative to $\hat\Gamma_1^{\rm HF}$, but, interestingly, in the Frobenius metric $\hat\Gamma_1^{\rm MRMF}$ achieves the smallest reduction. By contrast, $\hat\Gamma_1^{\rm MRMF}$ is the best-performing estimator in the intrinsic metric (\cref{fig:Gamma1_AI_LE}).
    }
    \label{fig:Gamma1_fro} 
\end{figure}

\begin{figure}
    \centering
    \captionsetup[subfigure]{justification=centering}
    \begin{subfigure}[t]{0.32\linewidth}
        \includegraphics[width=\linewidth]{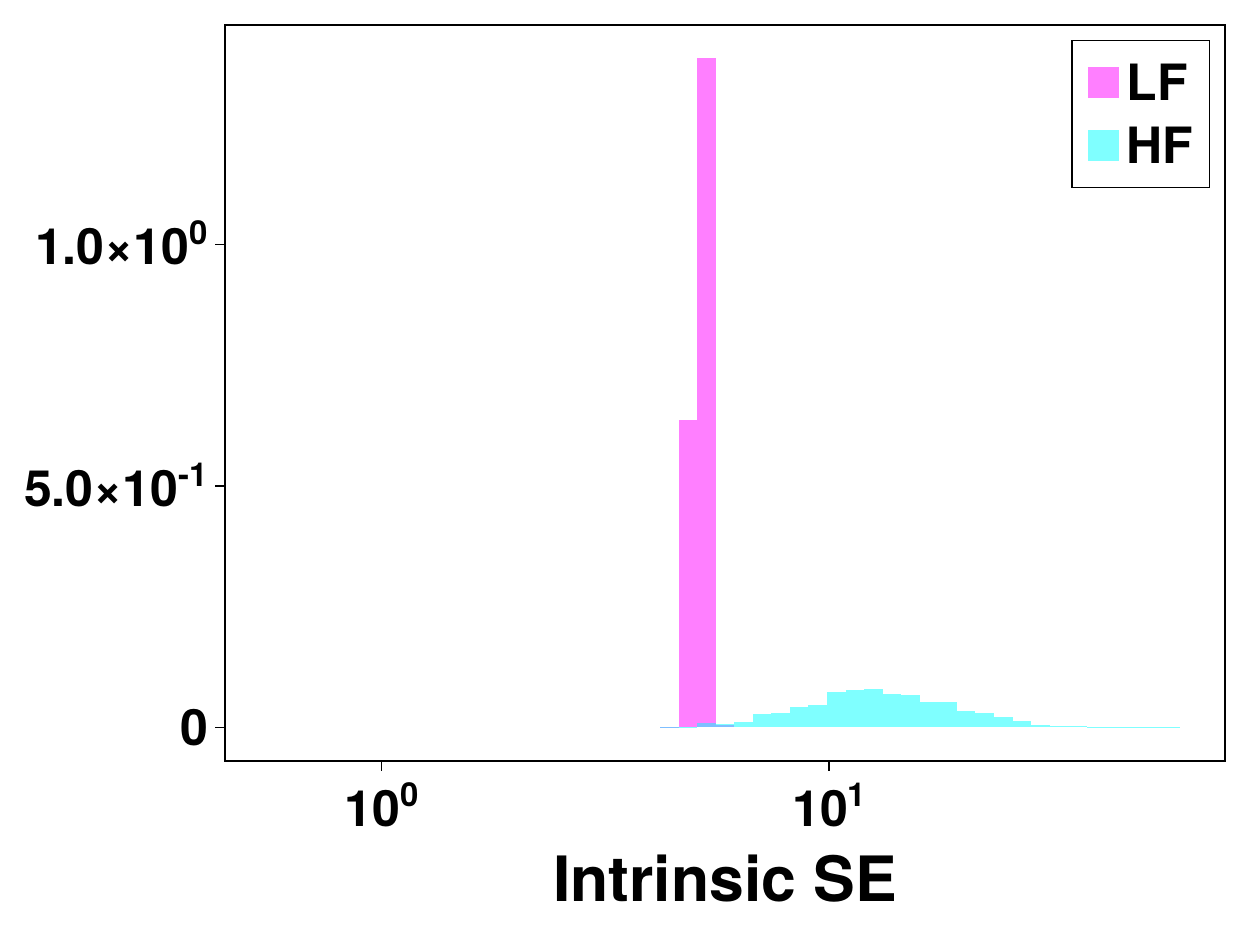}
        \subcaption*{Low-fidelity: 68\% decrease in intrinsic MSE}
    \end{subfigure}
    \begin{subfigure}[t]{0.32\linewidth}
        \includegraphics[width=\linewidth]{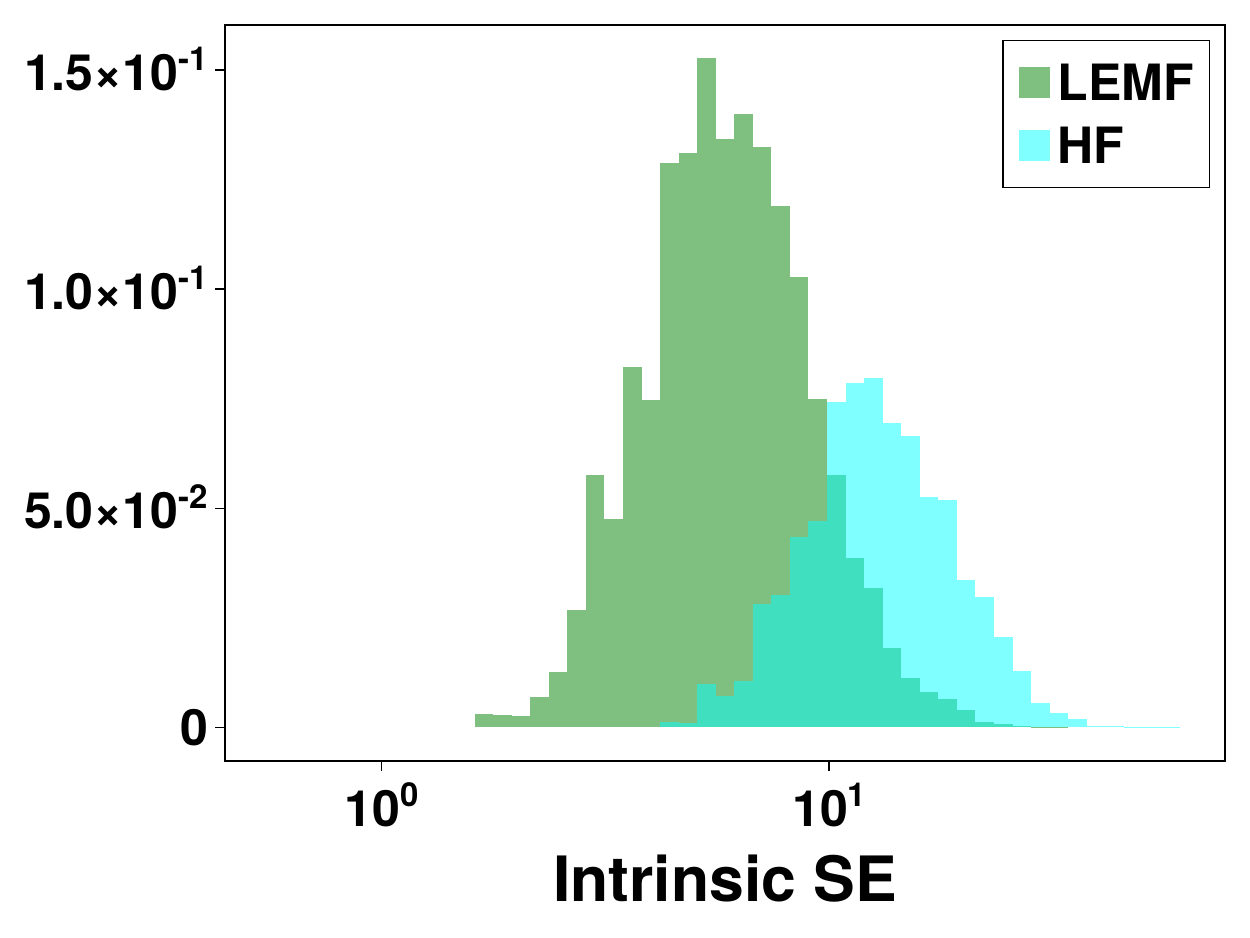}
        \subcaption*{LEMF: 51\% {decrease} in intrinsic MSE}
    \end{subfigure}
    \begin{subfigure}[t]{0.32\linewidth}
        \includegraphics[width=\linewidth]{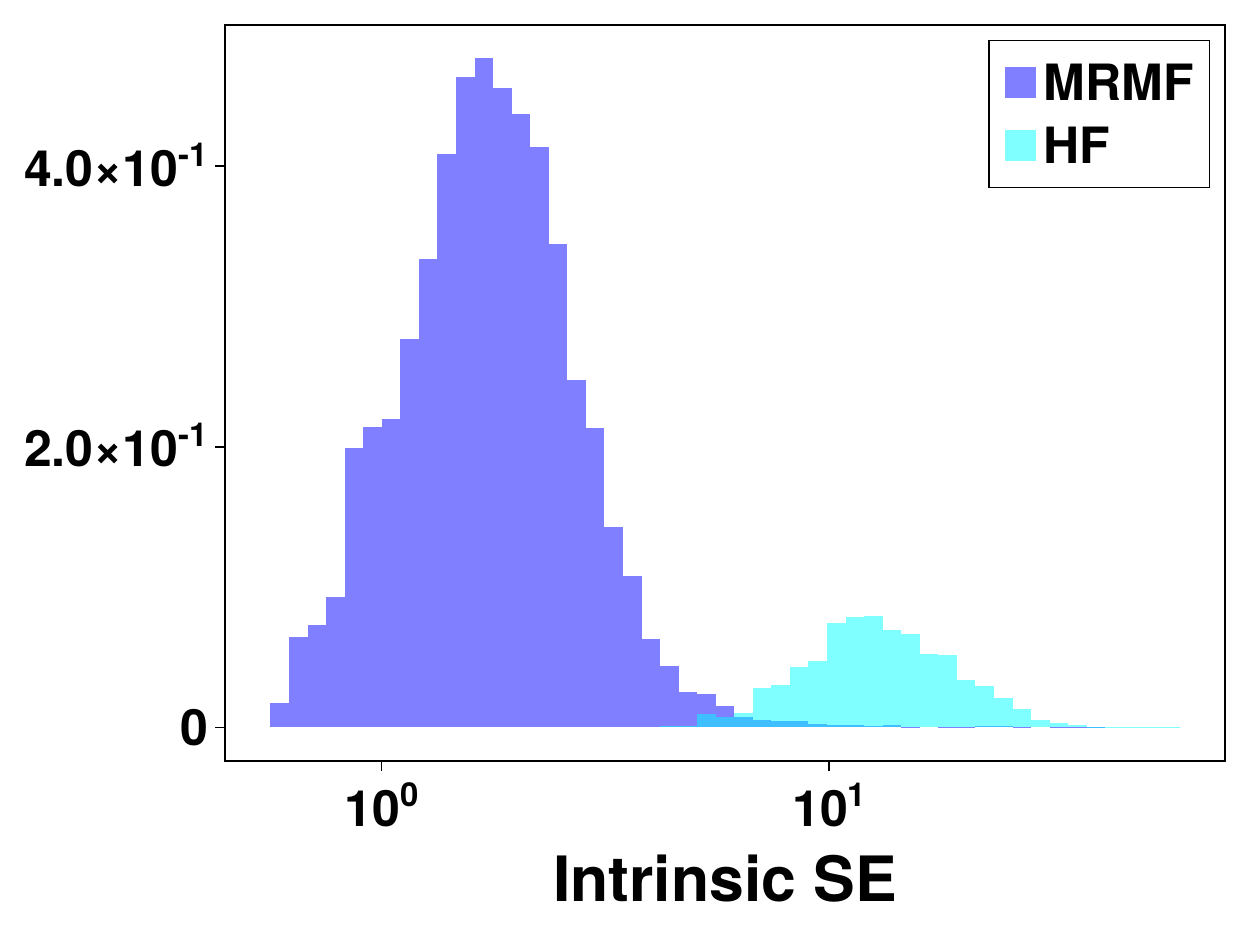}
        \caption*{Regression: 83\% decrease in intrinsic MSE}
    \end{subfigure}
    \\
    \begin{subfigure}[t]{0.32\linewidth}
        \includegraphics[width=\linewidth]{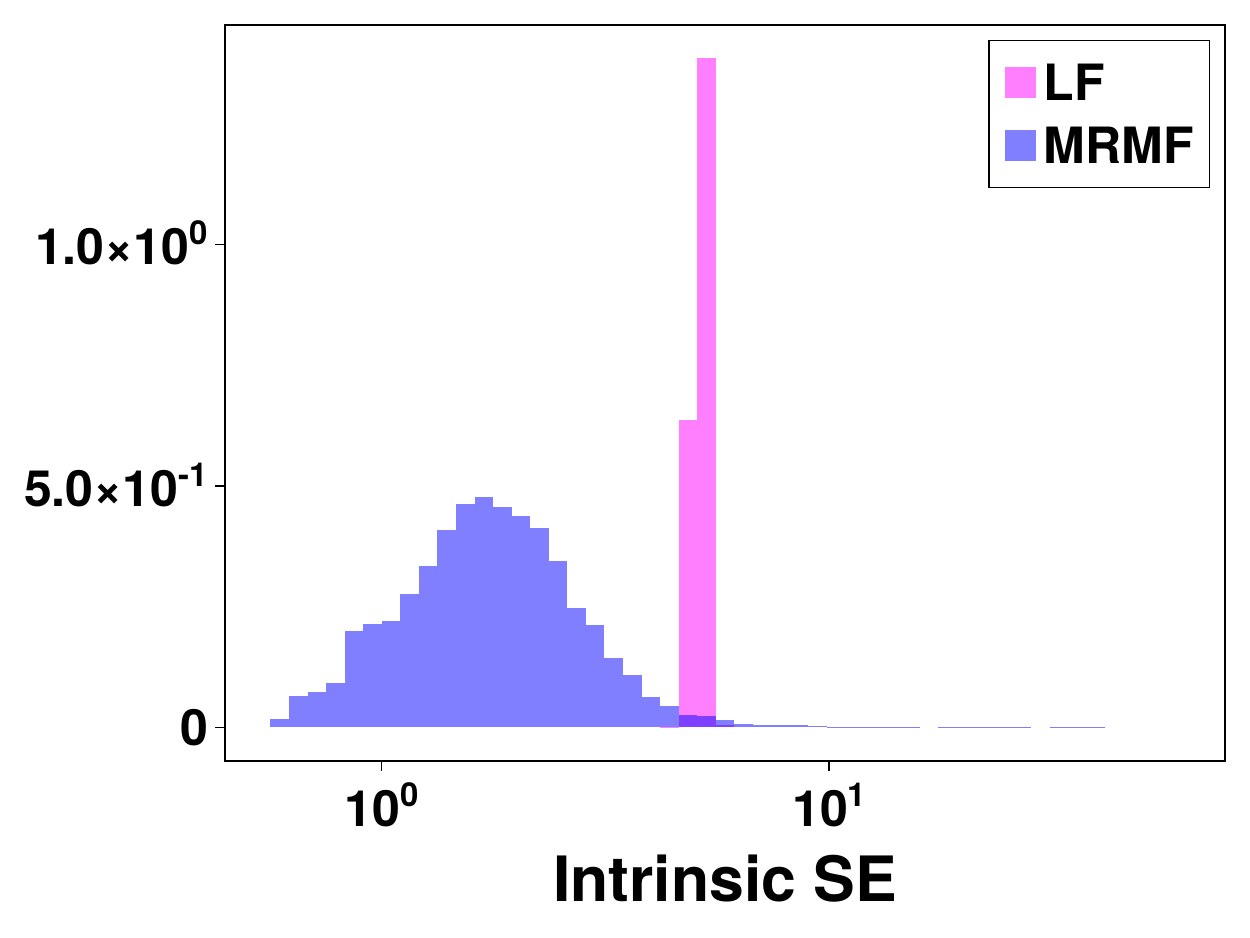}
    \end{subfigure}
    \begin{subfigure}[t]{0.32\linewidth}
        \includegraphics[width=\linewidth]{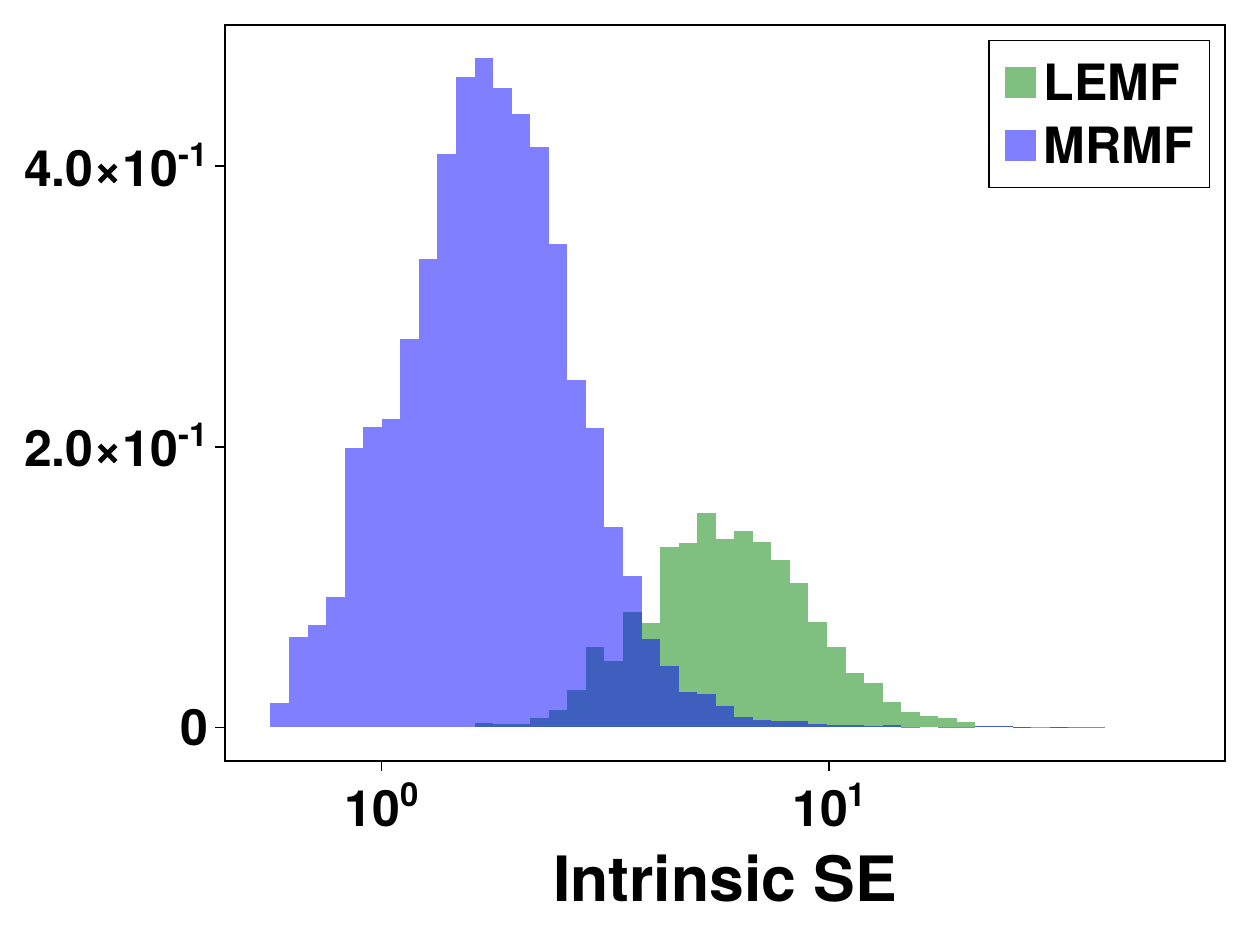}
    \end{subfigure}
    
    \caption{\textbf{Top:} Intrinsic squared error histograms corresponding to $\hat\Gamma_1^{\rm LF}$ (left), $\hat\Gamma_1^{\rm LEMF}$ (center), and $\hat\Gamma_1^{\rm MRMF}$ (right). Squared error histograms corresponding to $\hat\Gamma_1^{\rm HF}$ are overlaid in cyan on each plot for comparison. Changes in MSE are relative to $\hat\Gamma_1^{\rm HF}$. \textbf{Bottom:} Intrinsic squared error histograms of $\hat\Gamma_1^{\rm LF}$ (left) and $\hat\Gamma_1^{\rm LEMF}$ (right) compared to that of $\hat\Gamma_1^{\rm MRMF}$, overlaid in blue. Use of $\hat\Gamma_1^{\rm MRMF}$ results in a 49\% decrease in intrinsic MSE relative to $\hat\Gamma_1^{\rm LF}$ and a 66\% decrease relative to $\hat\Gamma_1^{\rm LEMF}$. We do not report results for $\hat\Gamma_{1}^{\rm EMF}$ in the intrinsic metric, which is only defined for SPD arguments, because 67\% of its  realizations are indefinite}
    \label{fig:Gamma1_AI_LE}
\end{figure}

In \cref{fig:Gamma1_fro,fig:Gamma1_AI_LE} we show squared error histograms corresponding to $\hat\Gamma_1^{\rm HF}$, $\hat\Gamma_1^{\rm LF}$, $\hat\Gamma_1^{\rm EMF}$, and $\hat\Gamma_1^{\rm MRMF}$. In general $\hat\Gamma_1^{\rm LF}$, $\hat\Gamma_1^{\rm EMF}$, $\hat\Gamma_1^{\rm LEMF}$, and $\hat\Gamma_1^{\rm MRMF}$ all yield substantial decreases in squared error relative to $\hat\Gamma_1^{\rm HF}$. In contrast to $\hat\Gamma_0^{\rm LEMF}$, $\hat\Gamma_1^{\rm LEMF}$ results in decreases, rather than increases, in MSE, relative to the high-fidelity-only estimator, but good performance in estimation of $\Gamma_1$ alone is not enough to ensure good estimates of $A_{\rm GMML}$. 
In a similar vein, the frequency with which $\hat\Gamma_1^{\rm EMF}$ is indefinite was only 67\%, a moderate decrease from the 82\% of $\hat\Gamma_0^{\rm EMF}$.
\clearpage 

\clearpage 
\section{SPD Product Manifolds}
\label{app:prodMan}
The product $\bbP_d^K = \bbP_d \times \cdots \times \bbP_d$ ($K$ times, $K\in \Z^+$) is a Riemannian manifold when equipped with tangent spaces and metric derived from $\bbP_d$. In this appendix we provide the relevant geometric and statistical definitions for $\bbP_d^K$, which in most cases follow directly from the properties of $\bbP_d$ discussed in \cref{sec:bg}.

\subsection{Geometry}
\label{app:prodMan_geom}
Let $\bfA = (A_1, \dots, A_K) \in \bbP_d^K$. The tangent space to $\bbP_d^K$ at $\bfA$ is 
\begin{equation*}
\rmT_\bfA \bbP_d^K = \mathrm{T}_{(A_1, \dots, A_K)}\bbP_d^K = \bigotimes_{k=1}^K \rmT_{A_k}\bbP_d. 
\end{equation*}
With this definition of tangent space, the Riemannian metric or \textbf{inner product} on $\bbP_d^K$ can be decomposed as follows: let $\bfU, \bfV \in \rmT_\bfA \bbP_d^K \subseteq \bbH_d^K$. We define $g_{\bfA}: \mathrm{T}_{\bfA}\bbP_d^K\times \mathrm{T}_{\bfA}\bbP_d^K \to \R$ via 
\begin{equation*}
g_{ \bfA }(\bfU, \bfV) = \langle \bfU, \; \bfV \rangle_\bfA = \sum_{k=1}^K \langle U_k,\; V_k\rangle_{A_k} = \sum_{k=1}^K \trace{U_kA_k\inv V_kA_k\inv}.
\end{equation*}
Corresponding to this inner product we have an \textbf{outer product} on $\rmT_\bfA\bbP_d^K$, 
\begin{equation*}
\bfU \otimes_{\bfA}\bfV = \begin{bmatrix} U_1 \\ \vdots \\ U_K \end{bmatrix} \otimes \begin{bmatrix}A_1\inv V_1 A_1\inv \\ \vdots \\ A_K\inv V_K A_K\inv\end{bmatrix},
\end{equation*}
where $\otimes$ is the Kronecker product and the result defines a linear mapping from $\rmT_\bfA\bbP_d^K$ to $\rmT_\bfA\bbP_d^K$. As in the $\bbP_d^1$ case, the trace of the $\bfA$ outer-product is equal to the $\bfA$ inner-product,
\begin{equation*}
\trace{\bfU \otimes_{\bfA} \bfV} = \langle \bfU,\bfV \rangle_{\bfA}. 
\end{equation*}
Bridging between $\bbP_d^K$ and $\rmT_\bfA\bbP_d^K$ we have the \textbf{logarithmic and exponential mappings} 
\begin{equation*}
\begin{aligned}
	\log_{\bfA}: \bbP_d^K \to \mathrm{T}_{\bfA}\bbP_d^K, \quad & \log_{\bfA}(\bfB) = (\log_{A_1}B_1,\dots, \log_{A_K} B_K) \\
	\exp_{\bfA}: \mathrm{T}_{\bfA}\bbP_d^K \to \bbP_d^K, \quad & \exp_{\bfA}(\bfX) = (\exp_{A_1} X_1,\dots, \exp_{A_K} X_K),
\end{aligned}
\end{equation*}
which are simply the logarithmic and exponential mappings on $\bbP_d^1$ applied elementwise to $\bfB \in \bbP_d^K$ and $\bfX \in \rmT_\bfA \bbP_d^K$ with the corresponding entries of $\bfA$.

Geodesics and distance on the product manifold $\bbP_d^K$ are given as follows:  
Let $\bfA, \bfB \in \bbP_d^K$.  If $\gamma_1(t), \dots, \gamma_K(t)$ are the geodesics from $A_1, \dots, A_K$ to $B_1, \dots, B_K$, respectively, on $\bbP_d$, then the \textbf{geodesic} from $\bfA $ to $\bfB$ on $\bbP_d^K$ is given by
\begin{equation}
	\begin{aligned}
		\gamma_{\bfA \to \bfB}(t) &= (\gamma_1(t),\dots, \gamma_K(t)) \\
		&= \left(A_1^{\frac{1}{2}}(A_1^{-\frac{1}{2}} B_1 A_1^{-\frac{1}{2}})^t A_1^{\frac{1}{2}},\dots, A_K^{\frac{1}{2}}(A_K^{-\frac{1}{2}} B_K A_K^{-\frac{1}{2}})^t A_K^{\frac{1}{2}} \right).
	\end{aligned}
	\label{eq:geodesicPnK}
\end{equation}

The \textbf{intrinsic distance} between $\bfA$ and $\bfB$ on $\bbP_d^K$ follows accordingly from \cref{eq:intrinsicdist},
\begin{equation*}
	d^2(\bfA, \bfB) = \sum_{k=1}^K d(A_k, B_k)^2
	= \sum_{k=1}^K \left\|\log(A_k^{-\frac{1}{2}}B_k A_k^{-\frac{1}{2}}) \right\|_F^2 
	= \sum_{k=1}^K \sum_{i=1}^d \log^2\lambda_i(A_k\inv B_k).
\end{equation*}

\subsection{Statistics}
\label{app:prodMan_stats}
Let $\bfS = (S_1, \dots, S_K)$ be a $\bbP_d^K$-valued random variable. In constructing our multifidelity covariance estimator we employ the following notions of statistics for $\bfS$ which are consistent with the general definitions in \cite{pennec2006intrinsic} but, due to the product manifold structure of $\bbP_d^K$ (\cref{app:prodMan_geom}), in many cases have nice decompositions to statistics on $\bbP_d^1$. 

To begin, the \textbf{expectation} of $\bfS$ is the element $\bfY \in \bbP_d^K$ which minimizes the expected squared distance to $\bfS$, 
\begin{equation}
\begin{aligned}
\bfE[S] &= \argmin_{\bfY \in \bbP_d^K} \E\left[d^2(\bfY, \bfS)\right]
= \argmin_{Y \in \bbP_d^K} \E\left[\sum_{k=1}^K d^2(Y_k, S_k) \right] 
= \argmin_{Y \in \bbP_d^K} \sum_{k=1}^K \E \left[d^2(Y_k, S_k)\right] \\
 &=  (\bfE[S_1], \dots, \bfE[S_K]) = (\Sigma_1, \dots, \Sigma_K) \equiv \bfSigma. 
\end{aligned}
\label{eq:prodman_mean}
\end{equation}
Because the squared distance between $\bfS$ and $\bfY$ decomposes into the sum of $K$ squared distances between the individual components of $\bfS$ and $\bfY$, to obtain the (Frechet) mean of $\bfS$ we simply take the Frechet mean of each component of $\bfS$. 

The \textbf{variance} of $\bfS$ is defined as the expected squared distance between $\bfS$ and its mean, 
\begin{equation*}
	\sigma^2_\bfS = \E[d^2(\bfS, \bfSigma)] = \E \left[ \sum_{k=1}^K d^2(S_k, \Sigma_k) \right] = \sum_{k=1}^K \sigma^2_{S_k},
\end{equation*}
where $\sigma^2_{S_k}$ is the variance of $S_k \in \bbP_d$, $k = 1, \dots, K$.  

As in the case of $\bbP_d^1$-valued random variables in \cref{sec:bg}, the \textbf{covariance} of $\bfS$ is the $\bfSigma$-outer-product of the ``vector difference'' $\log_\bfSigma \bfS$ with itself, 
\begin{equation*}
\Cov[\bfS] = \E[\log_\bfSigma \bfS \otimes_\bfSigma \log_\bfSigma \bfS] \equiv \Gamma_\bfS.
\end{equation*}
$\Gamma_\bfS$ is a symmetric positive semidefinite linear operator from $\bbH_d^K = \rmT_\bfSigma \bbP_d^K$ to $\rmT_\bfSigma \bbP_d^K = \bbH_d^k$. As on $\bbP_d^1$, we have $\trace{\Gamma_\bfS} = \sigma^2_\bfS$.  

Finally given $\bfS \sim (\Sigma, \Gamma_\bfS)$ \rev{with $\Gamma_\bfS$ invertible}, we define the \textbf{Mahalanobis distance} between $\bfS$ and a deterministic point $\bfY \in \bbP_d^K$
\begin{equation} 
d^2_\bfS(\bfY) = \langle \log_\bfSigma \bfY, \; \Gamma_\bfS\inv \log_\bfSigma \bfY \rangle_\bfSigma. 
\label{eq:prodMan}
\end{equation} 
The Mahalanobis distance is a $\Gamma_\bfS\inv$-weighted version of the intrinsic distance between $\bfSigma$ and $\bfY$, and, unless $\Gamma_\bfS$ is ``diagonal'' (in the appropriate sense), in general \textit{cannot} be decomposed into a sum of pairwise distances between $\Sigma_k$ and $Y_k$. The $\Gamma_\bfS\inv$ weighting introduces interaction between separate components of $\log_\bfSigma \bfY$.

\bibliographystyle{siamplain}
\makeatletter\@input{xx.tex}\makeatother
\bibliography{references}